\documentclass[11pt,a4paper]{article}%
\usepackage{amsfonts}
\usepackage{amsmath}
\usepackage{amssymb}
\usepackage{sectsty}
\usepackage[round,authoryear]{natbib}
\usepackage[margin=1.45in,a4paper]{geometry}
\usepackage[sc,center,compact,noindentafter,medium]{titlesec}
\usepackage[onehalfspacing,nodisplayskipstretch]{setspace}
\usepackage{graphicx}
\usepackage{paralist}
\usepackage[colorlinks,bookmarksnumbered]{hyperref}
\usepackage{array}
\usepackage{float}
\usepackage{latexsym}
\usepackage{epsfig}
\usepackage{sw20elba}
\usepackage{mathrsfs}
\usepackage{fancyhdr}%
\allowdisplaybreaks
\graphicspath{{C:/temp/}}
\newtheorem{theorem}{\normalfont\scshape Theorem}[section]
\newtheorem{corollary}{\normalfont\scshape Corollary}[section]
\newtheorem{proposition}{\normalfont\scshape Proposition}[section]
\newtheorem{lemma}{\normalfont\scshape Lemma}[section]
\newtheorem{condition}{\normalfont\scshape Assumption}

\newtheorem{remark}{\normalfont\scshape Remark}[section]

\expandafter
\def \expandafter \normalsize \expandafter{\normalsize \setlength \abovedisplayskip{10pt plus 2pt minus 7pt}}
\expandafter
\def \expandafter \normalsize \expandafter{\normalsize \setlength \abovedisplayshortskip{0pt plus 2pt}}
\expandafter
\def \expandafter \normalsize \expandafter{\normalsize \setlength \belowdisplayskip{10pt plus 2pt minus 7pt}}
\expandafter
\def \expandafter \normalsize \expandafter{\normalsize \setlength \belowdisplayshortskip{5pt plus 2pt minus 3pt}}

\numberwithin{equation}{section}
\pltopsep=\medskipamount
\begin{document}

\begin{center}
{\LARGE \textsc{bootstrap diagnostic tests}}%

%TCIMACRO{\TeXButton{Acknowledgement}{\renewcommand{\thefootnote}{}
%\footnote{
%\hspace{-7.2mm}
%\noindent$^{a}$ Department of Economics, University of Bologna, Italy.
%\newline$^{b}$ Department of Economics, University of Exeter, UK.
%\newline
%We thank for important comments and suggestions Aureo De Paula, Xavier D'Haultfoeuille, Samuel Engle, Patrik Guggenberger,
%Bruce Hansen, Jessie Li, Adam McCloskey, Simone Martinalli, Marcelo Moreira, Morten Nielsen, Guillaume Pouliot,
%Anders Rahbek, Jesse Shapiro, Michael Wolf, Edoardo Zanelli, seminar participants at ESSEC,
%LSE, Nova SBE, UCL, U Copenhagen, U Lisbon
%as well as participants at the Aarhus Workshop in Econometrics IV (U Aarhus), the 2025 Bristol Econometric Study Group
%Annual Conference, the `Econometrics in Rio' conference (FGV Rio, 2024), the 35th (EC)$^2$ Conference
%(University of Amsterdam), the 5th High Voltage Econometrics workshop (Collegio Carlo Alberto),
%the 11th Italian Congress of Econometrics and Empirical Economics (U Palermo)
%and the 2025 IAAE Conference (U Torino). We also thank Filippo Nardoni for research assistance.
%Cavaliere and Fanelli gratefully acknowledge financial support from the Italian Ministry of University and Research
%(PRIN 2020 Grant 2020B2AKFW and PRIN 2022 Grant 20229PFAX5).
%\newline Correspondence to: Giuseppe Cavaliere, Department of
%Economics, University of Bologna; email: giuseppe.cavaliere@unibo.it.
%}
%\addtocounter{footnote}{-1}
%\renewcommand{\thefootnote}{\arabic{footnote}}}}%
%BeginExpansion
\renewcommand{\thefootnote}{}
\footnote{
\hspace{-7.2mm}
\noindent$^{a}$ Department of Economics, University of Bologna, Italy.
\newline$^{b}$ Department of Economics, University of Exeter, UK.
\newline
We thank for important comments and suggestions Aureo De Paula, Xavier D'Haultfoeuille, Samuel Engle, Patrik Guggenberger,
Bruce Hansen, Jessie Li, Adam McCloskey, Simone Martinalli, Marcelo Moreira, Morten Nielsen, Guillaume Pouliot,
Anders Rahbek, Jesse Shapiro, Michael Wolf, Edoardo Zanelli, seminar participants at ESSEC,
LSE, Nova SBE, UCL, U Copenhagen, U Lisbon
as well as participants at the Aarhus Workshop in Econometrics IV (U Aarhus), the 2025 Bristol Econometric Study Group
Annual Conference, the `Econometrics in Rio' conference (FGV Rio, 2024), the 35th (EC)$^2$ Conference
(University of Amsterdam), the 5th High Voltage Econometrics workshop (Collegio Carlo Alberto),
the 11th Italian Congress of Econometrics and Empirical Economics (U Palermo)
and the 2025 IAAE Conference (U Torino). We also thank Filippo Nardoni for research assistance.
Cavaliere and Fanelli gratefully acknowledge financial support from the Italian Ministry of University and Research
(PRIN 2020 Grant 2020B2AKFW and PRIN 2022 Grant 20229PFAX5).
\newline Correspondence to: Giuseppe Cavaliere, Department of
Economics, University of Bologna; email: giuseppe.cavaliere@unibo.it.
}
\addtocounter{footnote}{-1}
\renewcommand{\thefootnote}{\arabic{footnote}}%
%EndExpansion
{\normalsize \vspace{0.1cm} }

{\large \textsc{Giuseppe Cavaliere}}$^{a,b}${\large \textsc{, Luca Fanelli}%
}$^{a}${\large \textsc{ and Iliyan Georgiev}}$^{a}${\normalsize \vspace
{0.2cm}\vspace{0.2cm}}

\bigskip

September 30th, 2025{\normalsize \vspace{0.2cm}\vspace{0.2cm}}

\bigskip

\vspace{-0.15cm}
\end{center}

Violation of the assumptions underlying classical (Gaussian) limit theory
often yields unreliable statistical inference. This paper shows that the
bootstrap can detect such violations by delivering simple and powerful
diagnostic tests that (a) induce no pre-testing bias, (b) use the same
critical values across applications, and (c) are consistent against deviations
from asymptotic normality. The tests compare the conditional distribution of a
bootstrap statistic with the Gaussian limit implied by valid specification and
assess whether the resulting discrepancy is large enough to indicate failure
of the asymptotic Gaussian approximation. The method is computationally
straightforward and only requires a sample of i.i.d. draws of the bootstrap
statistic. We derive sufficient conditions for the randomness in the data to
mix with the randomness in the bootstrap repetitions in a way such that (a),
(b) and (c) above hold. We demonstrate the practical relevance and broad
applicability of bootstrap diagnostics by considering several scenarios where
the asymptotic Gaussian approximation may fail, including weak instruments,
non-stationarity, parameters on the boundary of the parameter space, infinite
variance data and singular Jacobian in applications of the delta method. An
illustration drawn from the empirical macroeconomic literature
concludes.\bigskip

\medskip\noindent\textsc{Keywords}:\ Bootstrap inference; Pre-testing bias;
Random bootstrap measures.\bigskip

\begin{center}

\newpage
\end{center}

\section{introduction}

\label{Sec intro}

\textsc{Consider a standardized estimator} $T_{n}:=n^{1/2}(\hat{\theta}%
_{n}-\theta_{0})/\hat{\sigma}_{n}$ based on a data sample $D_{n}$, where
$\hat{\sigma}_{n}^{2}$ is an estimator of the asymptotic variance. Classical
asymptotic theory is usually based on a set of assumptions guaranteeing that,
in large samples, the distribution of $T_{n}$ is well-approximated, to the
first order, by some standard distribution, usually the normal one. That is,
$T_{n}\overset{d}{\rightarrow}Z$, $Z\sim%
%TCIMACRO{\TeXButton{N}{\mathscr{N}\!}}%
%BeginExpansion
\mathscr{N}\!%
%EndExpansion
(0,1)$, with `$\overset{d}{\rightarrow}$' denoting convergence in
distribution. When $\hat{\theta}_{n}$ is an extremum estimator, provided the
objective function can be expanded around a (pseudo-~)~true value $\theta_{0}%
$, such assumptions are usually related to (i) existence of moments, (ii)
stationarity and ergodicity, (iii) non-singularity of the Hessian (or, for
transformations of the original estimator, full rank of the implied Jacobian),
(iv) (pseudo-) true parameter in the interior of the parameter space; see,
e.g., Newey and McFadden (1994). For extremum estimators based on instrumental
variables [IV], (v) assumptions on the strength of the instruments are usually
required. Examples of estimators requiring assumptions such as (i)--(v) to
hold are, inter alia, (quasi) ML estimators, GMM estimators, nonlinear least
squares estimators and minimum distance estimators. With `valid specification'
we mean that such assumptions are met.

The detection of invalid specifications is crucial in applications. A key
challenge to proper statistical tests of valid specification is that they
typically induce a `pre-testing' bias in subsequent inferences. While in some
cases the pre-testing bias may be associated with conservative inference under
the null hypothesis (see, e.g., de Chaisemartin and D'Haultfoeuille, 2024),
tests conditional on the non-rejection of correct specification may be
severely oversized, even asymptotically.

In this paper, we show the novel result that the bootstrap delivers, as a
by-product, consistent (diagnostic) tests of invalid specification which do
not induce any pre-testing bias into subsequent inference procedures when the
null of valid specification is not rejected. That is, post-test (conditional)
inference is asymptotically \emph{exact} when conditioning is upon the
bootstrap tests not rejecting the null hypothesis of valid specification.

Our approach starts from the observation that -- usually under mild additional
requirements -- a bootstrap analog of $T_{n}$, say $T_{n}^{\ast}:=n^{1/2}%
(\hat{\theta}_{n}^{\ast}-\hat{\theta}_{n})/\hat{\sigma}_{n}$ or $T_{n}^{\ast
}:=n^{1/2}(\hat{\theta}_{n}^{\ast}-\hat{\theta}_{n})/\hat{\sigma}_{n}^{\ast}$,
will also be asymptotically normal under valid specification; i.e.,
$T_{n}^{\ast}\overset{d^{\ast}}{\rightarrow}_{p}Z$, with `$\overset{d^{\ast}%
}{\rightarrow}_{p}$' denoting convergence in distribution conditionally on
$D_{n}$; see Appendix~\ref{Appendix notation} for a formal definition. In
contrast, under invalid specification, $T_{n}^{\ast}$ may no longer be
asymptotically normal; rather, it usually has a non-Gaussian, random limiting
distribution; see Cavaliere and Georgiev (2020) and the references therein.
For instance, randomness of the limiting bootstrap measure may arise when (i)
the score contributions have infinite variance (Athreya, 1987; Knight, 1989;
Cavaliere, Georgiev and Taylor, 2016), (ii) when the data are non-stationary
(Basawa, Mallik, McCormick and Taylor, 1991; Cavaliere, Nielsen and Rahbek,
2015), (iii) when the Hessian or the Jacobian are (near-) rank deficient
(Datta, 1995; Angelini, Cavaliere and Fanelli, 2022; see also Han and
McCloskey, 2019), (iv) when the (pseudo-) true value lies near or on the
boundary of the parameter space (Andrews, 2000); (v) for IV\ estimators, when
the instruments are weak or irrelevant (see Section~\ref{Sec Example weak IV}).

This different asymptotic behavior of the bootstrap distribution of
$T_{n}^{\ast}$ can be exploited to detect invalid specification. To see why,
consider the discrepancy between the cumulative distribution function [cdf] of
$T_{n}^{\ast}$ conditional on $D_{n}$, say $\hat{G}_{n}$, and the limiting
standard Gaussian cdf $\Phi$, measured as $\hat{d}_{n}:=\parallel\hat{G}%
_{n}-\Phi\parallel$, where $\parallel\cdot\parallel$ is a user-chosen norm or
seminorm on the space of distribution functions. The distance $\hat{d}_{n}$ is
expected to shrink to zero at a specific rate when asymptotic normality holds,
while converging to a random limit when the assumptions fail to hold. Hence, a
simple test of valid specification could assess whether the realized
discrepancy $\hat{d}_{n}$ is large enough to reject the null.

As is standard in the bootstrap literature, $\hat{d}_{n}$ can be treated as a
known function of the data $D_{n}$, since $\hat{G}_{n}$ can be approximated
with any desired precision via the empirical distribution function [edf]
$\hat{G}_{n,m}^{\ast}$ of $m$ simulated realizations of $T_{n}^{\ast}$, with
$m$ arbitrarily large; that is, for any fixed $n$, as $m\rightarrow\infty$,
with probability one, $\hat{G}_{n,m}^{\ast}\rightarrow\hat{G}_{n}$ uniformly,
and hence $\hat{d}_{n,m}^{\ast}:=\lVert\hat{G}_{n,m}^{\ast}-\Phi
\rVert\rightarrow\hat{d}_{n}$ a.s. if $\parallel\cdot\parallel$ is continuous
with respect to uniform convergence.

While it could be tempting to construct diagnostic tests based on $\hat{d}%
_{n}$ (or a properly normalized version, such as $\sqrt{n}\hat{d}_{n}$), such
tests would suffer from at least two important drawbacks. First, their
large-$n$ asymptotic properties would depend on the particular bootstrap
application of interest, and would often be very difficult to derive, as they
require studying higher-order asymptotic expansions of $\hat{G}_{n}$. Second,
since $\hat{d}_{n}$ is a function of the data, these tests might give rise to
a pre-testing bias.

We show that these drawbacks disappear if inference, rather than being based
on $\hat{d}_{n}$, is based on its approximation $\hat{d}_{n,m}^{\ast}$, where
$m$ and $n$ diverge \emph{jointly }under the requirement that $m$ cannot be
too large when $n$ is finite. This asymptotic regime differs from the standard
sequential bootstrap asymptotics, where $m\rightarrow\infty$ first (such that
$\hat{G}_{n,m}^{\ast}-\hat{G}_{n}\approx0$) followed by $n\rightarrow\infty$
(such that $\hat{G}_{n}-\Phi\approx0$); see Andrews and Buchinsky (2000). In
particular, we show that when $m$ and $n$ diverge jointly, with $m$ diverging
at a proper rate relative to $n$, a test with a known asymptotic distribution
and based on the bootstrap statistic $\hat{d}_{n,m}^{\ast}$ can be designed to
assess specification invalidity. Moreover, this approach is computationally
straightforward, as it just requires to use the set of $m$ bootstrap
repetitions to compute $\hat{d}_{n,m}^{\ast}$. Put differently, it is
equivalent to the application of distance-based normality tests to the set of
$m$ bootstrap repetitions, with different normality tests corresponding to
different choices of the employed \mbox{(semi)}norm $\parallel\cdot\parallel$.
Finally, it can be performed using the same critical values in a broad range
of applications, and\ consistently detects deviations from asymptotic Gaussianity.

The role of the rate condition on $m$ relative to $n$ is to ensure that, under
the null hypothesis of valid specification, the test decision becomes, in the
limit, stochastically independent of the original data. This fact guarantees
that a bootstrap test based on $\hat{d}_{n,m}^{\ast}$ does not induce
pre-testing bias in large samples, in contrast to standard pre-tests for
specification (in)validity (e.g., tests of finite variance, stationarity
tests, pre-tests on the instrument strengths, and so forth). Instead, under a
set of relevant alternatives the\textbf{ }test is consistent as it exploits,
in the limit, the information in the data alone. In summary, according to the
validity of either the null or an alternative, the data choose asymptotically
between acceptance based on an independent random device or rejection with
probability approaching one. Interestingly, in a recent paper, de Chaisemartin
and D'Haultf\oe uille (2024)\ show that certain specification
tests\footnote{De Chaisemartin and D'Haultf\oe uille define `valid
specification'\ as the scenario in which the probability distribution
generating $D_{n}$ is such that a target estimator is consistent and
asymptotically normal. They consider pre-tests that can detect when such an
estimator becomes inconsistent. This definition differs from the one employed
here, where `valid specification'\ means that the underlying conditions
ensuring the applicability of standard asymptotic inference are satisfied in
the estimated model.}, when used as pre-tests, lead to conservative post-test
inference under the null of valid specification. We complement their result by
showing that the bootstrap delivers tests of specification validity leading to
\emph{exact }post-test inference as the sample size diverges.

This diagnostic potential of the bootstrap has not been developed in the
extant literature. Beran (1997) was the first to suggest examining the
bootstrap distribution to diagnose bootstrap failure. Davidson (2017) proposes
simulation-based diagnostics in order to determine when a given bootstrap
procedure works well or not in finite samples. B\aa rdsen and Fanelli (2015)
provide prima facie evidence that non-Gaussian bootstrap distributions may be
linked to weak identification in~DSGE~models; see also Angelini, Cavaliere and
Fanelli (2022, 2024), and Zhan (2018). Within the problem of statistical
reporting in a Bayesian communication framework, Andrews and Shapiro (2025)
show that the (Bayesian) bootstrap distribution can serve as a surrogate
posterior, and propose comparing it with the Gaussian approximation, using
distance measures such as the signed Kolmogorov metric, to assess whether the
conventional report adequately conveys uncertainty. A related approach is
found in Wang (2025), who proposes detecting violations of asymptotic
normality for GMM and extremum estimators by comparing quasi-Bayesian
posterior distributions with the Gaussian benchmark. Our contribution extends
this literature by demonstrating how distances between bootstrap and Gaussian
distributions can be used to construct formal specification tests that avoid
pre-testing bias. In terms of econometric theory, a further novelty lies in
the asymptotic regime we adopt, where both the sample size $n$ and the number
of bootstrap repetitions $m$ pass to infinity simultaneously. This setting is
rarely considered in the literature; a notable exception is Andrews and
Buchinsky (2000), who exploit this joint asymptotic regime to guide the choice
of $m$ in applied work.

To demonstrate the practical relevance and broad applicability of the
bootstrap in the detection of specification invalidity, we discuss its use in
the five scenarios (i)--(v) mentioned above. As regards case (i) of possible
weak instruments, we also present an empirical illustration where we revisit,
through the lens of bootstrap diagnostics, K\"{a}nzig's (2021) empirical
strategy for identifying the macroeconomic effects of a structural oil supply
news shock.

\paragraph*{structure of the paper.}

The paper is organized as follows. In Section~\ref{Sec Example weak IV} we
introduce a running example based on instrumental variable estimation.
Section~\ref{Sec main} contains our general results.
Section~\ref{Sec post diagnostics inference} establishes the key result that
the bootstrap procedure induces no pre-test bias in large samples. Additional
results and extensions are reported in Section~\ref{Sec extensions}, while
Section~\ref{Sec Examples} illustrates four further applications. An empirical
example is presented in Section~\ref{Sec Empirical}, and
Section~\ref{Sec conc} concludes. Notation and definitions used throughout the
paper can be found in Appendix~\ref{Appendix notation}. Proofs are collected
in Appendices~\ref{Appendix main proofs} and~\ref{Appendix additional proofs}.

\section{an example based on instrumental variables}

\label{Sec Example weak IV}

Consider the following linear IV\ regression with one endogenous regressor:%
\[
y_{i}=\beta x_{i}+u_{i}\text{, }x_{i}=\pi^{\prime}z_{i}+v_{i}%
\]
where the $k\times1$ vector of instruments $z_{i}$ is non-stochastic and the
errors $(u_{i},v_{i})^{\prime}$ are i.i.d. with mean zero and
variance-covariance matrix $\Sigma$; to simplify, the diagonal elements of
$\Sigma$ are set to $\sigma_{u}^{2}=\sigma_{v}^{2}=1$ and the off-diagonal
elements to $\rho_{uv}\in(0,1)$. Without loss of generality, we also set
$S_{zz}=I_{k}$, using the generic notation $S_{ab}:=n^{-1}\sum_{i=1}^{n}%
a_{i}b_{i}^{\prime}$. Given a sample of $n$ observations, the 2SLS estimator
of $\beta$ is $\hat{\beta}_{n}:=S_{xx.z}^{-1}S_{xy.z}=(S_{xz}S_{zy}%
)/(S_{xz}S_{zx})$.

Consider further a (Gaussian) parametric bootstrap where the instruments are
fixed in the bootstrap world. The bootstrap data are generated as
\[
y_{i}^{\ast}=\hat{\beta}_{n}x_{i}^{\ast}+u_{i}^{\ast}\text{, }x_{i}^{\ast
}=\hat{\pi}_{n}^{\prime}z_{i}+v_{i}^{\ast}%
\]
where $\hat{\pi}_{n}:=S_{zx}$ is the OLS\ estimator from the (first-stage)
regression of $x_{i}$ on $z_{i}$. For simplicity, assume that $\Sigma$ is
known, such that $(u_{i}^{\ast},v_{i}^{\ast})^{\prime}$ can be taken as i.i.d.
$%
%TCIMACRO{\TeXButton{N}{\mathscr{N}\!}}%
%BeginExpansion
\mathscr{N}\!%
%EndExpansion
\left(  0,\Sigma\right)  $ conditionally on the data.

Let first $\pi\neq0$ be fixed (i.e., the instruments be `strong'), a case that
we label a `valid specification'. Then, under standard assumptions for the
central limit theorem (CLT),%
\[
T_{n}:=\sqrt{n}\frac{\hat{\beta}_{n}-\beta}{\omega}=\frac{\sqrt{n}\pi^{\prime
}S_{zu}}{(\pi^{\prime}\pi)^{1/2}}+o_{p}(1)\overset{d}{\rightarrow}%
%TCIMACRO{\TeXButton{N}{\mathscr{N}\!}}%
%BeginExpansion
\mathscr{N}\!%
%EndExpansion
(0,1)
\]
where $\omega^{2}:=(\pi^{\prime}\pi)^{-1}$. Moreover, as $\hat{\pi}%
_{n}^{\prime}\hat{\pi}_{n}\rightarrow_{p}\pi^{\prime}\pi\neq0$, conditionally
on the data,
\begin{equation}
T_{n}^{\ast}:=\sqrt{n}\frac{\hat{\beta}_{n}^{\ast}-\hat{\beta}_{n}}%
{\hat{\omega}_{n}}=\frac{\sqrt{n}\hat{\pi}_{n}^{\prime}S_{zu^{\ast}}}%
{(\hat{\pi}_{n}^{\prime}\hat{\pi}_{n})^{1/2}}+O_{p}^{\ast}(n^{-1/2})\sim%
%TCIMACRO{\TeXButton{N}{\mathscr{N}\!}}%
%BeginExpansion
\mathscr{N}\!%
%EndExpansion
(0,1)+o_{p}^{\ast}(1), \label{eq bootstrap IV stat under strong IV}%
\end{equation}
with $\hat{\omega}_{n}^{2}:=(\hat{\pi}_{n}^{\prime}\hat{\pi}_{n})^{-1}$.
Therefore $T_{n}^{\ast}\overset{d^{\ast}}{\rightarrow}_{p}%
%TCIMACRO{\TeXButton{N}{\mathscr{N}\!}}%
%BeginExpansion
\mathscr{N}\!%
%EndExpansion
(0,1)$.

Instead, suppose next that the instruments are weak as in Staiger and Stock
(1997), i.e., $\pi=\lambda n^{-1/2}$ for some vector $\lambda$; see also
Bound, Jaeger and Baker (1995). This is an instance of an invalid
specification. Assuming that
\[
\sqrt{n}\binom{S_{zu}}{S_{zv}}=n^{-1/2}\sum_{i=1}^{n}\binom{u_{i}}{v_{i}%
}\otimes z_{i}\overset{d}{\rightarrow}\zeta:=(\zeta_{u}^{\prime},\zeta
_{v}^{\prime})^{\prime}%
\]
with $\zeta$ a zero-mean Gaussian vector with variance-covariance matrix
$\Sigma\otimes I_{k}$, we have
\begin{equation}
\hat{\beta}_{n}-\beta=\frac{S_{xz}S_{zu}}{S_{xz}S_{zx}}=\frac{\left(  \sqrt
{n}\pi+\sqrt{n}S_{zv}\right)  ^{\prime}\sqrt{n}S_{zu}}{\left(  \sqrt{n}%
\pi+\sqrt{n}S_{zv}\right)  ^{\prime}\left(  \sqrt{n}\pi+\sqrt{n}S_{zv}\right)
}\overset{d}{\rightarrow}\xi(\lambda,\zeta):=\frac{\left(  \lambda+\zeta
_{v}\right)  ^{\prime}\zeta_{u}}{\left(  \lambda+\zeta_{v}\right)  ^{\prime
}(\lambda+\zeta_{v})}. \label{eq IV betahat-betaweak case}%
\end{equation}
Similarly, the bootstrap analog of $\hat{\beta}_{n}-\beta$ can be written as%
\[
\hat{\beta}_{n}^{\ast}-\hat{\beta}_{n}=\frac{S_{x^{\ast}z}S_{zu^{\ast}}%
}{S_{x^{\ast}z}S_{zx^{\ast}}}=\frac{\left(  \sqrt{n}\hat{\pi}_{n}+\sqrt
{n}S_{zv^{\ast}}\right)  ^{\prime}\sqrt{n}S_{zu^{\ast}}}{\left(  \sqrt{n}%
\hat{\pi}_{n}+\sqrt{n}S_{zv^{\ast}}\right)  ^{\prime}\left(  \sqrt{n}\hat{\pi
}_{n}+\sqrt{n}S_{zv^{\ast}}\right)  }%
\]
where, conditionally on the data, $n^{1/2}(S_{zu^{\ast}}^{\prime},S_{zv^{\ast
}}^{\prime})^{\prime}\sim\zeta^{\ast}$, with $\zeta^{\ast}:=(\zeta_{u}%
^{\ast\prime},\zeta_{v}^{\ast\prime})^{\prime}$ distributed as $\zeta$. Since,
jointly with (\ref{eq IV betahat-betaweak case}), $\sqrt{n}\hat{\pi}%
_{n}\rightarrow_{d}\ell:=\lambda+\zeta_{v}$, we find using arguments as in
Cavaliere and Georgiev (2020, proof of Theorem 4.1) that $\hat{\beta}%
_{n}^{\ast}-\hat{\beta}_{n}\overset{d^{\ast}}{\rightarrow}_{w}\xi(\ell
,\zeta^{\ast})|\ell$ and\ that
\begin{equation}
T_{n}^{\ast}:=\sqrt{n}\frac{\hat{\beta}_{n}^{\ast}-\hat{\beta}_{n}}%
{\hat{\omega}_{n}}=\frac{\hat{\beta}_{n}^{\ast}-\hat{\beta}_{n}}{\hat{\omega
}_{n}/\sqrt{n}}\overset{d^{\ast}}{\rightarrow}_{w}\frac{\xi(\ell,\zeta^{\ast
})}{\sqrt{\ell^{\prime}\ell}}%
%TCIMACRO{\TeXButton{\big|}{\big|}}%
%BeginExpansion
\big|%
%EndExpansion
\ell; \label{eq IV weak instrument bootstrap}%
\end{equation}
see Appendix~\ref{Appendix notation} for the definition of $\overset{d^{\ast}%
}{\rightarrow}_{w}$. Hence, the bootstrap distribution has a random limit and,
as $n\rightarrow\infty$, the bootstrap cdf satisfies $\hat{G}_{n}%
(x)\rightarrow_{w}%
%TCIMACRO{\TeXButton{G}{\mathscr{G}\!}}%
%BeginExpansion
\mathscr{G}\!%
%EndExpansion
(x):=\mathbb{P(}\xi(\ell,\zeta^{\ast})/\sqrt{\ell^{\prime}\ell}\leq x|\ell)$,
$x\in\mathbb{R}$, on $\mathscr{D}_{\mathbb{R}}$, where the limit differs from
the Gaussian cdf.

\begin{figure}[t]
\makebox[\textwidth][c]{
\includegraphics[width=1.25\textwidth]{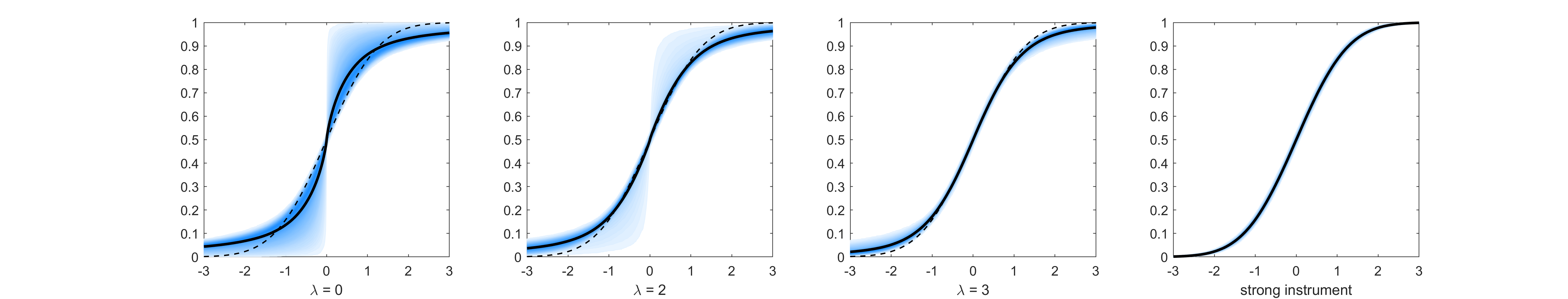} }
\vspace{-20pt}\caption{Fan chart of $M=1{,}000$ i.i.d.\ realizations of the
(conditional) cdf $\hat{G}_{n}$ of $T_{n}^{\ast}$ under weak and strong
instruments ($n = 1{,}000$).}%
\label{figure_fan_chart}%
\end{figure}

\begin{remark}
\label{Remark simulated Ghat}The limiting randomness of the bootstrap
distribution $\hat{G}_{n}$ under weak instruments is illustrated in Figure
\ref{figure_fan_chart}, where we report the fan chart of $M=1,000$ i.i.d.
realizations of $\hat{G}_{n}$ for $k=1$, $n=1,000$ and using a standard
Gaussian parametric bootstrap with $z_{i}=1$ and $\rho_{uv}=0.9$. The
randomness and non-normality in the bootstrap cdf $\hat{G}_{n}$, quite evident
for small values of $\lambda$, ameliorates as $\lambda$ increases and (up to
sampling error due to finite $n$) disappears in the strong-instrument case
where $\pi\neq0$ is fixed.\hfill$\square$
\end{remark}

\begin{remark}
Notice that under weak instruments the bootstrap does not replicate the
asymptotic distribution of the original statistic, given by $\xi(\lambda
,\zeta)$ of (\ref{eq IV betahat-betaweak case}), essentially because in the
limiting bootstrap experiment $\lambda$ is replaced by the random vector
$\ell$, where $\ell\neq\lambda$ with probability one. This result explains the
inconsistency of the bootstrap in the weak IV\ framework, as previously
documented in the literature (see, e.g., Davidson and MacKinnon, 2010), in
terms of randomness of the limit bootstrap measure.\hfill$\square$
\end{remark}

\begin{remark}
\label{Remark IV nonparametric bs}The results in this section do not
substantially change if $z_{i}$ is random and resampled along with
$\varepsilon_{i}^{\ast}$ and $u_{i}^{\ast}$, and/or if $\{\varepsilon
_{i}^{\ast},u_{i}^{\ast}\}_{i=1}^{n}$ are i.i.d. draws from the (centered)
residuals $\{\hat{\varepsilon}_{i},\hat{u}_{i}\}_{i=1}^{n}$ as in standard
non-parametric bootstrap designs (in the latter case, the random asymptotic
distribution in (\ref{eq IV weak instrument bootstrap}) is more
involved).\hfill$\square$
\end{remark}

\section{testing specification validity}

\label{Sec main}

Assume that $\theta\in\mathbb{R}$ and consider a bootstrap statistic
$T_{n}^{\ast}$ that is asymptotically $%
%TCIMACRO{\TeXButton{N}{\mathscr{N}\!}}%
%BeginExpansion
\mathscr{N}\!%
%EndExpansion
(0,1)$ under valid specification. $T_{n}^{\ast}$ can be, e.g., of the form
$T_{n}^{\ast}:=(\hat{\theta}_{n}^{\ast}-\hat{\theta}_{n})/\operatorname*{se}%
(\hat{\theta}_{n}^{\ast})$ or $T_{n}^{\ast}:=(\hat{\theta}_{n}^{\ast}%
-\hat{\theta}_{n})/\operatorname*{se}(\hat{\theta}_{n})$ (non-studentized
statistics of the form $T_{n}^{\ast}:=\sqrt{n}(\hat{\theta}_{n}^{\ast}%
-\hat{\theta}_{n})$ can also be considered). As we shall see in this section,
a test of valid specification, which does not induce pre-testing bias, can be
obtained by assessing whether the bootstrap replicates the Gaussian asymptotic
distribution. As an initial step, we quantify the discrepancy between the
bootstrap cdf and the asymptotic Gaussian cdf.

\subsection{preliminaries}

Let $\hat{G}_{n}(\cdot):=\mathbb{P}^{\ast}(T_{n}^{\ast}\leq\cdot)$ be the cdf
of $T_{n}^{\ast}$, conditional on the data $D_{n}$. We consider measuring the
discrepancy between $\hat{G}_{n}$ and the $%
%TCIMACRO{\TeXButton{N}{\mathscr{N}\!}}%
%BeginExpansion
\mathscr{N}\!%
%EndExpansion
(0,1)$ cdf $\Phi$ by $\hat{d}_{n}:=\parallel\hat{G}_{n}-\Phi\parallel$ where
$\Vert\cdot\Vert:\mathscr{D}_{\mathbb{R}}\rightarrow\lbrack0,\infty)$ is a
user-chosen norm or seminorm; alternative discrepancy measures are discussed
in Section~\ref{Sec further discrepancy measures}. For instance, setting
$\Vert\cdot\Vert=\Vert\cdot\Vert_{\infty}$, i.e., the sup norm on
$\mathscr{D}_{\mathbb{R}}$, delivers the well known Kolmogorov-Smirnov [%
%TCIMACRO{\TeXButton{\textsf{KS}}{\textsf{KS}}}%
%BeginExpansion
\textsf{KS}%
%EndExpansion
] distance%
\begin{equation}
\hat{d}_{n}=\hat{d}_{n}^{%
%TCIMACRO{\TeXButton{\textsf{KS}}{\textsf{KS}}}%
%BeginExpansion
\textsf{KS}%
%EndExpansion
}:=\Vert\hat{G}_{n}-\Phi\Vert_{\infty}=\sup_{u\in\mathbb{R}}|\hat{G}%
_{n}(u)-\Phi(u)|. \label{eq KS distance}%
\end{equation}
Under the bootstrap consistency hypothesis $T_{n}^{\ast}\overset{d^{\ast}%
}{\rightarrow}_{p}%
%TCIMACRO{\TeXButton{N}{\mathscr{N}\!}}%
%BeginExpansion
\mathscr{N}\!%
%EndExpansion
(0,1)$, by Polya's theorem,
\begin{equation}
\hat{d}_{n}=\Vert\hat{G}_{n}-\Phi\Vert_{\infty}\overset{p}{\rightarrow}0,
\label{eq bs consistency}%
\end{equation}
and similarly, for any (semi)norm $\Vert\cdot\Vert\ $that is continuous on
$\mathscr{D}_{\mathbb{R}}$ with the uniform metric;\ examples are the signed
%TCIMACRO{\TeXButton{\textsf{KS}}{\textsf{KS}} }%
%BeginExpansion
\textsf{KS}
%EndExpansion
(Andrews and Shapiro, 2025) and the Cramer-von Mises norms. However, should
bootstrap consistency for the Gaussian limit fail, also
(\ref{eq bs consistency}) will fail to hold. In particular, for the
IV\ example of Section~\ref{Sec Example weak IV} as well as for all the
examples in Section~\ref{Sec Examples}, it holds that $\hat{G}_{n}\overset
{w}{\rightarrow}%
%TCIMACRO{\TeXButton{G}{\mathscr{G}\!}}%
%BeginExpansion
\mathscr{G}\!%
%EndExpansion
$ in $\mathscr{D}_{\mathbb{R}}$, where $%
%TCIMACRO{\TeXButton{G}{\mathscr{G}\!}}%
%BeginExpansion
\mathscr{G}\!%
%EndExpansion
$ is a (random) cdf such that $%
%TCIMACRO{\TeXButton{G}{\mathscr{G}\!}}%
%BeginExpansion
\mathscr{G}\!%
%EndExpansion
\neq\Phi$ (a.s.). By the continuous mapping theorem [CMT], provided the
transformation $\Vert\cdot\Vert$ is continuous on $\mathscr{D}_{\mathbb{R}}%
$,\footnote{If $%
%TCIMACRO{\TeXButton{G}{\mathscr{G}\!}}%
%BeginExpansion
\mathscr{G}\!%
%EndExpansion
$ is sample-path continuous, continuity of $\Vert\cdot\Vert$ on
$\mathscr{D}(\mathbb{R})$ equipped with the uniform metric suffices.} $\hat
{d}_{n}$ itself has a (possibly) random limit:%
\begin{equation}
\hat{d}_{n}:=\Vert\hat{G}_{n}-\Phi\Vert\overset{d}{\rightarrow}%
%TCIMACRO{\TeXButton{Y}{\mathscr{Y}}}%
%BeginExpansion
\mathscr{Y}%
%EndExpansion
:=\Vert%
%TCIMACRO{\TeXButton{G}{\mathscr{G}\!}}%
%BeginExpansion
\mathscr{G}\!%
%EndExpansion
-\Phi\Vert\label{eq dn random limit}%
\end{equation}
where $%
%TCIMACRO{\TeXButton{Y}{\mathscr{Y}}}%
%BeginExpansion
\mathscr{Y}%
%EndExpansion
>0$ a.s. if $\Vert\cdot\Vert$ is a norm.\footnote{For seminorms, e.g.,
$\hat{d}_{n}(A):=\sup_{u\in A\subset\mathbb{R}}|\hat{G}_{n}(u)-\Phi(u)|$, also
$\mathbb{P}(%
%TCIMACRO{\TeXButton{Y}{\mathscr{Y}}}%
%BeginExpansion
\mathscr{Y}%
%EndExpansion
=0)>0$ may hold; for an example see the parameter-on-the-boundary case of
Section~\ref{Sec Example boundary}, where the positivity of $%
%TCIMACRO{\TeXButton{Y}{\mathscr{Y}}}%
%BeginExpansion
\mathscr{Y}%
%EndExpansion
$ for $\hat{d}_{n}(A)$ depends on the choice of the set $A$.} The different
asymptotic behaviors in (\ref{eq bs consistency}) and
(\ref{eq dn random limit}) will be exploited to develop bootstrap tests for
valid specification.

As is standard, the cdf $\hat{G}_{n}$, as well as the discrepancy $\hat{d}%
_{n}$, can be approximated using a sample $T_{n:i}^{\ast}$, $i=1,...,m$, of
conditionally independent copies of $T_{n}^{\ast}$ (obtained by simulation):%
\begin{equation}
\hat{G}_{n,m}^{\ast}(\cdot)\text{$:=$}\frac{1}{m}\sum\nolimits_{i=1}%
^{m}\mathbb{I}_{\mathbb{\{}T_{n:i}^{\ast}\leq\cdot\}}\text{,\hspace{0.5cm}%
}\hat{d}_{n,m}^{\ast}:=\Vert\hat{G}_{n,m}^{\ast}-\Phi\Vert
,\label{eq def G*n,m and d*_m,n}%
\end{equation}
with $m$ user-chosen. Both $\hat{d}_{n,m}^{\ast}$ and $\hat{G}_{n,m}^{\ast}$,
the latter being usually employed in applications of the bootstrap to compute
\textit{p}-values and confidence sets, are key to assess specification
validity without inducing pre-testing bias.

\subsection{asymptotic regimes}

Consider $\hat{G}_{n,m}^{\ast}$ and $\hat{d}_{n,m}^{\ast}$ of
(\ref{eq def G*n,m and d*_m,n}). By the Glivenko-Cantelli theorem, for any $n$
and as $m\rightarrow\infty$, $\Vert\hat{G}_{n,m}^{\ast}-\hat{G}_{n}%
\Vert_{\infty}\overset{a.s.^{\ast}}{\rightarrow}0$ (a.s.) Then, for any $n$,
also $\hat{d}_{n,m}^{\ast}\overset{a.s.^{\ast}}{\rightarrow}\hat{d}_{n}$
(a.s.), provided that $\Vert\cdot\Vert$ is continuous on
$\mathscr{D}_{\mathbb{R}}$ with the Skorokhod $J_{1}$ or the $\sup$ metric.
Hence, for practical purposes $\hat{G}_{n}$ and $\hat{d}_{n}$ can be treated
as known and asymptotic inference based on transformations of $\hat{G}%
_{n}-\Phi$, like $\hat{d}_{n}$, is in principle feasible.

As anticipated in Section~\ref{Sec intro}, however, it turns out that
inference based on such transformations is unattractive in practice, as
(i)\ their null asymptotic distribution as $n\rightarrow\infty$ is in general
not only very hard to obtain, but also application-specific (see Section
\ref{Sec Example boundary} for an example involving $\hat{d}_{n}$); and (ii) a
problem of post-diagnostic test bias emerges.

A closer look shows that these issues arise under an implicit sequential
asymptotic regime -- which we label as $(m,n\rightarrow\infty)_{\text{seq }}$
using the notation in Phillips and Moon (1999) -- where first $m\rightarrow
\infty$ in order for $\hat{d}_{n,m}^{\ast}$ to collapse to $\hat{d}_{n}$, and
then $n\rightarrow\infty$. Standard bootstrap asympotic theory, which would
discuss $\hat{d}_{n}$ as $n\rightarrow\infty$ rather than the actually
computed $\hat{d}_{n,m}^{\ast}$, can be interpreted as employing
$(m,n\rightarrow\infty)_{\text{seq }}$ asymptotics.

We now ask whether a different asymptotic regime can be chosen such that
(i)\ the asymptotic distributions of test statistics are invariant across a
wide range of applications under the null of valid specification, (ii) no
post-diagnostic bias is present under the null, and (iii) the tests are
consistent against a relevant class of alternatives. Two candidate asymptotic
regimes are as follows.

First, consider the sequential regime $(n,m\rightarrow\infty)_{\text{seq }}$,
where $n\rightarrow\infty$ followed by $m\rightarrow\infty$. In the case of
the $\sup$ norm, $\hat{d}_{n,m}^{\ast}=\Vert\hat{G}_{n,m}^{\ast}-\Phi
\Vert_{\infty}$ can be written as $\hat{d}_{n,m}^{\ast}=\phi_{m}(T_{n:1}%
^{\ast},...,T_{n:m}^{\ast})$ for a continuous $\phi_{m}\nolinebreak%
:\nolinebreak\mathbb{R}^{m}\nolinebreak\rightarrow\nolinebreak\mathbb{R}$.
Because under the null of valid specification $T_{n}^{\ast}\overset{d^{\ast}%
}{\rightarrow}_{p}Z\sim%
%TCIMACRO{\TeXButton{N}{\mathscr{N}\!}}%
%BeginExpansion
\mathscr{N}\!%
%EndExpansion
(0,1)$, it holds that
\[
\hat{d}_{n,m}^{\ast}=\phi_{m}(T_{n:1}^{\ast},...,T_{n:m}^{\ast})\overset
{d^{\ast}}{\rightarrow}_{p}d_{m}^{\ast}:=\phi_{m}(Z_{1},...,Z_{m})
\]
as $n\rightarrow\infty$, with the $Z_{i}$'s being independent $%
%TCIMACRO{\TeXButton{N}{\mathscr{N}\!}}%
%BeginExpansion
\mathscr{N}\!%
%EndExpansion
(0,1)$, $i=1,...,m$. Now, let $m\rightarrow\infty$ after $n\rightarrow\infty$.
Under the null, this yields by standard empirical process theory that, with
$W$ denoting a standard Brownian bridge, $\sqrt{m}d_{m}^{\ast}\overset
{d}{\rightarrow}\Vert W(\Phi)\Vert_{\infty}$, i.e., the Kolmogorov
distribution. Therefore, if $(n,m\rightarrow\infty)_{\text{seq}}$,
\[%
%TCIMACRO{\TeXButton{T}{\mathscr{T}}}%
%BeginExpansion
\mathscr{T}%
%EndExpansion
_{n,m}^{\ast}:=\sqrt{m}\hat{d}_{n,m}^{\ast}=\sqrt{m}\Vert\hat{G}_{n,m}^{\ast
}-\Phi\Vert_{\infty}\overset{d^{\ast}}{\rightarrow}_{p}\Vert W(\Phi
)\Vert_{\infty}.
\]
In contrast, under alternatives such that $\hat{G}_{n}\overset{w}{\rightarrow}%
%TCIMACRO{\TeXButton{G}{\mathscr{G}\!}}%
%BeginExpansion
\mathscr{G}\!%
%EndExpansion
\neq\Phi$ a.s., it holds that $%
%TCIMACRO{\TeXButton{T}{\mathscr{T}}}%
%BeginExpansion
\mathscr{T}%
%EndExpansion
_{n,m}^{\ast}\overset{p^{\ast}}{\rightarrow}_{p}\infty$ as $(n,m\rightarrow
\infty)_{\text{seq }}$, yielding consistent tests.

Although the sequential `$m$ after $n$' asymptotic regime leads to tractable
derivations, it provides little justification for conducting inference based
on $\Vert W(\Phi)\Vert_{\infty}$. Indeed, in practice, one has more control on
$m$ (number of bootstrap repetitions) rather than $n$ (sample length). Hence,
we also consider an asymptotic regime where $m$ and $n$ diverge jointly,
denoted as $(n,m\rightarrow\infty)$. Under the null and under an additional
rate condition relating $m$ to $n$, this regime allows to achieve asymptotic
distributions that are invariant across applications (e.g., the $\Vert
W(\Phi)\Vert_{\infty}$ limit seen previously for the $\sup$ norm), whereas
under relevant alternatives the resulting tests are consistent. Focusing on
the statistic $%
%TCIMACRO{\TeXButton{T}{\mathscr{T}}}%
%BeginExpansion
\mathscr{T}%
%EndExpansion
_{n,m}^{\ast}:=\sqrt{m}\hat{d}_{n,m}^{\ast}$, the rate condition makes sure
that $\hat{G}_{n,m}^{\ast}-\hat{G}_{n}$ is dominant in the derivation of $%
%TCIMACRO{\TeXButton{T}{\mathscr{T}}}%
%BeginExpansion
\mathscr{T}%
%EndExpansion
_{n,m}^{\ast}$'s asymptotic distribution under the null, whereas under
alternatives of interest, $\hat{G}_{n}-\Phi$ is dominant in the limit.

Notice that both the $(n,m\rightarrow\infty)_{\text{seq}}$ and
$(n,m\rightarrow\infty)$ regimes are intended for justifying inference
employing $\hat{G}_{n,m}^{\ast}$ rather than true $\hat{G}_{n}$. Although this
choice implies a loss of information for every fixed $n$, as $\hat{G}%
_{n,m}^{\ast}$ is used before it has converged to $\hat{G}_{n}$, it is the
bootstrap randomness in $\hat{G}_{n,m}^{\ast}-\hat{G}_{n}$ that can make the
asymptotic distribution of $\hat{d}_{n,m}^{\ast}$ invariant across
applications and can eliminate the post-diagnostic bias. Indeed, for both
regimes no post-diagnostic test bias is present asymptotically, as the proof
of Theorem~\ref{Thm pretest} below for the joint regime goes through also for
the sequential `$m$ after $n$' regime.

\subsection{the diagnostic test}

\label{sec the diagnostic test}

Redefine $%
%TCIMACRO{\TeXButton{T}{\mathscr{T}}}%
%BeginExpansion
\mathscr{T}%
%EndExpansion
_{n,m}^{\ast}$ introduced above using a generic (semi)norm, i.e., $%
%TCIMACRO{\TeXButton{T}{\mathscr{T}}}%
%BeginExpansion
\mathscr{T}%
%EndExpansion
_{n,m}^{\ast}:=m^{1/2}\hat{d}_{n,m}^{\ast}$, $\hat{d}_{n,m}^{\ast}:=\Vert
\hat{G}_{n,m}^{\ast}-\Phi\Vert$ with $\hat{G}_{n,m}^{\ast}(\cdot):=m^{-1}%
\sum\nolimits_{i=1}^{m}\mathbb{I}_{\mathbb{\{}T_{n:i}^{\ast}\leq\cdot\}}$ and
the $T_{n:i}^{\ast}$'s being i.i.d. copies of $T_{n}^{\ast}$ conditionally on
the data $D_{n}$. We note the following.

First, $%
%TCIMACRO{\TeXButton{T}{\mathscr{T}}}%
%BeginExpansion
\mathscr{T}%
%EndExpansion
_{n,m}^{\ast}$ depends on the data $D_{n}$ as well as on $m$ auxiliary
variates, say $W_{n}^{\ast}$, used to generate the $m$ bootstrap draws
$T_{n:i}^{\ast}$, $i=1,\ldots,m$. For instance $W_{n}^{\ast}$, which is
defined jointly with $D_{n}$ on a possibly extended probability space, could
be thought of as a vector of $m$ i.i.d. $%
%TCIMACRO{\TeXButton{U}{\mathscr{U}\!}}%
%BeginExpansion
\mathscr{U}\!%
%EndExpansion
_{[0,1]}$ r.v.s, independent of $D_{n}$, such that $T_{n:i}^{\ast}=\hat{G}%
_{n}^{-1}(W_{n,i}^{\ast})$, $i=1,\ldots,m$, with $\hat{G}_{n}^{-1}$ the
generalized inverse of $\hat{G}_{n}$.

Second, $%
%TCIMACRO{\TeXButton{T}{\mathscr{T}}}%
%BeginExpansion
\mathscr{T}%
%EndExpansion
_{n,m}^{\ast}$ can be written as
\begin{equation}%
%TCIMACRO{\TeXButton{T}{\mathscr{T}}}%
%BeginExpansion
\mathscr{T}%
%EndExpansion
_{n,m}^{\ast}=\mathcal{Z}_{n,m}^{\ast}+a_{n,m}^{\ast}\text{, }\mathcal{Z}%
_{n,m}^{\ast}:=m^{1/2}\Vert\hat{G}_{n,m}^{\ast}-\hat{G}_{n}\Vert,
\label{eq Z* and a*}%
\end{equation}
with $a_{n,m}^{\ast}$ implicitly defined. Here, $\mathcal{Z}_{n,m}^{\ast}$ is
a rescaled measure of the distance between the estimator $\hat{G}_{n,m}^{\ast
}$ and $\hat{G}_{n}$, while $a_{n,m}^{\ast}$ is related to the distance
between $\hat{G}_{n}$ and the Gaussian cdf $\Phi$. In particular, for any $n$,
$|a_{n,m}^{\ast}|\leq\sqrt{m}\Vert\hat{G}_{n}-\Phi\Vert$ by Lemma
\ref{Lemma an,b*} in Appendix~\ref{Appendix main proofs}.

We now provide conditions such that, under the valid specification hypothesis,
the asymptotic distribution of $%
%TCIMACRO{\TeXButton{T}{\mathscr{T}}}%
%BeginExpansion
\mathscr{T}%
%EndExpansion
_{n,m}^{\ast}$ can be derived and hence used to perform a proper statistical
test. In particular, we consider the following assumption.

\begin{condition}
\label{Assn 1}The (semi)norm $\Vert\cdot\Vert$ is continuous on
$\mathscr{D}_{\mathbb{R}}$. Moreover, $\Vert\hat{G}_{n}-\Phi\Vert
=O_{p}(n^{-\alpha})$ for some $\alpha>0$.
\end{condition}

Assumption~\ref{Assn 1} is a condition on the rate of convergence of the
bootstrap cdf to the Gaussian cdf under valid specification. It is satisfied
with $\alpha=1/2$ if $\Vert\cdot\Vert\leq C\Vert\cdot\Vert_{\infty}$ for some
finite constant $C$ (as is the case of, e.g., the signed
%TCIMACRO{\TeXButton{\textsf{KS}}{\textsf{KS}} }%
%BeginExpansion
\textsf{KS}
%EndExpansion
norm or the Cramer-von Mises norm), and the bootstrap statistic has a one-term
Edgeworth expansion of the form $\hat{G}_{n}(x)=\Phi(x)+\hat{q}_{n}%
(x)n^{-1/2}+o_{p}(n^{-1/2})$ uniformly in $x$, as is usually the case; see,
e.g., Hall (1992). For some symmetric statistics, Assumption~\ref{Assn 1} can
hold with $\alpha=1$. For Gaussian parametric bootstraps, Assumption
\ref{Assn 1} may be even satisfied with arbitrary $\alpha>0$; see, e.g., the
case in Section~\ref{Sec Example boundary}.

The following result holds under Assumption~\ref{Assn 1}.

\begin{theorem}
\label{Thm 1}Let $T_{n}^{\ast}\overset{d^{\ast}}{\rightarrow}_{p}%
%TCIMACRO{\TeXButton{N}{\mathscr{N}\!}}%
%BeginExpansion
\mathscr{N}\!%
%EndExpansion
(0,1)$. Under Assumption~\ref{Assn 1}, if $(n,m\rightarrow\infty)$ and
$m/n^{2\alpha}\rightarrow0$,

\begin{description}
\item (i) $%
%TCIMACRO{\TeXButton{T}{\mathscr{T}}}%
%BeginExpansion
\mathscr{T}%
%EndExpansion
_{n,m}^{\ast}=\mathcal{Z}_{n,m}^{\ast}+o_{p}(1)\overset{d^{\ast}}{\rightarrow
}_{p}%
%TCIMACRO{\TeXButton{K}{\mathscr{K}\!}}%
%BeginExpansion
\mathscr{K}\!%
%EndExpansion
:=\Vert W(\Phi)\Vert$, where $W$ is standard Brownian bridge.

\item (ii) $p_{n,m}^{\ast}:=1-H(%
%TCIMACRO{\TeXButton{T}{\mathscr{T}}}%
%BeginExpansion
\mathscr{T}%
%EndExpansion
_{n,m}^{\ast})\overset{w^{\ast}}{\rightarrow}_{p}%
%TCIMACRO{\TeXButton{U}{\mathscr{U}\!}}%
%BeginExpansion
\mathscr{U}\!%
%EndExpansion
_{[0,1]}$ if the cdf $H$ of $%
%TCIMACRO{\TeXButton{K}{\mathscr{K}\!}}%
%BeginExpansion
\mathscr{K}\!%
%EndExpansion
$ is continuous.
\end{description}
\end{theorem}

Theorem~\ref{Thm 1} suggests that a simple diagnostic procedure can be
obtained by comparing $%
%TCIMACRO{\TeXButton{T}{\mathscr{T}}}%
%BeginExpansion
\mathscr{T}%
%EndExpansion
_{n,m}^{\ast}$ with critical values from the distribution of $%
%TCIMACRO{\TeXButton{K}{\mathscr{K}\!}}%
%BeginExpansion
\mathscr{K}\!%
%EndExpansion
$, or through the associated asymptotic \textit{p}-value $p_{n,m}^{\ast}$.
Provided both $m$ and $n$ grow to infinity at a proper relative rate, Theorem
\ref{Thm 1} guarantees that the procedure is asymptotically correctly sized.

\begin{remark}
The logic of the proof can be easily followed in the case of the signed
pointwise discrepancy $\hat{d}_{n}(x):=\hat{G}_{n}(x)-\Phi(x)$ for some fixed
$x\in\mathbb{R}$. In this case, $\mathcal{Z}_{n,m}^{\ast}$ and $a_{n,m}^{\ast
}$ of (\ref{eq Z* and a*}) are given by $\mathcal{Z}_{n,m}^{\ast}=m^{1/2}%
(\hat{G}_{n,m}^{\ast}(x)-\hat{G}_{n}(x))$ and $a_{n,m}^{\ast}=m^{1/2}(\hat
{G}_{n}(x)-\Phi(x))$. If $\hat{d}_{n}(x)$ satisfies Assumption~\ref{Assn 1}
for some $\alpha>0$, it holds that $a_{n,m}^{\ast}\rightarrow_{p}0$ as
$(n,m\rightarrow\infty)$ with $m/n^{2\alpha}\rightarrow0$. Moreover, with
$\xi_{i}^{\ast}:=\mathbb{I}_{\mathbb{\{}T_{n:i}^{\ast}\leq x\}}-\mathbb{E}%
^{\ast}[\mathbb{I}_{\mathbb{\{}T_{n:i}^{\ast}\leq x\}}]$, such that
$\mathcal{Z}_{n,m}^{\ast}=m^{-1/2}\sum_{i=1}^{m}\xi_{i}^{\ast}$, by a standard
Berry-Esseen bound we have that $\mathcal{\tilde{Z}}_{n,m}^{\ast}%
:=\mathbb{E}^{\ast}[\xi_{i}^{\ast}{}^{2}]^{-1/2}\mathcal{Z}_{n,m}^{\ast}$
satisfies, for some $C<\infty$,%
\[
\parallel\mathbb{P}^{\ast}\mathbb{(}\mathcal{\tilde{Z}}_{n,m}^{\ast}\leq
\cdot)-\Phi(\cdot)\parallel_{\infty}\leq Cm^{-1/2}\mathbb{E}^{\ast}[|\xi
_{i}^{\ast}|^{3}]\leq Cm^{-1/2}%
\]
whenever $\mathbb{E}^{\ast}[\xi_{i}^{\ast2}]=\hat{G}_{n}(x)(1-\hat{G}%
_{n}(x))\neq0$, which occurs with probability approaching one as
$n\rightarrow\infty$ since $\mathbb{E}^{\ast}[\xi_{i}^{\ast2}]\rightarrow
_{p}v^{2}(x):=\Phi\left(  x\right)  (1-\Phi(x))$ under the hypothesis that
$T_{n}^{\ast}\overset{d^{\ast}}{\rightarrow}_{p}%
%TCIMACRO{\TeXButton{N}{\mathscr{N}\!}}%
%BeginExpansion
\mathscr{N}\!%
%EndExpansion
(0,1)$ as $n\rightarrow\infty$. We conclude that $\mathcal{Z}_{n,m}^{\ast
}=v(x)\mathcal{\tilde{Z}}_{n,m}^{\ast}+o_{p}^{\ast}(1)\overset{d^{\ast}%
}{\rightarrow}_{p}%
%TCIMACRO{\TeXButton{N}{\mathscr{N}\!}}%
%BeginExpansion
\mathscr{N}\!%
%EndExpansion
\left(  0,v(x)\right)  \sim W(\Phi(x))$ as $(n,m\rightarrow\infty)$%
.\hfill$\square$
\end{remark}

\begin{remark}
For commonly used norms, the distribution of $%
%TCIMACRO{\TeXButton{K}{\mathscr{K}\!}}%
%BeginExpansion
\mathscr{K}\!%
%EndExpansion
$ is well known; for example, with $\Vert\cdot\Vert=\Vert\cdot\Vert_{\infty}$,
$%
%TCIMACRO{\TeXButton{K}{\mathscr{K}\!}}%
%BeginExpansion
\mathscr{K}\!%
%EndExpansion
$ follows the Kolmogorov distribution which has a continuous cdf. In general,
critical values (as well as \textit{p}-values) can be determined by Monte
Carlo simulation with arbitrary accuracy.\hfill$\square$
\end{remark}

The test based on $%
%TCIMACRO{\TeXButton{T}{\mathscr{T}}}%
%BeginExpansion
\mathscr{T}%
%EndExpansion
_{n,m}^{\ast}$ also has non-trivial asymptotic power against bootstrap
inconsistency for the standard Gaussian distribution. This is shown next.

\begin{theorem}
\label{Thm 2}Suppose that $\hat{G}_{n}\overset{w}{\rightarrow}%
%TCIMACRO{\TeXButton{G}{\mathscr{G}\!}}%
%BeginExpansion
\mathscr{G}\!%
%EndExpansion
$ in $%
%TCIMACRO{\TeXButton{curly D}{\mathscr{D}}}%
%BeginExpansion
\mathscr{D}%
%EndExpansion
_{\mathbb{R}}$ as $n\rightarrow\infty$.\ Suppose further that $\Vert\cdot
\Vert$ is continuous on $%
%TCIMACRO{\TeXButton{curly D}{\mathscr{D}}}%
%BeginExpansion
\mathscr{D}%
%EndExpansion
_{\mathbb{R}}$ and $\Vert%
%TCIMACRO{\TeXButton{G}{\mathscr{G}\!}}%
%BeginExpansion
\mathscr{G}\!%
%EndExpansion
-\Phi\Vert>0$ a.s. Then, for any $c\in\left(  0,\infty\right)  $, it holds
that $\mathbb{P}^{\ast}(%
%TCIMACRO{\TeXButton{T}{\mathscr{T}}}%
%BeginExpansion
\mathscr{T}%
%EndExpansion
_{n,m}^{\ast}\geq c)\overset{p}{\rightarrow}1$ as $(n,m\rightarrow\infty)$.
\end{theorem}

An inspection of the proof of Theorem~\ref{Thm 2} reveals that, as expected,
for large $m$ the power of the test is determined by the realized value of
$\hat{d}_{n}:=\Vert\hat{G}_{n}-\Phi\Vert$, which is asymptotically distributed
as the r.v. $%
%TCIMACRO{\TeXButton{Y}{\mathscr{Y}}}%
%BeginExpansion
\mathscr{Y}%
%EndExpansion
:=\Vert%
%TCIMACRO{\TeXButton{G}{\mathscr{G}\!}}%
%BeginExpansion
\mathscr{G}\!%
%EndExpansion
-\Phi\Vert>0$ a.s. Such realization depends on the original data $D_{n}$ only,
and not on the $m$ bootstrap repetitions used to generate the bootstrap
statistic. Larger outcomes of $\hat{d}_{n}$ correspond -- ceteris paribus --
to larger power.

\subsection{an example based on instrumental variables (cont'd)}

\label{Sec Example IV reprise}

When $\pi\neq0$ is fixed, such that the instruments are strong, for the
parametric bootstrap described in Section~\ref{Sec Example weak IV} it holds,
without additional assumptions, that $\Vert\hat{G}_{n}-\Phi\Vert_{\infty
}=O_{p}(n^{-\alpha})$ for any $\alpha\in(0,\frac{1}{2})$; this is shown in
Appendix~\ref{Appendix additional proofs} by using the machinery of parametric
tail estimates. Hence, Assumption~\ref{Assn 1} is verified with $\alpha
\in(0,\frac{1}{2})$ for the $\sup$ norm and its dominated norms, and Theorem
\ref{Thm 1} applies to them. Using uniform Edgeworth expansions, also
$\Vert\hat{G}_{n}-\Phi\Vert_{\infty}=O_{p}(n^{-1/2})$ has been shown to hold
for the non-parametric i.i.d. bootstrap (where $u_{i}^{\ast}$ and $v_{i}%
^{\ast}$ are resampled from the residuals $\hat{u}_{i}:=y_{i}-\hat{\beta}%
_{n}x_{i}$ and $\hat{v}_{i}:=x_{i}-\hat{\pi}_{n}^{\prime}z_{i}$), under mild
regularity conditions on $(u_{i},v_{i})$ (essentially, existence of
higher-order moments); see Moreira, Porter and Suarez (2009, Theorem 3).

Under weak instruments, we found previously that the bootstrap cdf of
$T_{n}^{\ast}$ satisfies $\hat{G}_{n}(x)\rightarrow_{w}%
%TCIMACRO{\TeXButton{G}{\mathscr{G}\!}}%
%BeginExpansion
\mathscr{G}\!%
%EndExpansion
(x):=\mathbb{P(}\xi(\ell,\zeta^{\ast})/\sqrt{\ell^{\prime}\ell}\leq x$ $|$
$\ell)$ on $\mathscr{D}_{\mathbb{R}}$, which is non-Gaussian with probability
one. Thus, Theorem~\ref{Thm 2} applies in this case.%

%TCIMACRO{\TeXButton{B}{\begin{table}[t] \centering}}%
%BeginExpansion
\begin{table}[t] \centering
%EndExpansion
\setlength{\tabcolsep}{2.6pt}%

\begin{tabular}
[c]{rrrrrrrrrcrrrrrrrr}%
\multicolumn{18}{c}{}\\
(\textsc{a}) & \multicolumn{8}{c}{\textsf{KS}} &  &
\multicolumn{8}{c}{\textsf{AD}}\\\cline{2-9}\cline{11-18}%
\multicolumn{1}{c}{$n\backslash\zeta$} & $0.50$ & $0.55$ & $0.60$ & $0.65$ &
$0.70$ & $0.80$ & $0.90$ & $1.00$ & \multicolumn{1}{r}{} & $0.50$ & $0.55$ &
$0.60$ & $0.65$ & $0.70$ & $0.80$ & $0.90$ & $1.00$\\\hline
$100$ & $4.9$ & $5.2$ & $5.4$ & $5.4$ & $5.7$ & $6.9$ & $8.0$ & $10.5$ &  &
$5.3$ & $5.6$ & $5.7$ & $5.6$ & $5.7$ & $6.8$ & $7.7$ & $9.6$\\
$200$ & $5.2$ & $5.4$ & $5.6$ & $5.6$ & $5.6$ & $7.0$ & $8.7$ & $10.4$ &  &
$5.6$ & $5.7$ & $5.7$ & $5.6$ & $5.8$ & $6.7$ & $7.5$ & $9.8$\\
$400$ & $5.0$ & $5.0$ & $5.3$ & $5.6$ & $6.0$ & $6.3$ & $7.9$ & $10.6$ &  &
$5.1$ & $5.1$ & $5.0$ & $5.4$ & $5.8$ & $6.2$ & $7.5$ & $9.3$\\
$800$ & $4.9$ & $4.9$ & $5.4$ & $5.6$ & $5.7$ & $6.6$ & $7.8$ & $10.2$ &  &
$5.1$ & $4.7$ & $5.2$ & $5.4$ & $5.7$ & $6.2$ & $7.4$ & $9.7$\\\hline
$\overset{}{\text{(\textsc{b})}}$ & \multicolumn{8}{c}{$\overset{\overset{}{}%
}{\text{\textsf{KS}}}$} &  & \multicolumn{8}{c}{\textsf{AD}}\\\cline{2-9}%
\cline{11-18}%
\multicolumn{1}{c}{$n\backslash\zeta$} & $0.50$ & $0.55$ & $0.60$ & $0.65$ &
$0.70$ & $0.80$ & $0.90$ & $1.00$ & \multicolumn{1}{r}{} & $0.50$ & $0.55$ &
$0.60$ & $0.65$ & $0.70$ & $0.80$ & $0.90$ & $1.00$\\\hline
$100$ & $29.6$ & $33.7$ & $39.2$ & $44.6$ & $51.8$ & $63.7$ & $74.3$ & $83.3$
&  & $30.7$ & $35.6$ & $41.9$ & $49.1$ & $56.5$ & $69.1$ & $80.1$ & $89.2$\\
$200$ & $37.4$ & $43.8$ & $50.2$ & $56.7$ & $62.8$ & $76.3$ & $86.4$ & $93.4$
&  & $40.0$ & $46.8$ & $55.1$ & $62.3$ & $68.8$ & $82.6$ & $91.5$ & $97.1$\\
$400$ & $46.1$ & $52.8$ & $60.9$ & $69.0$ & $75.9$ & $87.5$ & $95.2$ & $98.4$
&  & $51.0$ & $58.1$ & $66.7$ & $75.3$ & $82.1$ & $93.0$ & $97.9$ & $99.6$\\
$800$ & $54.9$ & $63.2$ & $71.6$ & $79.3$ & $85.4$ & $94.7$ & $98.3$ & $99.6$
&  & $60.4$ & $69.0$ & $77.9$ & $85.4$ & $91.0$ & $97.7$ & $99.5$ &
$100.0$\\\hline
\end{tabular}
\vspace{-0.5em}
\caption{IV Regression -- Empirical rejection frequencies of the bootstrap diagnostic tests based on the Kolmogorov-Smirnov (\textsf{KS}) and Anderson-Darling (\textsf{AD}) norms with $m=\lfloor n^\zeta \rfloor$,
$\zeta \in [0.5,1.0]$. Panel (\textsc{a}): strong instrument; Panel (\textsc{b}): weak instrument.}\label{Table IV}%
%TCIMACRO{\TeXButton{E}{\end{table}}}%
%BeginExpansion
\end{table}%
%EndExpansion

We conclude by reporting in Table~\ref{Table IV} the (percentage) empirical
rejection probabilities (ERPs) of the bootstrap test based on the sup norm
(\textsf{KS}); for comparison, we also consider the test based on the
Anderson-Darling norm (\textsf{AD}). The data generating process is as in
Remark~\ref{Remark simulated Ghat} with $z_{t}\sim$i.i.d.$%
%TCIMACRO{\TeXButton{N}{\mathscr{N}}}%
%BeginExpansion
\mathscr{N}%
%EndExpansion
(0,1)$ and the bootstrap is non-parametric (see Remark
\ref{Remark IV nonparametric bs}) with $z_{t}$ fixed in the bootstrap world; a
constant is included in estimation. The upper panel corresponds to the strong
instrument case ($\pi=1$), while the lower panel considers the weak instrument
case with $\pi=\lambda n^{-1/2}$ and $\lambda=2$. To assess how well the
asymptotic theory $(n,m\rightarrow\infty,\;m/n\rightarrow0)$ approximates
finite-sample behavior, we consider different choices of $m$, namely
$m=\lfloor n^{\zeta}\rfloor$ with $\zeta\in\{0.50,0.55,\ldots
,0.70,0.80,0.90,1.00\}$. ERPs are computed using $10,000$ Monte Carlo
replications; the nominal level is $5\%$.

In the strong instrument case (upper panel), the ERPs remain relatively close
to the nominal level, particularly for lower values of $m$ relative to $n$.
They increase as $m$ grows; for \textsf{KS} they never exceed $11\%$, while
for \textsf{AD} they can rise up to about $10\%$. Consistently with the
theoretical expectation, under the weak instrument scenario (lower panel), the
ERPs increase with $m,n$.

\section{post-diagnostics inference}

\label{Sec post diagnostics inference}

The bootstrap approach developed in the previous section can be used in
diagnostic pre-testing of specification validity, with standard inference
carried out when the diagnostic procedure does not reject. An important
question is whether post-diagnostic inferences are biased by the outcome of
the pre-test, in the sense that conditionally on a correct non-rejection by
the pre-test, the rejection probabilities of post-diagnostic tests are
affected even asymptotically. The answer is no.

The key for this result is that, under valid specification, the bootstrap
statistic $%
%TCIMACRO{\TeXButton{T}{\mathscr{T}}}%
%BeginExpansion
\mathscr{T}%
%EndExpansion
_{n,m}^{\ast}$ becomes independent of the original data $D_{n}$ under the
joint asymptotic regime $(n,m\rightarrow\infty)$ and the rate condition of
Theorem~\ref{Thm 1}. That is, in the limit $%
%TCIMACRO{\TeXButton{T}{\mathscr{T}}}%
%BeginExpansion
\mathscr{T}%
%EndExpansion
_{n,m}^{\ast}$ only depends on the bootstrap variates used to generate
$\hat{G}_{n,m}^{\ast}$ and no longer on the data $D_{n}$, thus eliminating any
post-diagnostic bias.

To see why, it suffices to note that the existence of a non-random limit for
the conditional law of a bootstrap statistic given the data is an asymptotic
independence property. It is this very property that the conditional law of
the bootstrap diagnostic statistic $%
%TCIMACRO{\TeXButton{T}{\mathscr{T}}}%
%BeginExpansion
\mathscr{T}%
%EndExpansion
_{n,m}^{\ast}$ enjoys under the conditions of Theorem~\ref{Thm 1}. The meaning
of the implied asymptotic independence is clarified in the next theorem, where
$%
%TCIMACRO{\TeXButton{T}{\mathscr{T}}}%
%BeginExpansion
\mathscr{T}%
%EndExpansion
_{n,m}^{\ast}$ can stand for any bootstrap statistic and $(n,m\rightarrow
\infty,R)$ means that $(n,m\rightarrow\infty)$ under a rate condition $R$
(such as the one given in Assumption 1).

\begin{theorem}
\label{Thm pretest}Let the conditional distribution of a bootstrap statistic $%
%TCIMACRO{\TeXButton{T}{\mathscr{T}}}%
%BeginExpansion
\mathscr{T}%
%EndExpansion
_{n,m}^{\ast}$ given the data $D_{n}$ converge in probability to a nonrandom
distribution as $(n,m\rightarrow\infty,R)$. Then, as $(n,m\rightarrow
\infty,R)$:

\begin{description}
\item (a) for measurable real functions $f$ and continuous bounded real
functions $g$ with matching domains,%
\[
\sup\nolimits_{\Vert f\Vert_{\infty}\leq1}\left\vert \mathbb{E[}f(D_{n})g(%
%TCIMACRO{\TeXButton{T}{\mathscr{T}}}%
%BeginExpansion
\mathscr{T}%
%EndExpansion
_{n,m}^{\ast})]-\mathbb{E[}f(D_{n})]\mathbb{E[}g(%
%TCIMACRO{\TeXButton{T}{\mathscr{T}}}%
%BeginExpansion
\mathscr{T}%
%EndExpansion
_{n,m}^{\ast})]\right\vert \rightarrow0\text{;}%
\]

\item (b)\ for non-negligible continuity sets $B$ of $%
%TCIMACRO{\TeXButton{T}{\mathscr{T}}}%
%BeginExpansion
\mathscr{T}%
%EndExpansion
_{n,m}^{\ast}$'s limit distribution,
\[
\sup\nolimits_{A\in\sigma(D_{n})}|\mathbb{P}(A|%
%TCIMACRO{\TeXButton{T}{\mathscr{T}}}%
%BeginExpansion
\mathscr{T}%
%EndExpansion
_{n,m}^{\ast}\in B)-\mathbb{P}(A)|\rightarrow0,
\]
where $\sigma(D_{n})$ is the $\sigma$-algebra generated by the data $D_{n}$.
\end{description}
\end{theorem}

A non-negligible continuity set is a set with positive probability whose
boundary has zero probability. For instance, if the limit distribution of $%
%TCIMACRO{\TeXButton{T}{\mathscr{T}}}%
%BeginExpansion
\mathscr{T}%
%EndExpansion
_{n,m}^{\ast}$ has a continuous cdf $H$, then $B=[0,t]$ is a non-negligible
continuity set whenever $t>0$ is such that $H(t)>0$. This gives rise to the
following corollary relevant for testing.

\begin{corollary}
\label{Corollary pretest}Consider any statistic $\hat{\rho}_{n}\in\mathbb{R}$,
measurable with respect to the data $D_{n}$ and with asymptotic cdf $F_{\rho}$
as $n\rightarrow\infty$. Let the conditional distribution of $%
%TCIMACRO{\TeXButton{T}{\mathscr{T}}}%
%BeginExpansion
\mathscr{T}%
%EndExpansion
_{n,m}^{\ast}$ given the data converge in probability to a nonrandom
distribution with a continuous cdf $H$ as $(n,m\rightarrow\infty,R)$. Then,
for every continuity point $s$ of $F_{\rho}$ any every $t$ with $H(t)>0$, it
holds that, as $(n,m\rightarrow\infty,R)$,%
\[
\sup_{s\in\mathbb{R}}|\mathbb{P}(\hat{\rho}_{n}\leq s|%
%TCIMACRO{\TeXButton{T}{\mathscr{T}}}%
%BeginExpansion
\mathscr{T}%
%EndExpansion
_{n,m}^{\ast}\leq t)-F_{\rho}(s)|\rightarrow0.
\]

\end{corollary}

Under the conditions of Theorem~\ref{Thm 1}, including $T_{n}^{\ast}%
\overset{d^{\ast}}{\rightarrow}_{p}%
%TCIMACRO{\TeXButton{N}{\mathscr{N}\!}}%
%BeginExpansion
\mathscr{N}\!%
%EndExpansion
(0,1)$, the diagnostic statistic $%
%TCIMACRO{\TeXButton{T}{\mathscr{T}}}%
%BeginExpansion
\mathscr{T}%
%EndExpansion
_{n,m}^{\ast}$ satisfies the assumptions of Corollary~\ref{Corollary pretest}
with $H=\Phi$ and $R$ being the rate condition in Assumption~\ref{Assn 1}.
Therefore, in large samples the finite-sample quantiles of $\hat{\rho}_{n}$
conditional on $%
%TCIMACRO{\TeXButton{T}{\mathscr{T}}}%
%BeginExpansion
\mathscr{T}%
%EndExpansion
_{n,m}^{\ast}$ being below (or above) a given critical value\ $t$ approximate
well the unconditional quantiles of $\hat{\rho}_{n}$'s asymptotic
distribution. Hence, should the diagnostics based on $%
%TCIMACRO{\TeXButton{T}{\mathscr{T}}}%
%BeginExpansion
\mathscr{T}%
%EndExpansion
_{n,m}^{\ast}$ correctly fail to reject specification validity, inference
based on $\hat{\rho}_{n}$ is free of pre-testing bias as $n\rightarrow\infty$.

\subsection{an example based on instrumental variables (cont'd)}

\label{Sec Example IV reprise pretest}

We now briefly show Theorem~\ref{Thm pretest} and
Corollary~\ref{Corollary pretest} in action by considering the 2SLS
$t$\nobreakdash-test for the null $\beta=\beta_{0}$, denoted $t_{\mathsf{IV}%
,n}$, in the setting of the IV\ example of Sections~\ref{Sec Example weak IV}
and \ref{Sec Example IV reprise}, assuming that $z_{t}$ is strong (i.e., a
valid specification), so that $t_{\mathsf{IV},n}\rightarrow_{d}%
%TCIMACRO{\TeXButton{N}{\mathscr{N}\!}}%
%BeginExpansion
\mathscr{N}\!%
%EndExpansion
(0,1)$. Indeed, an extensive literature on IV regression documents substantial
bias in pre-tests of instrument strength (see, e.g., Andrews, Stock, and Sun,
2019), making it natural to ask whether the bootstrap diagnostic test
introduces any such bias.

We compare two different scenarios. In the first one, the researcher runs the
$t$\nobreakdash-test without pre-testing valid specification. In the second
one, the researcher initially (pre-)tests for valid specification using the
bootstrap diagnostic test $%
%TCIMACRO{\TeXButton{T}{\mathscr{T}}}%
%BeginExpansion
\mathscr{T}%
%EndExpansion
_{n,m}^{\ast}$, and then runs the $t$\nobreakdash-test only if the bootstrap
test does not reject valid specification. According to
Corollary~\ref{Corollary pretest} applied with $\hat{\rho}_{n}=t_{\mathsf{IV}%
,n}$ and $F_{\rho}=\Phi$, the \emph{unconditional }rejection probabilities of
the $t$\nobreakdash-test (computed without pre-testing valid specification)
and the \emph{conditional} rejection probabilities (computed conditionally on
the bootstrap diagnostic test failing to reject) should coincide as
$(n,m\rightarrow\infty)$ at proper relative rates, see
Section~\ref{Sec Example IV reprise}. These probabilities are estimated in
Table~\ref{Table IV pretest} using the same Monte Carlo design as in
Section~\ref{Sec Example IV reprise}. The unconditional and the conditional
ERPs are virtually indistinguishable, thus supporting the results in Corollary
\ref{Corollary pretest}.%

%TCIMACRO{\TeXButton{B}{\begin{table}[t] \centering}}%
%BeginExpansion
\begin{table}[t] \centering
%EndExpansion
\setlength{\tabcolsep}{2.0pt}%

\begin{tabular}
[c]{rrrrrrrrrrcrrrrrrrr}%
\multicolumn{19}{c}{}\\
&  & \multicolumn{8}{c}{\textsf{KS}} &  & \multicolumn{8}{c}{\textsf{AD}%
}\\\cline{3-10}\cline{12-19}%
\multicolumn{1}{c}{$n\backslash\zeta$} & unc. & $0.50$ & $0.55$ & $0.60$ &
$0.65$ & $0.70$ & $0.80$ & $0.90$ & $1.00$ & \multicolumn{1}{r}{} & $0.50$ &
$0.55$ & $0.60$ & $0.65$ & $0.70$ & $0.80$ & $0.90$ & $1.00$\\\hline
$100$ & \multicolumn{1}{c}{$\mathit{4.9}$} & $4.9$ & $4.9$ & $4.9$ & $4.9$ &
$4.9$ & $4.9$ & $4.9$ & $4.9$ &  & $4.9$ & $4.9$ & $4.9$ & $4.9$ & $4.9$ &
$4.8$ & $4.8$ & $4.9$\\
$200$ & \multicolumn{1}{c}{$\mathit{5.3}$} & $5.3$ & $5.3$ & $5.4$ & $5.3$ &
$5.4$ & $5.3$ & $5.3$ & $5.4$ &  & $5.2$ & $5.2$ & $5.3$ & $5.3$ & $5.4$ &
$5.2$ & $5.2$ & $5.3$\\
$400$ & \multicolumn{1}{c}{$\mathit{4.7}$} & $4.7$ & $4.7$ & $4.8$ & $4.8$ &
$4.8$ & $4.7$ & $4.7$ & $4.7$ &  & $4.7$ & $4.7$ & $4.8$ & $4.8$ & $4.8$ &
$4.7$ & $4.7$ & $4.7$\\
$800$ & \multicolumn{1}{c}{$\mathit{5.1}$} & $5.1$ & $5.1$ & $5.1$ & $5.1$ &
$5.1$ & $5.1$ & $5.2$ & $5.1$ &  & $5.0$ & $5.1$ & $5.1$ & $5.1$ & $5.1$ &
$5.1$ & $5.1$ & $5.1$\\\hline
\end{tabular}
\vspace{-0.5em}
\caption{IV Regression -- Empirical rejection probabilities of the $t_{\textsf{IV},n}$ test. Results are unconditional (in italics) and conditional on the bootstrap tests not rejecting specification validity.
Bootstrap tests computed with $m=\lfloor n^\zeta \rfloor$, $\zeta \in [0.5,1.0]$.}
\label{Table IV pretest}%
%TCIMACRO{\TeXButton{E}{\end{table}}}%
%BeginExpansion
\end{table}%
%EndExpansion

\section{extensions}

\label{Sec extensions}

\subsection{diagnostics reporting}

\label{sec extension diagnostics repo}

An intrinsic feature of any diagnostics based on $%
%TCIMACRO{\TeXButton{T}{\mathscr{T}}}%
%BeginExpansion
\mathscr{T}%
%EndExpansion
_{n,m}^{\ast}$ is that the associated \textit{p}-value, say $p_{n,m}^{\ast}$,
is not $D_{n}$-measurable, as it depends also on the realization of the
auxiliary bootstrap variates $W_{n}^{\ast}$ used to generate $%
%TCIMACRO{\TeXButton{T}{\mathscr{T}}}%
%BeginExpansion
\mathscr{T}%
%EndExpansion
_{n,m}^{\ast}$; see Section~\ref{sec the diagnostic test}. This differs from
ordinary bootstrap inference, where the bootstrap \textit{p}-values are
usually regarded as measurable with respect to the data $D_{n}$, as is the
case where $(m,n\rightarrow\infty)_{\text{seq}}$. Moreover, under the
conditions of Theorem~\ref{Thm 1}, the dependence of $p_{n,m}^{\ast}$ on the
bootstrap variates does not vanish as $(n,m\rightarrow\infty)$, in the sense
that $p_{n,m}^{\ast(1)}-p_{n,m}^{\ast(2)}$ need not go to zero for two
conditionally independent copies $p_{n,m}^{\ast(1)}\ $and $p_{n,m}^{\ast(2)}$
of $p_{n,m}^{\ast}$.

As a consequence, if the bootstrap sample is not given ex ante but, rather, is
generated by the practitioner, then for any given data $D_{n}$ different
researchers may end up obtaining different \textit{p}-values $p_{n,m}^{\ast}$.
This may create issues in terms of reproducibility, as it is unclear which
value of $p_{n,m}^{\ast}$ should be used for decision-making and
reporting.\footnote{In fact, in some applications the set of bootstrap draws
used to compute $%
%TCIMACRO{\TeXButton{T}{\mathscr{T}}}%
%BeginExpansion
\mathscr{T}%
%EndExpansion
_{n,m}^{\ast}$ is given as an input for the diagnostic procedure. This, for
instance, may happen when a replication package includes the set of bootstrap
repetitions used to compute a bootstrap statistic but not the code used to
generate them. In such cases, unless the set of repetitions are split into
subsamples, only one realization of $p_{n,B}^{\ast}$ can be computed.}

A viable way to mitigate this issue is to use the convergence fact
$p_{n,m}^{\ast}\overset{d^{\ast}}{\rightarrow}_{p}%
%TCIMACRO{\TeXButton{U}{\mathscr{U}\!}}%
%BeginExpansion
\mathscr{U}\!%
%EndExpansion
_{[0,1]}$ under specification validity, see Theorem~\ref{Thm 1}. This
convergence and Theorem~\ref{Thm 2} have the following corollary in terms of
$\hat{\pi}_{n,m}(\eta):=\mathbb{P}^{\ast}(p_{n,m}^{\ast}\leq\eta)$, $\eta
\in(0,1)$.

\begin{corollary}
Under the assumptions of Theorem~\ref{Thm 1}, $\hat{\pi}_{n,m}(\eta
)\rightarrow_{p}\eta$ for any \mbox{$\eta\in(0,1)$}. In contrast, under the
assumptions of Theorem~\ref{Thm 2}, $\hat{\pi}_{n,m}(\eta)\rightarrow_{p}1$.
\end{corollary}

Hence, rather than reporting a single draw $p_{n,m}^{\ast}$, the diagnostic
procedure could additionally include an informal check of whether $\hat{\pi
}_{n,m}(\eta)$ substantially exceeds the user-chosen significance level $\eta
$. In practice $\hat{\pi}_{n,m}(\eta)$ is not known but, as is standard, it
can be approximated with any desired precision by drawing an arbitrarily large
number $K$ of i.i.d. (conditionally on the data) realizations of
$p_{n,m}^{\ast}$, say $p_{n,m:k}^{\ast}$, $k=1,\ldots K$, and letting
$\hat{\pi}_{n,m,K}^{\ast}:=K^{-1}\sum_{k=1}^{K}\mathbb{I}_{\{p_{n,m:k}^{\ast
}\leq\eta\}}$.

In principle, the significance of $\hat{\pi}_{n,m,K}^{\ast}(\eta)-\eta$ could
be assessed formally by means of a test, using the (pointwise) standard error
$(\hat{\pi}_{n,m,K}^{\ast}(\eta)(1-\hat{\pi}_{n,m,K}^{\ast}(\eta))/K)^{1/2}$
or, with the null imposed, $(\eta\left(  1-\eta\right)  /K)^{1/2}$. Also a
uniform test over $\eta$ could be performed, using the convergence $\sqrt
{K}\sup_{\eta\in\lbrack0,1]}|(\hat{\pi}_{n,m,K}^{\ast}(\eta)-\eta
)|\overset{w^{\ast}}{\rightarrow}_{p}\sup_{[0,1]}|W|$ as $(n,m,K\rightarrow
\infty)$ implied by Proposition \ref{prop diagnostics} below. Under valid
specification, however, such tests would reproduce the problem of outcome
dependence on the bootstrap variates employed, here $W_{n,k}^{\ast}$
($k=1,...,K$), and results would again differ across researchers.

\begin{proposition}
\label{prop diagnostics}Under the assumptions of Theorem~\ref{Thm 1},
$\sqrt{K}(\hat{\pi}_{n,m,K}^{\ast}(\cdot)-(\cdot))\overset{w^{\ast}%
}{\rightarrow}_{p}W(\cdot)$ on $\mathscr{D}[0,1]$ as $(n,m,K\rightarrow
\infty)$, where $W$ denotes a standard Brownian bridge.
\end{proposition}

\subsection{alternative discrepancy measures}

\label{Sec further discrepancy measures}

The discussion so far has focused on the
%TCIMACRO{\TeXButton{\textsf{KS}}{\textsf{KS}} }%
%BeginExpansion
\textsf{KS}
%EndExpansion
or uniform norm, and the norms it dominates. Apart from, e.g., the mentioned
signed
%TCIMACRO{\TeXButton{\textsf{KS}}{\textsf{KS}} }%
%BeginExpansion
\textsf{KS}
%EndExpansion
and Cramer-von Mises norms, also the seminorm $\sup_{A}|\cdot|$ for an
interval $A$ (e.g., $A:=[1.96,\infty)$), leading to $\hat{d}_{n}(A)=\sup_{u\in
A}|\hat{G}_{n}(u)-\Phi(u)|$, is dominated by the
%TCIMACRO{\TeXButton{\textsf{KS}}{\textsf{KS}} }%
%BeginExpansion
\textsf{KS}
%EndExpansion
norm, and similarly $\hat{d}_{n}(x):=|\hat{G}_{n}(x)-\Phi(x)|$ obtained by
focusing on a single point in the support (e.g., $x=1.96$). They all satisfy
the rate condition in Assumption~\ref{Assn 1} whenever the
%TCIMACRO{\TeXButton{\textsf{KS}}{\textsf{KS}} }%
%BeginExpansion
\textsf{KS}
%EndExpansion
norm does so, and are covered by the theory.

In addition, range-based, quantile-based discrepancy measures or measures
focusing on specific moments of the bootstrap distributions could be used. For
instance, a moment-based discrepancy based on the third and fourth moments can
be defined as%
\[
\hat{d}_{n}:=\Vert v_{n}\Vert_{\Omega}:=v_{n}^{\prime}\Omega v_{n}\text{,
}v_{n}:=\int_{\mathbb{R}}(u^{3},u^{4})^{\prime}(d\hat{G}_{n}(u)-d\Phi
(u))=\mathbb{E}^{\ast}(T_{n}^{\ast3},T_{n}^{\ast4}-3)^{\prime}%
\]
with $\Omega$ a symmetric p.d. matrix. Such discrepancy measures need not be
continuous on $\mathscr{D}_{\mathbb{R}}$ and a strengthening of
(\ref{eq bs consistency}) along the lines discussed, e.g., in Hahn and Liao
(2021) may be required for their convergence to zero in probability.

A minimal justification for the use of $\Vert\cdot\Vert_{\Omega}$ in
applications is that the ensuing $\hat{d}_{n,m}^{\ast}$ can be written as
$\hat{d}_{n,m}^{\ast}=\psi_{m}(T_{n:1}^{\ast},...,T_{n:m}^{\ast})$ for a
continuous $\psi_{m}\nolinebreak:\nolinebreak\mathbb{R}^{m}\nolinebreak%
\rightarrow\nolinebreak\mathbb{R}$. Under the null that $T_{n}^{\ast}%
\overset{d^{\ast}}{\rightarrow}_{p}Z\sim%
%TCIMACRO{\TeXButton{N}{\mathscr{N}\!}}%
%BeginExpansion
\mathscr{N}\!%
%EndExpansion
(0,1)$, it holds that $\hat{d}_{n,m}^{\ast}\overset{d^{\ast}}{\rightarrow}%
_{p}d_{m}^{\ast}:=\psi_{m}(Z_{1},...,Z_{m})$ as $n\rightarrow\infty$, with the
$Z_{i}$'s independent $%
%TCIMACRO{\TeXButton{N}{\mathscr{N}\!}}%
%BeginExpansion
\mathscr{N}\!%
%EndExpansion
(0,1)$, $i=1,...,m$. Therefore, for large $n$, tests could be conducted using
the finite-$m$ quantiles of $\psi_{m}(Z_{1},...,Z_{m})$, or even its large $m$
asymptotic quantiles. Similar considerations apply to the popular
Shapiro-Wilks statistic.

\subsection{alternative null hypotheses}

\label{Sec alternative null hypotheses}

The discrepancy measures discussed so far compare the bootstrap cdf with the $%
%TCIMACRO{\TeXButton{N}{\mathscr{N}\!}}%
%BeginExpansion
\mathscr{N}\!%
%EndExpansion
(0,1)$ cdf. Suppose, however, that interest is in assessing the convergence
$T_{n}^{\ast}\overset{d^{\ast}}{\rightarrow}_{p}%
%TCIMACRO{\TeXButton{N}{\mathscr{N}\!}}%
%BeginExpansion
\mathscr{N}\!%
%EndExpansion
(0,\sigma^{2})$ for some unspecified $\sigma^{2}>0$. This could be done by
redefining the reference statistic as $\tilde{T}_{n}^{\ast}:=T_{n}^{\ast}%
/\hat{\sigma}_{n}$, $\hat{\sigma}_{n}^{2}:=\mathbb{V}^{\ast}[T_{n}^{\ast}]$,
provided $\hat{\sigma}_{n}^{2}$ is consistent\footnote{The condition
$\mathbb{E}^{\ast}|T_{n}^{\ast}|^{2+\epsilon}=O_{p}(1)$ for some $\epsilon>0$
suffices.} for $\sigma^{2}$, such that $\tilde{T}_{n}^{\ast}\overset{d^{\ast}%
}{\rightarrow}_{p}%
%TCIMACRO{\TeXButton{N}{\mathscr{N}\!}}%
%BeginExpansion
\mathscr{N}\!%
%EndExpansion
(0,1)$. Similarly, assessing the convergence $T_{n}^{\ast}\overset{d^{\ast}%
}{\rightarrow}_{p}%
%TCIMACRO{\TeXButton{N}{\mathscr{N}\!}}%
%BeginExpansion
\mathscr{N}\!%
%EndExpansion
(\mu,\sigma^{2})$ with unknown $\mu$ and $\sigma^{2}>0$ can be done by
considering $\check{T}_{n}^{\ast}:=(T_{n}^{\ast}-\hat{\mu}_{n})/\hat{\sigma
}_{n},$ $\hat{\mu}_{n}:=\mathbb{E}^{\ast}[T_{n}^{\ast}]$, such that $\check
{T}_{n}^{\ast}\overset{d^{\ast}}{\rightarrow}_{p}%
%TCIMACRO{\TeXButton{N}{\mathscr{N}\!}}%
%BeginExpansion
\mathscr{N}\!%
%EndExpansion
(0,1)$ if $(\hat{\mu}_{n},\hat{\sigma}_{n}^{2})$ is consistent for
$(\mu,\sigma^{2})$. Here both $\hat{\mu}_{n}$ and $\hat{\sigma}_{n}^{2}$ can
be calculated with arbitrary precision by using a sufficiently large number
$M\gg m$ of bootstrap repetitions.

Because $\hat{\mu}_{n}\ $and $\hat{\sigma}_{n}^{2}$ are functions of the data
(i.e., $D_{n}$-measurable), the theory of Section~\ref{Sec main} can be
applied to $\tilde{T}_{n}^{\ast}$ and $\check{T}_{n}^{\ast}$, such that under
the conditions of Theorem~\ref{Thm 1} their asymptotic distributions are not
affected by uncertainty due to the estimation of $\sigma^{2}$ (and $\mu$).
Specifically, maintaining the notation $\hat{G}_{n}$ for the bootstrap cdf of
$T_{n}^{\ast}$, the rate condition in Assumption~\ref{Assn 1} boils down to
$||\hat{G}(\cdot\hat{\sigma}_{n})-\Phi\left(  \cdot\right)  ||=O_{p}%
(n^{-\alpha})$ and $||\hat{G}(\cdot\hat{\sigma}_{n}+\hat{\mu}_{n})-\Phi\left(
\cdot\right)  ||=O_{p}(n^{-\alpha})$ respectively for $\tilde{T}_{n}^{\ast}$
and $\check{T}_{n}^{\ast}$. The condition can be checked either directly or by
using the estimates%
\begin{align*}
||\hat{G}(\cdot\hat{\sigma}_{n})-\Phi\left(  \cdot\right)  ||  &  \leq
||\hat{G}(\cdot\sigma)-\Phi\left(  \cdot\right)  ||+|\hat{\sigma}_{n}%
^{2}-\sigma^{2}|O_{p}(1),\\
||\hat{G}(\cdot\hat{\sigma}_{n}+\hat{\mu}_{n})-\Phi\left(  \cdot\right)  ||
&  \leq||\hat{G}(\cdot\sigma+\mu)-\Phi\left(  \cdot\right)  ||+(|\hat{\sigma
}_{n}^{2}-\sigma^{2}|+|\hat{\mu}_{n}-\mu|)O_{p}(1)\text{,}%
\end{align*}
which hold whenever $\hat{\sigma}_{n}^{2}$ and $\hat{\mu}_{n}$ are consistent.
Then, if $||\hat{G}(\cdot\sigma)-\Phi\left(  \cdot\right)  ||=O_{p}%
(n^{-\alpha})$ and $\hat{\sigma}_{n}^{2}-\sigma^{2}=O_{p}(n^{-1/2})$, it
follows that $||\hat{G}(\cdot\hat{\sigma}_{n})-\Phi\left(  \cdot\right)
||=O_{p}(n^{-\min\{\alpha,1/2\}})$, and similarly in the case of additional centering.

Finally, statistics such as $%
%TCIMACRO{\TeXButton{T}{\mathscr{T}}}%
%BeginExpansion
\mathscr{T}%
%EndExpansion
_{n,m}^{\ast}:=m^{1/2}\breve{d}_{n,m}^{\ast}$, $\breve{d}_{n,m}^{\ast}%
:=\Vert\breve{G}_{n,m}^{\ast}-\Phi\Vert$, could also be implemented, with
$\breve{G}_{n,m}^{\ast}$ the edf of the standardized bootstrap sample
$(T_{n:i}^{\ast}-m_{n,m}^{\ast})/s_{n,m}^{\ast}$, $i=1,\ldots,m$, and
$m_{n,m}^{\ast}$ and $s_{n,m}^{\ast}$ the sample mean and standard deviation
of $T_{n:1}^{\ast},\ldots,T_{n,m}^{\ast}$, respectively. The resulting tests,
which would be similar to Lilliefors' normality test, are not covered by the
theory in Section~\ref{Sec main} because $(T_{n:i}^{\ast}-m_{n,m}^{\ast
})/s_{n,m}^{\ast}$ are not conditionally i.i.d. given the data. The continuity
considerations of Section~\ref{Sec further discrepancy measures} carry over, however.

\section{applications}

\label{Sec Examples}

We introduce here some additional applications where the bootstrap can be used
to detect specification invalidity. These applications, which are kept
deliberately simple, focus on detecting failures of the following assumptions:
(i) stationarity, (ii) parameter in the interior of the parameter space, (iii)
finite variance, and (iv)\ non-singular Jacobian in applications of the delta
method. For each application, we discuss the set up and conditions ensuring
that the main theory results from Sections~\ref{Sec main} and
\ref{Sec post diagnostics inference} hold.

\subsection{%
%TCIMACRO{\QTR{frametitle}{non-stationarity}}%
%BeginExpansion
non-stationarity%
%EndExpansion
}

\label{Sec Example Non-stationary}

Assume that the data are generated by a standard autoregressive recursion:
\[
y_{t}=\phi_{0}y_{t-1}+\varepsilon_{t}\text{, }t=1,\ldots,n
\]
where $\{\varepsilon_{t}\}_{t=1}^{\infty}$ is an i.i.d. sequence r.v.'s with
$\mathbb{E[}\varepsilon_{t}]=0$ and $\sigma^{2}:=\mathbb{E[}\varepsilon
_{t}^{2}]=1$; $y_{0}$ is assumed to be fixed. The least squares estimator of
$\phi_{0}$ is $\hat{\phi}_{n}:=\sum_{t=1}^{n}y_{t}y_{t-1}/\sum_{t=1}%
^{n}y_{t-1}^{2}$; under the stability condition $|\phi_{0}|<1$, it holds that
\[
T_{n}:=\frac{\hat{\phi}_{n}-\phi_{0}}{\operatorname*{se}(\hat{\phi}_{n}%
)}\overset{d}{\rightarrow}Z\sim%
%TCIMACRO{\TeXButton{N}{\mathscr{N}\!}}%
%BeginExpansion
\mathscr{N}\!%
%EndExpansion
(0,1)
\]
where $\operatorname*{se}(\hat{\phi}_{n}):=\hat{\sigma}_{n}(\sum_{t=1}%
^{n}y_{t-1}^{2})^{-1/2}$, $\hat{\sigma}_{n}^{2}$ being the sample variance of
$\hat{\varepsilon}_{t}:=y_{t}-\hat{\phi}_{n}y_{t-1}$. A simple parametric,
recursive bootstrap generates the bootstrap data as follows:
\[
y_{t}^{\ast}=\hat{\phi}_{n}y_{t-1}^{\ast}+\varepsilon_{t}^{\ast},\text{
}\varepsilon_{t}^{\ast}\sim\text{i.i.d.}%
%TCIMACRO{\TeXButton{N}{\mathscr{N}\!}}%
%BeginExpansion
\mathscr{N}\!%
%EndExpansion
(0,1)\text{, }t=1,\ldots n,
\]
initialized at $y_{0}^{\ast}:=y_{0}$, with $\{\varepsilon_{t}^{\ast}%
\}_{t=1}^{\infty}$ independent of the original data. The bootstrap analog of
$T_{n}$ satisfies, as $n\rightarrow\infty$,%
\[
T_{n}^{\ast}:=\frac{\hat{\phi}_{n}^{\ast}-\hat{\phi}_{n}}{\operatorname*{se}%
(\hat{\phi}_{n}^{\ast})}\overset{d^{\ast}}{\rightarrow}_{p}Z\sim%
%TCIMACRO{\TeXButton{N}{\mathscr{N}\!}}%
%BeginExpansion
\mathscr{N}\!%
%EndExpansion
(0,1),
\]
$\hat{\phi}_{n}^{\ast}$ and $\operatorname*{se}(\hat{\phi}_{n}^{\ast})$ being
the bootstrap analogs of $\hat{\phi}_{n}$ and $\operatorname*{se}(\hat{\phi
}_{n})$, respectively; see Bose (1988).

Suppose now that $\phi_{0}=1$ or, more generally, that $\phi_{0}=\phi
_{n}=1+\lambda n^{-1}$, $\lambda\in\mathbb{R}$, such that the data are
non-stationary. Then%
\[
T_{n}\overset{d}{\rightarrow}\xi(\lambda,B):=\Big(\int_{0}^{1}J_{\lambda
}(u)^{2}du\Big)^{-1/2}\int_{0}^{1}J_{\lambda}(u)dB(u)
\]
where $B$ is a standard Brownian motion on $\mathscr{D}{}_{[0,1]}$ and
$J_{\lambda}$ the Ornstein-Uhlenbeck process on $\mathscr{D}{}_{[0,1]}$
satisfying the stochastic differential equation $dJ=\lambda J+dB$ (see, e.g.,
Phillips, 1987, or Andrews and Guggenberger, 2009). An implication is that the
bootstrap statistic $T_{n}^{\ast}$ has a non-Gaussian, random limit:
\begin{equation}
T_{n}^{\ast}\overset{d^{\ast}}{\rightarrow}_{w}\xi(\ell,B^{\ast}%
)|\ell\label{eq ar(1) boot limit}%
\end{equation}
where the random term $\ell\sim\int_{0}^{1}J_{\lambda}(u)dB(u)/\int_{0}%
^{1}J_{\lambda}(u)^{2}du$ arises from the weak convergence $n(\hat{\phi}%
_{n}-\phi_{0})\overset{d}{\rightarrow}\ell$ and $B^{\ast}$ is a standard
Brownian motion on $[0,1]$, independent of $\ell$; see Basawa et al. (1991).
In terms of cdf's, (\ref{eq ar(1) boot limit}) is equivalent to the weak
convergence in $\mathscr{D}{}_{\mathbb{R}}$%
\begin{equation}
\hat{G}_{n}(x):=\mathbb{P}^{\ast}(T_{n}^{\ast}\leq x)\overset{w}{\rightarrow}%
%TCIMACRO{\TeXButton{G}{\mathscr{G}\!}}%
%BeginExpansion
\mathscr{G}\!%
%EndExpansion
(x):=\mathbb{P}\left(  \xi(\ell,B^{\ast})\leq x|\ell\right)  .
\label{eq limit G for UR example}%
\end{equation}
Again, $%
%TCIMACRO{\TeXButton{G}{\mathscr{G}\!}}%
%BeginExpansion
\mathscr{G}\!%
%EndExpansion
\neq\Phi$ a.s. as asymptotic normality of the bootstrap statistic fails.

\paragraph*{validity of the diagnostic procedure.}

Bose (1988) provides sufficient conditions for the nonparametric bootstrap
based on least-squares residuals to admit an Edgeworth expansion. These
primarily require that $\mathbb{E}[\varepsilon_{t}^{8}]<\infty$ and that the
autoregressive characteristic roots (i.e., $1/\phi_{0}$) lie outside the unit
circle; see his conditions (A.1)--(A.3). In particular, these conditions imply
that $\parallel\hat{G}_{n}-\Phi\parallel_{\infty}=O_{p}(n^{-1/2})$; hence,
Assumption%
%TCIMACRO{\TeXButton{~}{~}}%
%BeginExpansion
~%
%EndExpansion
\ref{Assn 1} is satisfied with $\alpha=1/2$ for the
%TCIMACRO{\TeXButton{\textsf{KS}}{\textsf{KS}} }%
%BeginExpansion
\textsf{KS}
%EndExpansion
norm and dominated norms. Consequently, by Theorem%
%TCIMACRO{\TeXButton{~}{~}}%
%BeginExpansion
~%
%EndExpansion
\ref{Thm 1} validity of the diagnostic procedure for the null hypothesis of
stationarity requires that $m/n\rightarrow0$ as $(n,m\rightarrow\infty)$. If,
as suggested earlier in this section, the Gaussian parametric bootstrap is
used instead, then (A.1) and (A.2) in Bose (1988)\ are automatically satisfied
on the bootstrap data, irrespective of the properties of the original
$\varepsilon_{t}$'s, provided that $\hat{\phi}_{n}$, which is used to generate
the bootstrap data recursively, is consistent.

Under non-stationarity ($\phi_{0}=\phi_{n}=1+\lambda n^{-1}$), it holds that
$\parallel\hat{G}_{n}-\Phi\parallel_{\infty}\overset{d}{\rightarrow}%
%TCIMACRO{\TeXButton{Y}{\mathscr{Y}}}%
%BeginExpansion
\mathscr{Y}%
%EndExpansion
:=\parallel%
%TCIMACRO{\TeXButton{G}{\mathscr{G}\!}}%
%BeginExpansion
\mathscr{G}\!%
%EndExpansion
-\Phi\parallel_{\infty}>0$ with $%
%TCIMACRO{\TeXButton{G}{\mathscr{G}\!}}%
%BeginExpansion
\mathscr{G}\!%
%EndExpansion
$ as in (\ref{eq limit G for UR example}), and Theorem~\ref{Thm 2} guarantees
that non-stationarity can be detected with probability approaching one by
tests employing the
%TCIMACRO{\TeXButton{\textsf{KS}}{\textsf{KS}} }%
%BeginExpansion
\textsf{KS}
%EndExpansion
or dominated norms.

\subsection{parameter on the boundary}

\label{Sec Example boundary}

As in Andrews (2000), consider an i.i.d. sample $D_{n}:=\{y_{i}\}_{i=1}^{n}$
from a population with $\mathbb{E[}y_{i}]=:\theta_{0}\in\Theta:=[0,\infty)$,
$\Theta$ being the parameter space, and $\mathbb{V[}y_{i}^{2}]=1$. The
Gaussian quasi-maximum likelihood estimator [QMLE], given by $\hat{\theta}%
_{n}:=\max\{0,\bar{y}_{n}\}$ with $\bar{y}_{n}:=n^{-1}%
%TCIMACRO{\tsum \nolimits_{i=1}^{n}}%
%BeginExpansion
{\textstyle\sum\nolimits_{i=1}^{n}}
%EndExpansion
y_{i}$, is such that, when $\theta_{0}\in\operatorname*{int}\Theta$,%
\begin{equation}
T_{n}:=\sqrt{n}(\hat{\theta}_{n}-\theta_{0})=\max\{-\sqrt{n}\theta_{0}%
,\sqrt{n}(\bar{y}_{n}-\theta_{0})\}\overset{d}{\rightarrow}Z\text{, }Z\sim%
%TCIMACRO{\TeXButton{N}{\mathscr{N}\!}}%
%BeginExpansion
\mathscr{N}\!%
%EndExpansion
(0,1). \label{eq asy for Tn no boundary}%
\end{equation}

Consider the Gaussian parametric bootstrap MLE, $\hat{\theta}_{n}^{\ast}%
:=\max\{0,\bar{y}_{n}^{\ast}\}$, where conditionally on the original data
$D_{n}$, $\bar{y}_{n}^{\ast}\sim%
%TCIMACRO{\TeXButton{N}{\mathscr{N}\!}}%
%BeginExpansion
\mathscr{N}\!%
%EndExpansion
(\hat{\theta}_{n},n^{-1/2})$. The bootstrap analog of $T_{n}$ is $T_{n}^{\ast
}:=\sqrt{n}(\hat{\theta}_{n}^{\ast}-\hat{\theta}_{n})=\max\{-\sqrt{n}%
\hat{\theta}_{n},$ $\sqrt{n}(\bar{y}_{n}^{\ast}-\hat{\theta}_{n})\}$ and
satisfies
\[
T_{n}^{\ast}|D_{n}\sim\max\{-\sqrt{n}\hat{\theta}_{n},Z^{\ast}\}\text{,
}Z^{\ast}\sim%
%TCIMACRO{\TeXButton{N}{\mathscr{N}\!}}%
%BeginExpansion
\mathscr{N}\!%
%EndExpansion
(0,1)
\]
with conditional cdf $\hat{G}_{n}(x)=\Phi(x)\mathbb{I}_{\mathbb{\{}x\geq
-\sqrt{n}\hat{\theta}_{n}\}}$. When $\theta_{0}$ is an interior point of
$\Theta$, since $\sqrt{n}\hat{\theta}_{n}\rightarrow-\infty$ (a.s.) as
$n\rightarrow\infty$, it holds that%
\begin{equation}
T_{n}^{\ast}\overset{d^{\ast}}{\rightarrow}_{p}Z^{\ast}\text{, }Z^{\ast}\sim%
%TCIMACRO{\TeXButton{N}{\mathscr{N}\!}}%
%BeginExpansion
\mathscr{N}\!%
%EndExpansion
(0,1) \label{eq Tn* with boundary}%
\end{equation}
and the bootstrap replicates the standard normal distribution.

Suppose now that $\theta_{0}$ is on the boundary ($\theta_{0}=0$) or `close'
to it ($\theta_{0}=\lambda n^{-1/2}$, $\lambda\in\lbrack0,\infty)$). In this
case, (\ref{eq asy for Tn no boundary}) no longer holds; instead
\[
T_{n}=\max\{-\sqrt{n}\theta_{0},\sqrt{n}(\bar{y}_{n}-\theta_{0})\}\overset
{d}{\rightarrow}\xi(\lambda):=\max\{-\lambda,Z\}
\]
which differs from a normal random variable. Similarly, and in contrast to
(\ref{eq Tn* with boundary}), the bootstrap cdf satisfies $\hat{G}_{n}%
(\cdot)=\Phi(\cdot)\mathbb{I}_{\mathbb{\{}\cdot\geq-\sqrt{n}\hat{\theta}%
_{n}\}}\rightarrow_{w}\Phi(\cdot)\mathbb{I}_{\mathbb{\{}\cdot\geq-\ell\}}$ in
$\mathscr{D}{}_{\mathbb{R}}$, which is established using the convergence fact
$\sqrt{n}\hat{\theta}_{n}=\sqrt{n}(\hat{\theta}_{n}-\theta_{0})+\sqrt{n}%
\theta_{0}\overset{d}{\rightarrow}\xi(\lambda)+\lambda=:\ell$. This is
equivalent to the weak convergence in distribution%
\[
T_{n}^{\ast}\overset{d^{\ast}}{\rightarrow}_{w}\xi^{\ast}(\ell)|\ell
\]
where $\xi^{\ast}(\ell):=\max\{-\ell,Z^{\ast}\}$, $Z^{\ast}\sim%
%TCIMACRO{\TeXButton{N}{\mathscr{N}\!}}%
%BeginExpansion
\mathscr{N}\!%
%EndExpansion
(0,1)$ independent of $\ell$. Since $\ell$ is finite a.s., the conditional cdf
of $T_{n}^{\ast}$ is non-Gaussian in the limit, with probability one.

\paragraph*{validity of the diagnostic procedure.}

As seen above, conditionally on the data the parametric bootstrap statistic
$T_{n}^{\ast}$ has conditional cdf $\hat{G}_{n}(x)=\Phi(x)\mathbb{I}%
_{\{x\geq-\sqrt{n}\hat{\theta}_{n}\}}$, and the associated
%TCIMACRO{\TeXButton{\textsf{KS}}{\textsf{KS}}}%
%BeginExpansion
\textsf{KS}%
%EndExpansion
\ distance $\hat{d}_{n}:=\parallel\hat{G}_{n}-\Phi\parallel_{\infty}$
satisfies%
\[
\hat{d}_{n}=\sup_{x\in\mathbb{R}}|\Phi(x)\mathbb{I}_{\{x\geq-\sqrt{n}%
\hat{\theta}_{n}\}}-\Phi(x)|=\sup_{x\in\mathbb{R}}|\Phi(x)\mathbb{I}%
_{\{x\leq-\sqrt{n}\hat{\theta}_{n}\}}|=\Phi(-\sqrt{n}\hat{\theta}_{n})\text{.}%
\]
Under the null that $\theta_{0}$ is a fixed interior point ($\theta_{0}>0$),
$\hat{d}_{n}\overset{p}{\rightarrow}0$ exponentially fast, and the
%TCIMACRO{\TeXButton{\textsf{KS}}{\textsf{KS}} }%
%BeginExpansion
\textsf{KS}
%EndExpansion
norm satisfies Assumption~\ref{Assn 1} for any $\alpha>0$. This implies that
any power growth rate of $m$ in terms of $n$ is allowed in Theorem~\ref{Thm 1}
for tests employing the
%TCIMACRO{\TeXButton{\textsf{KS}}{\textsf{KS}}}%
%BeginExpansion
\textsf{KS}%
%EndExpansion
\ distance.

In contrast, when $\theta_{0}$ is on (or near) the boundary, $\hat{d}_{n}$
satisfies%
\[
\hat{d}_{n}=\Phi(-\sqrt{n}\hat{\theta}_{n})\overset{d}{\rightarrow}%
%TCIMACRO{\TeXButton{Y}{\mathscr{Y}}}%
%BeginExpansion
\mathscr{Y}%
%EndExpansion
=\Phi\left(  -\ell\right)  ,
\]
with $\ell$ as previously defined. Thus, $\hat{d}_{n,m}^{\ast}$ diverges at
the rate of $\sqrt{m}$ as $(n,m\rightarrow\infty)$.

\subsection{infinite variance}

\label{sec example InfV}

As in Section~\ref{Sec Example boundary}, assume that $D_{n}:=\{y_{i}%
\}_{i=1}^{n}$ where the $y_{i}$'s are i.i.d. with $\theta_{0}:=\mathbb{E[}%
y_{i}]$ and $\sigma^{2}:=\mathbb{V[}y_{i}^{2}]\in(0,\infty)$. This assumption
implies that the sample mean $\hat{\theta}_{n}:=n^{-1}\sum_{i=1}^{n}y_{i}$
satisfies $T_{n}:=\sqrt{n}(\hat{\theta}_{n}-\theta_{0})/\sigma\overset
{d}{\rightarrow}Z$, $Z\sim%
%TCIMACRO{\TeXButton{N}{\mathscr{N}\!}}%
%BeginExpansion
\mathscr{N}\!%
%EndExpansion
(0,1)$, as $n\rightarrow\infty$. With $\{y_{i}^{\ast}\}_{i=1}^{n}$ denoting a
sample of $n$ draws, independent conditionally on $D_{n}$, from the edf of
$\{y_{i}\}_{i=1}^{n}$, the i.i.d. bootstrap counterpart of $T_{n}$ is
$T_{n}^{\ast}:=\sqrt{n}(\hat{\theta}_{n}^{\ast}-\hat{\theta}_{n})/\hat{\sigma
}_{n}$, where $\hat{\theta}_{n}^{\ast}:=n^{-1}\sum_{i=1}^{n}y_{i}^{\ast}$,
$\hat{\sigma}_{n}^{2}:=n^{-1}\sum_{i=1}^{n}(y_{i}-\hat{\theta}_{n})^{2}$. As
is known, as $n\rightarrow\infty$ (e.g., Singh, 1981), $T_{n}^{\ast}%
\overset{d^{\ast}}{\rightarrow}_{p}Z$, $Z\sim%
%TCIMACRO{\TeXButton{N}{\mathscr{N}\!}}%
%BeginExpansion
\mathscr{N}\!%
%EndExpansion
(0,1)$; in terms of cdfs, $\hat{G}_{n}(x):=\mathbb{P}^{\ast}(T_{n}^{\ast}\leq
x)\overset{p}{\rightarrow}\Phi(x)$.

Consider now the case where $\mathbb{E[}y_{t}^{2}]=\infty$. Specifically,
assume that $y_{t}$ is in the domain of attraction of a symmetric stable law
with tail index $\nu\in(1,2)$, denoted as $%
%TCIMACRO{\TeXButton{S}{\mathscr{S}\!\!}}%
%BeginExpansion
\mathscr{S}\!\!%
%EndExpansion
$\thinspace$(\nu)$. In this case the CLT fails to hold; instead, for some
diverging real sequence $\{a_{n}\}$,
\[
a_{n}n^{1/2}T_{n}=a_{n}n(\hat{\theta}_{n}-\theta_{0})\overset{d}{\rightarrow}%
%TCIMACRO{\TeXButton{S}{\mathscr{S}\!\!}}%
%BeginExpansion
\mathscr{S}\!\!%
%EndExpansion
\,(\nu);
\]
see, e.g., Feller (1971). Note that $%
%TCIMACRO{\TeXButton{S}{\mathscr{S}\!\!}}%
%BeginExpansion
\mathscr{S}\!\!%
%EndExpansion
\,\left(  \nu\right)  $ can be written as $%
%TCIMACRO{\TeXButton{S}{\mathscr{S}\!\!}}%
%BeginExpansion
\mathscr{S}\!\!%
%EndExpansion
$\thinspace$(\nu)\sim\sum\nolimits_{k=1}^{\infty}\delta_{k}Z_{k}$, where the
$\delta_{k}$'s are i.i.d. Rademacher and $Z_{k}^{1/\nu}:=\sum_{i=1}^{k}E_{i}$
with $\{E_{i}\}_{i=1}^{\infty}$ an i.i.d. sequence of exponential r.v.'s with
$\mathbb{E}[E_{i}]=1$; see Lepage, Woodroofe and Zinn (1981). Athreya (1987)
and Knight (1989) show that it this case also the i.i.d. bootstrap counterpart
of $T_{n}$ is not asymptotically Gaussian; precisely, the bootstrap measure is
random in the limit:
\begin{equation}
T_{n}^{\ast}:=a_{n}n(\hat{\theta}_{n}^{\ast}-\hat{\theta}_{n})\overset
{d^{\ast}}{\rightarrow}_{w}\frac{\sum\nolimits_{k=1}^{\infty}\delta_{k}%
Z_{k}(M_{k}^{\ast}-1)}{(\sum\nolimits_{k=1}^{\infty}Z_{k}^{2})^{1/2}}%
%TCIMACRO{\TeXButton{\Big|}{\Big|}}%
%BeginExpansion
\Big|%
%EndExpansion
\{\delta_{k},Z_{k}\}\text{,} \label{eq infv 1}%
\end{equation}
where the $\delta_{k}$'s and $Z_{k}$'s are as previously defined. The
$M_{k}^{\ast}$'s, which are i.i.d. $%
%TCIMACRO{\TeXButton{P}{\mathscr{P}\!}}%
%BeginExpansion
\mathscr{P}\!%
%EndExpansion
(1)$ r.v.'s, independent of $\{\delta_{k},Z_{k}\}$, induce the randomness in
the limit bootstrap measure; see Theorem 2 in Knight (1989). If a wild
bootstrap with Rademacher multipliers is used instead, that is, $\hat{\theta
}_{n}^{\ast}:=\hat{\theta}_{n}+n^{-1}\sum_{i=1}^{n}(y_{i}-\hat{\theta}%
_{n})w_{i}^{\ast}$, with the $w_{i}^{\ast}$'s being i.i.d. Rademacher random
variables conditionally on $D_{n}$, then $T_{n}^{\ast}\overset{d^{\ast}%
}{\rightarrow}_{w}\sum\nolimits_{k=1}^{\infty}\delta_{k}^{\ast}Z_{k}%
(\sum\nolimits_{k=1}^{\infty}Z_{k}^{2})^{-1/2}%
%TCIMACRO{\TeXButton{\big|}{\big|}}%
%BeginExpansion
\big|%
%EndExpansion
\{Z_{k}\}$, see Cavaliere, Georgiev and Taylor (2013, 2016). For both
bootstrap schemes, the asymptotic bootstrap measure is random and a.s. non-Gaussian.

\paragraph*{validity of the diagnostic procedure.}

Consider first the case of valid specification where the $y_{t}$'s have finite
variance. As seen above, $\hat{G}_{n}(x):=\mathbb{P}^{\ast}(T_{n}^{\ast}\leq
x)\overset{p}{\rightarrow}\Phi(x)$. The rate of the previous convergence,
required as an input in Assumption~\ref{Assn 1}, can be established under
slightly more than finite second moments.

Specifically, if $\mathbb{E[}|y_{t}|^{2\kappa}]<\infty$ for some $\kappa>1$,
then by the Berry-Esseen bound%
\[
\Vert\hat{G}_{n}-\Phi\Vert_{\infty}\leq\frac{C}{n^{1/2}}\mathbb{E}^{\ast
}[|y_{i}^{\ast}-\hat{\theta}_{n}|^{3}]=\frac{C}{n^{1/2}}\frac{1}{n}\sum
_{i=1}^{n}|y_{i}-\hat{\theta}_{n}|^{3}%
\]
For $\kappa\geq\frac{3}{2}$ (such that $\mathbb{E}|y_{i}|^{3}<\infty$), the
r.h.s. of the previous equation is $O_{a.s.}(n^{-1/2})$, while for $\kappa
\in\left(  1,3/2\right)  $ it is $o_{a.s.}(n^{-\frac{3(\kappa-1)}{2\kappa}})$
by the Marcinkiewicz-Zigmund strong law. Hence, the
%TCIMACRO{\TeXButton{\textsf{KS}}{\textsf{KS}} }%
%BeginExpansion
\textsf{KS}
%EndExpansion
norm satisfies Assumption~\ref{Assn 1} with $\alpha=\min\{\frac{3(\kappa
-1)}{2\kappa},\frac{1}{2}\}$ and Theorem~\ref{Thm 1} applies. Agnostically,
one may select $m$ as growing at a logarithmic rate, e.g., $m=\ln(n)$, which
would satisfy the requirement in Theorem~\ref{Thm 1} for any $\kappa>1$.

Consider now the case where $\mathbb{E[}y_{i}^{2}]=\infty$. Then, $T_{n}%
^{\ast}$ has the random limit given in (\ref{eq infv 1}); by Theorem 3 in
Knight (1989), its random limiting cdf, $%
%TCIMACRO{\TeXButton{G}{\mathscr{G}\!}}%
%BeginExpansion
\mathscr{G}\!%
%EndExpansion
$ say, is a.s. sample-path continuous. Hence, (\ref{eq infv 1}) implies that
$\Vert\hat{G}_{n}-\Phi\Vert_{\infty}\rightarrow_{d}%
%TCIMACRO{\TeXButton{Y}{\mathscr{Y}}}%
%BeginExpansion
\mathscr{Y}%
%EndExpansion
:=\Vert%
%TCIMACRO{\TeXButton{G}{\mathscr{G}\!}}%
%BeginExpansion
\mathscr{G}\!%
%EndExpansion
-\Phi\Vert_{\infty}>0$ a.s.; Theorem~\ref{Thm 2} applies and the diagnostic
test based on the
%TCIMACRO{\TeXButton{\textsf{KS}}{\textsf{KS}} }%
%BeginExpansion
\textsf{KS}
%EndExpansion
norm rejects with probability approaching one.

\subsection{near-singular Jacobian}

\label{Sec Example Jaco}

Suppose that an estimator $\hat{\theta}_{n}$ of an unknown parameter
$\theta_{0}$ satisfies
\begin{equation}
Z_{n}:=\sqrt{n}(\hat{\theta}_{n}-\theta_{0})/\sigma\overset{d}{\rightarrow
}Z\sim%
%TCIMACRO{\TeXButton{N}{\mathscr{N}\!}}%
%BeginExpansion
\mathscr{N}\!%
%EndExpansion
(0,1), \label{eq example Jaco initial est}%
\end{equation}
and a bootstrap analog $Z_{n}^{\ast}:=\sqrt{n}(\hat{\theta}_{n}^{\ast}%
-\hat{\theta}_{n})/\hat{\sigma}_{n}$ is available such that $Z_{n}^{\ast
}\overset{d^{\ast}}{\rightarrow}_{p}Z^{\ast}\sim%
%TCIMACRO{\TeXButton{N}{\mathscr{N}\!}}%
%BeginExpansion
\mathscr{N}\!%
%EndExpansion
(0,1)$ jointly with (\ref{eq example Jaco initial est}), with $Z^{\ast}$
independent of $Z$ and $\hat{\sigma}_{n}^{2}\overset{p}{\rightarrow}%
\tilde{\sigma}^{2}>0$ not necessarily equal to $\sigma^{2}$.

Consider inference on $\tau_{0}:=g(\theta_{0})$, where $g$ is twice
continuously differentiable. With $\hat{\tau}_{n}:=g(\hat{\theta}_{n})$, let
$T_{n}:=\sqrt{n}(\hat{\tau}_{n}-\tau_{0})$. By a second-order expansion around
$\theta_{0}$,
\[
T_{n}=\dot{g}Z_{n}+\tfrac{1}{2}g^{\prime\prime}(\bar{\theta}_{n}%
)n^{-1/2}\sigma^{2}Z_{n}^{2}%
\]
where $\bar{\theta}_{n}$ is on the line segment between $\theta_{0}$ and
$\hat{\theta}_{n}$, and $\dot{g}:=g^{\prime}(\theta_{0})$. Provided $\dot
{g}\neq0$, it holds that $T_{n}\overset{d}{\rightarrow}\sigma\dot{g}Z$.

Now, consider the bootstrap analog of $T_{n}$. With $\hat{\tau}_{n}^{\ast
}:=g(\hat{\theta}_{n}^{\ast})$ and $\hat{\tau}_{n}:=g(\hat{\theta}_{n})$, we
have
\[
T_{n}^{\ast}:=\sqrt{n}(\hat{\tau}_{n}^{\ast}-\hat{\tau}_{n})=g^{\prime}%
(\hat{\theta}_{n})\hat{\sigma}_{n}Z_{n}^{\ast}+\tfrac{1}{2}g^{\prime\prime
}(\bar{\theta}_{n}^{\ast})n^{-1/2}\hat{\sigma}_{n}^{2}Z_{n}^{\ast2}%
\]
for some $\bar{\theta}_{n}^{\ast}$ between $\hat{\theta}_{n}$ and $\hat
{\theta}_{n}^{\ast}$. As for $T_{n}$, if $\dot{g}\neq0$ then $T_{n}^{\ast
}\overset{d^{\ast}}{\rightarrow}_{p}\tilde{\sigma}\dot{g}Z$; hence,
$T_{n}^{\ast}$ is asymptotically normal.

Suppose instead that $\dot{g}:=g^{\prime}(\theta_{0})=\lambda n^{-1/2}$,
$\lambda\in\lbrack0,\infty)$, such that the Jacobian is singular ($\lambda=0$)
or nearly singular ($\lambda\in\left(  0,\infty\right)  $). Then $T_{n}%
=o_{p}(1)$ while, provided $\ddot{g}:=g^{\prime\prime}(\theta_{0})\neq0$,
$\sqrt{n}T_{n}$ has a chi-square-type limit distribution:
\[
\sqrt{n}T_{n}=\lambda\sigma Z_{n}+\tfrac{1}{2}g^{\prime\prime}(\hat{\theta
}_{n}^{\ast})\sigma^{2}Z_{n}^{2}\overset{d}{\rightarrow}\lambda\sigma
Z+\tfrac{1}{2}\ddot{g}\sigma^{2}Z^{2}.
\]
Similarly, using the convergence fact
\[
\sqrt{n}g^{\prime}(\hat{\theta}_{n})=\sqrt{n}g^{\prime}(\theta_{0}%
)+g^{\prime\prime}(\theta_{0})\sqrt{n}(\hat{\theta}_{n}-\theta_{0}%
)+o_{p}(1)\overset{d}{\rightarrow}\ell:=\lambda+\ddot{g}\sigma Z,
\]
which holds jointly with the convergence of $Z_{n}^{\ast}$, it follows that
$T_{n}^{\ast}$ satisfies%
\begin{equation}
\sqrt{n}T_{n}^{\ast}=n(\hat{\tau}_{n}^{\ast}-\hat{\tau}_{n})=\sqrt{n}%
g^{\prime}(\hat{\theta}_{n})\hat{\sigma}_{n}Z_{n}^{\ast}+\tfrac{1}{2}%
g^{\prime\prime}(\bar{\theta}_{n}^{\ast})\hat{\sigma}_{n}^{2}Z_{n}^{\ast
2}\overset{d^{\ast}}{\rightarrow}_{w}\left.  \ell\tilde{\sigma}Z^{\ast}%
+\tfrac{1}{2}\ddot{g}\tilde{\sigma}^{2}Z^{\ast2}\right\vert \ell
\label{eq jaco bootstrap asy}%
\end{equation}
where $Z^{\ast}\sim%
%TCIMACRO{\TeXButton{N}{\mathscr{N}\!}}%
%BeginExpansion
\mathscr{N}\!%
%EndExpansion
(0,1)$ is independent of $\ell$. The right hand side of
(\ref{eq jaco bootstrap asy}) defines a random, non-Gaussian distribution,
which we denote by $%
%TCIMACRO{\TeXButton{G}{\mathscr{G}\!}}%
%BeginExpansion
\mathscr{G}\!%
%EndExpansion
$.

\paragraph*{validity of the diagnostic procedure.}

Assume first that the Jacobian is non-singular; i.e., $\ddot{g}\neq0$. If
$Z_{n}^{\ast}$ admits a standard Edgeworth expansion, such that its
conditional cdf, say $\hat{F}_{n}$, satisfies $\parallel\hat{F}_{n}%
-\Phi\parallel_{\infty}=O_{p}(n^{-1/2})$, then also $\hat{G}_{n}$, the
conditional cdf of $T_{n}^{\ast}$, satisfies $\parallel\hat{G}_{n}(g^{\prime
}(\hat{\theta}_{n})\hat{\sigma}_{n}\cdot)-\Phi(\cdot)\parallel_{\infty}%
=O_{p}(n^{-1/2})$; see Appendix~\ref{Appendix additional proofs}. As $\hat
{G}_{n}(g^{\prime}(\hat{\theta}_{n})\hat{\sigma}_{n}\cdot)\ $is the edf of the
bootstrap t-ratio $\tilde{T}_{n}^{\ast}:=T_{n}^{\ast}/(g^{\prime}(\hat{\theta
}_{n})\hat{\sigma}_{n})$, which is well-defined with probability approaching
one, it follows that Assumption~\ref{Assn 1} is verified with $\alpha=1/2$ for
$\tilde{T}_{n}^{\ast}$ and the
%TCIMACRO{\TeXButton{\textsf{KS}}{\textsf{KS}} }%
%BeginExpansion
\textsf{KS}
%EndExpansion
norm, and Theorem~\ref{Thm 1} applies for diagnostics based on $\tilde{T}%
_{n}^{\ast}$.

In contrast, when the Jacobian is near zero, $\sqrt{n}g^{\prime}(\hat{\theta
}_{n})\overset{d}{\rightarrow}\ell$ and (\ref{eq jaco bootstrap asy}) hold
jointly. As a result, it is shown in Appendix~\ref{Appendix additional proofs}
that
\[
\hat{d}_{n}:=\Vert\hat{G}_{n}(g^{\prime}(\hat{\theta}_{n})\hat{\sigma}%
_{n}\cdot)-\Phi\Vert_{\infty}\rightarrow1-%
%TCIMACRO{\TeXButton{G}{\mathscr{G}\!}}%
%BeginExpansion
\mathscr{G}\!%
%EndExpansion
(0)\mathbb{I}_{\{\ell\geq0\}}>0\text{ a.s.}%
\]
Further, the conclusions of Theorem~\ref{Thm 2} are valid for $\tilde{T}%
_{n}^{\ast}$-based diagnostics employing the
%TCIMACRO{\TeXButton{\textsf{KS}}{\textsf{KS}} }%
%BeginExpansion
\textsf{KS}
%EndExpansion
norm; see again Appendix~\ref{Appendix additional proofs}.

\section{an empirical illustration}

\label{Sec Empirical}

To illustrate the usefulness and potential of our bootstrap diagnostic
approach, in this section we consider an example from the empirical
macroeconomic framework. We focus on the strategy employed by K\"{a}nzig
(2021) to identify a structural oil supply news shock within a structural VAR
identified with an external instrument (SVAR-IV), see Stock and Watson (2018).
The novelty of K\"{a}nzig's (2021) analysis lies in his construction of an
external instrument, $z_{t}$, for the oil supply news shock, $\varepsilon
_{\text{oil},t}$, by extending the high-frequency (HF) approach, originally
introduced by Gertler and Karadi (2015), outside the monetary policy
framework. Specifically, in Kanzig's (2021) baseline model, the instrument
$z_{t}$ is the first principal component of six time series which reflect
variations in oil price futures with different maturities around OPEC
production announcements.

K\"{a}nzig's (2021) model includes six variables ($g=6$): real oil prices,
world oil production, world oil inventories, world industrial production, US
industrial production, and the US consumer price index (CPI). These variables,
observed monthly over the period 1974M1--2017M12 ($n=528$), are collected in
the $g\times1$ vector $y_{t}$ and modeled through a VAR system with $p=12$
lags. Interest is in the dynamic responses of $y_{t+h}$ at horizon
$h\in\{0,1,\ldots\}$ to an oil supply news shock of magnitude $\varepsilon
_{\text{oil},t}:=\mathsf{x}$, measured as%
\begin{equation}%
%TCIMACRO{\TeXButton{\textsf{IRF}}{\textsf{IRF}}}%
%BeginExpansion
\textsf{IRF}%
%EndExpansion
(h):=\mathbb{E[}y_{t+h}\mid\varepsilon_{\text{oil},t}=\mathsf{x}%
,\varepsilon_{2,t}=0,\mathcal{I}_{t-1}]-\mathbb{E[}y_{t+h}\mid\varepsilon
_{t}=0\mathsf{,}\mathcal{I}_{t-1}]=(R^{\prime}\mathcal{C}_{\Pi}^{h}%
R)b\mathsf{x} \label{eq IRFs population}%
\end{equation}
where $\varepsilon_{t}:=(\varepsilon_{\text{oil},t},\varepsilon_{2,t}^{\prime
})^{\prime}$, $\varepsilon_{2,t}$ being the vector of latent non-target (not
of interest) shocks, $\mathcal{I}_{t-1}$ is the information set at time $t-1$,
$\mathcal{C}_{\Pi}$ the\ VAR\ companion matrix, $R$ a selection matrix such
that $R^{\prime}R=I_{g}$, and $b$ a $g\times1$ vector of structural parameters
capturing the instantaneous response of the variables to the shock (note that
$%
%TCIMACRO{\TeXButton{\textsf{IRF}}{\textsf{IRF}}}%
%BeginExpansion
\textsf{IRF}%
%EndExpansion
(0)=b\mathsf{x}$).

The dynamic causal effects in (\ref{eq IRFs population}) are identified and
estimated using $z_{t}$ as instrument for $\varepsilon_{\text{oil},t}$. Under
relevance and exogeneity of $z_{t}$, i.e., $\mathbb{E}[z_{t}\varepsilon
_{\text{oil},t}]=\phi\neq0$ and $\mathbb{E}[z_{t}\varepsilon_{2,t}^{\prime
}]=0$,
\begin{equation}
\mathbb{E}[u_{t}z_{t}]=\phi b\text{, }b=(b_{1},b_{2}^{\prime})^{\prime
}\text{;} \label{monent_conditions2}%
\end{equation}
here, $b$ is partitioned conformably with the vector of structural shocks. The
moment condition (\ref{monent_conditions2}) is the key ingredient for the
estimation of (\ref{eq IRFs population}). With $u_{t}:=(u_{1,t},u_{2,t}%
^{\prime})^{\prime}$ (and $y_{t}:=(y_{1,t},y_{2,t}^{\prime})^{\prime}$) also
partitioned conformably, (\ref{monent_conditions2}) implies the
representation:%
\begin{equation}
\beta:=b_{2}/b_{1}=\mathbb{E[}u_{2,t}z_{t}\mathbb{]}/\mathbb{E[}u_{1,t}%
z_{t}\mathbb{]} \label{relative_effect}%
\end{equation}
where $\beta$ captures the (relative) instantaneous responses of the variables
in $y_{2,t}$ to a supply news shock of magnitude $\varepsilon_{\text{oil}%
,t}=\mathsf{x}=b_{1}^{-1}$, i.e., such that the instantaneous response of
$y_{1,t}$ to $\varepsilon_{\text{oil},t}$ equals $1$ (the so-called `unit
effect' normalization). The IV estimator of $\beta$ is given by $\hat{\beta
}:=S_{\hat{u}_{1}z}^{-1}S_{\hat{u}_{2}z}$, where $\hat{u}_{t}:=(\hat{u}%
_{1,t},\hat{u}_{2,t}^{\prime})^{\prime}$, $t=1,...,n$, are the VAR residuals.
For $\mathsf{x}=b_{1}^{-1}$, the IRFs (\ref{eq IRFs population})\ are
estimated by replacing $\beta$ with $\hat{\beta}$ and the VAR\ companion
matrix $\mathcal{C}_{\Pi}$ with the corresponding OLS estimate.

Similarly to the IV regression framework of Section~\ref{sec example InfV},
Montiel Olea, Stock and Watson (2020) show that if $z_{t}$ is a weak
instrument, the estimator of $%
%TCIMACRO{\TeXButton{\textsf{IRF}}{\textsf{IRF}}}%
%BeginExpansion
\textsf{IRF}%
%EndExpansion
(h)$ is inconsistent and asymptotically non-Gaussian. As is typical in the
literature, K\"{a}nzig (2021) pre-tests the strength of the instrument $z_{t}$
using an F-test from a first-stage regression of $\hat{u}_{1,t}$ on $z_{t}$.
The reported `regular' F-statistic in his Table 1 is $22.7$, which is above
the homoskedastic threshold of $10.3$ (Stock and Yogo, 2005). In addition,
K\"{a}nzig (2021) reports a `robust' F-statistic of $10.6$. Based on this
evidence, which K\"{a}nzig (2021) considers supportive of $z_{t}$ being a
strong instrument, the IRFs are estimated as described above, and their
uncertainty is assessed using $68\%$ and $90\%$ moving block bootstrap
confidence intervals (Jentsch and Lunsford, 2019, 2022), based on $10,000$
bootstrap replications. These confidence intervals, shown in K\"{a}nzig's
Figure 3, appear broadly in line with the uncertainty commonly encountered by
applied macroeconomists.

\begin{figure}[t]
\centering
\includegraphics[width=1.0\textwidth]{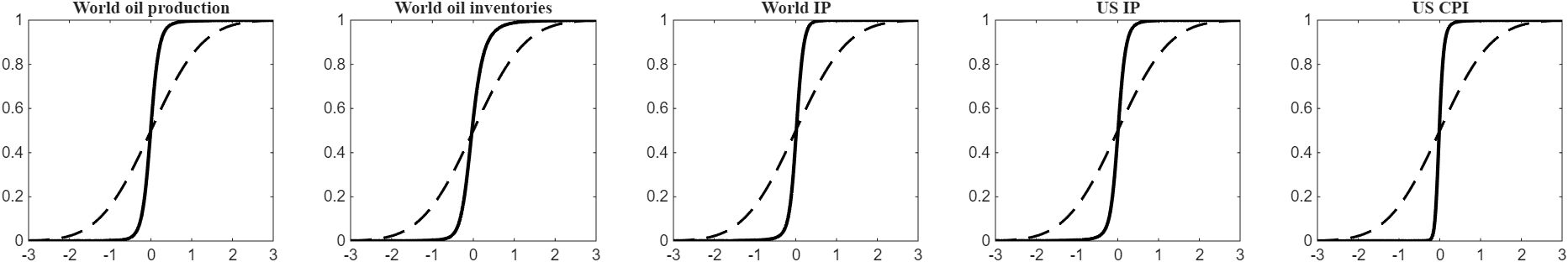}
\par
\vspace{10pt}
%Add vertical space between panels
\par
\includegraphics[width=1.0\textwidth]{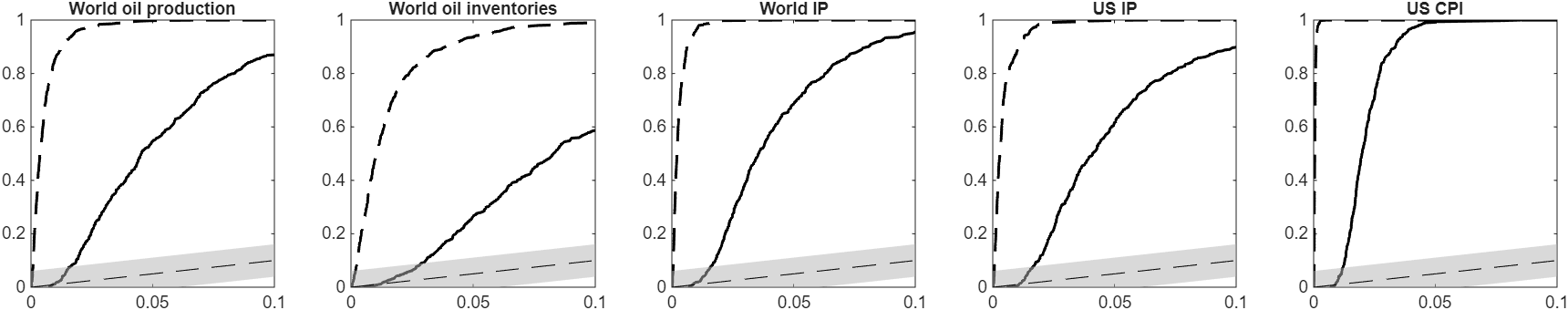}
\par
\vspace{-10pt}\caption{Bootstrap diagnostics. Upper panel: bootstrap
(conditional) cdfs for the on-impact IRFs (solid lines) against the Gaussian
distribution (dashed lines). Lower panel: average fractions of rejections
($\hat{\pi}_{n,m,K}^{\ast}(\eta)$) for nominal levels $\eta\in(0,0.10)$,
$m=10$ (solid lines) and $m=20$ (dashed lines).}%
\label{fig:kanzig_combined}%
\end{figure}

%\begin{figure}[t]
%\makebox[\textwidth][c]{
%\includegraphics[width=1.00\textwidth]{p-values_kanzig.png} } \vspace
%{-20pt}\caption{Average fractions of rejections ($\hat{\pi}_{n,m,K}^{*}%
%(\alpha)$) for nominal levels $\alpha\in(0,1/2)$. Solid line: KS test; dashed
%line: Lilliefors.}%
%\label{figure_p-values_kanzig}%
%\end{figure}

By exploiting K\"{a}nzig's bootstrap computations, we are able to reassess
specification validity of his SVAR-IV model through the lens of bootstrap
diagnostics. In this case, assuming correct specification of the VAR\ model,
the null hypothesis of valid specification corresponds to the instrument
$z_{t}$ being sufficiently strong for the target shock $\varepsilon
_{\text{oil},t}$, a condition that ensures that IRFs are estimated
consistently and standard asymptotic inference is valid.

Using K\"{a}nzig's (2021) \textsf{Matlab} code, we first estimate the $5$
bootstrap distributions $\hat{G}_{n}^{(j)}(\cdot):=\mathbb{P}^{\ast}%
(\hat{\beta}_{j}^{\ast}-\hat{\beta}_{j}\leq\cdot)$, $j=1,\ldots,5$, using
$10,000$ moving block bootstrap replications of the vector $\hat{\beta}^{\ast
}$ (properly standardized, see Section \ref{Sec alternative null hypotheses}).
The $\hat{G}_{n}^{(j)}$'s, reported in the upper panel of
Figure~\ref{fig:kanzig_combined}, are substantially non-Gaussian. For each
$j=1,\ldots,5$, we run $K=500$ tests based on independent samples of
$m\in\{10,20\}$ bootstrap replications; $m$ is deliberately small with respect
to $n$. The lower panel of Figure \ref{fig:kanzig_combined} reports, for
significance levels $\eta\in\left(  0,0.1\right)  $, the average rejection
rates $\hat{\pi}_{n,m,K}^{\ast}(\eta):=K^{-1}\sum_{k=1}^{K}\mathbb{I}%
_{\mathbb{\{}p_{n,m:k}^{\ast}\leq\eta\}}$, where $p_{n,m:k}^{\ast},$
$k=1,\ldots,K$ are the \textit{p}-values associated to the $K$ tests; see
Section~\ref{sec extension diagnostics repo}. Note that due to use of the
moving block bootstrap, these results are robust to the presence of
conditional heteroskedasticity of unknown form in the data (Jentsch and
Lunsford, 2022).

It can be observed that for all components of $\beta$, the null hypothesis is
strongly rejected at any conventional significance level. Larger values of $m$
reinforce this conclusion. This result\footnote{Results are robust to (i) the
choice of the horizon $h$, (ii) the choice of the norm and (iii)\ the
standardization of the bootstrap estimator.} likely indicates that the SVAR-IV
estimation is based on a weak instrument. Interestingly, neither K\"{a}nzig's
(2021) $68\%$ and $90\%$ bootstrap confidence intervals nor the F-tests fully
capture this phenomenon, which the bootstrap diagnostic test detects effectively.

\section{concluding remarks}

\label{Sec conc}

This paper shows that the bootstrap delivers, as a by-product, a versatile
diagnostic procedure for detecting invalid specifications, i.e., violations of
the assumptions underlying standard asymptotic theory. The procedure, which is
based on measures of the discrepancy between the conditional distribution of a
bootstrap statistic and the Gaussian distribution, provides a flexible and
computationally straightforward alternative to traditional misspecification
tests. A key advantage of this approach is that it does not induce pre-testing
bias, thereby improving the reliability of post-test statistical inference.
That is, under the null of valid specification, inference conditional on not
rejecting the bootstrap tests is asymptotically \emph{exact}.

A further major feature of this procedure is its flexibility. On the one hand,
it can be tailored to test the validity of a specific assumption, such as
stationarity of the data or relevance of a set of instrumental variables,
hence allowing to construct tests specifically designed to have power against
alternatives of interest. On the other hand, it can function as a more general
diagnostic test for bootstrap consistency; that is, to test whether the
distribution of a given bootstrap statistic is close to a theoretical Gaussian
limit. This dual capability makes the method broadly applicable, whether
testing for specific irregularities or assessing the validity of bootstrap
procedures in general.

Our paper also contributes to the literature on bootstrap inference. The usual
approach in bootstrap theory is to analyze the asymptotic properties of a
bootstrap cdf, say $\hat{G}_{n}$, as the sample size $n$ diverges. In
practice, however, $\hat{G}_{n}$ is estimated using the edf $\hat{G}%
_{n,m}^{\ast}$ of $m$ (conditionally i.i.d.) realization of the bootstrap
statistic. Therefore, standard bootstrap theory is implicitly based on letting
$m\rightarrow\infty$ first (such that the estimation error $\hat{G}%
_{n,m}^{\ast}-\hat{G}_{n}$ disappears), followed by $n\rightarrow\infty$. Our
paper complements the standard approach by exploring the case where
$n,m\rightarrow\infty$ jointly, rather than sequentially. We show that, under
suitable conditions $\hat{G}_{n,m}^{\ast}$ has an asymptotic distribution
which is known under a wide range of applications and, crucially, becomes
asymptotically independent of the original data. This asymptotic independence
result is key to avoid that the bootstrap test of valid specification distorts
post-test inference.

The results in this paper can be extended in several directions. For instance,
we have not discussed the role of the choice of the norm in determining the
finite sample behavior and the power properties of the proposed tests. Second,
an open question is how to efficiently choose $m$ in practice, such that an
optimal balance of size and power is achieved. Third, all the examples
considered here assume that the reference statistical model is of low
dimension. It is of interest to establish whether our approach could be used
to detect validity of the Gaussian asymptotic approximations in
high-dimensional cases. Fourth, to what extent the properties of the bootstrap
highlighted in this paper are shared by other methods such as the jackknife
(see, e.g., Hansen, 2025), permutation tests (Young, 2019), subsampling
(Politis, Romano and Wolf, 1999)\ or (quasi) Bayesian methods (as in Wang,
2025)\ are open questions. These are left to future research.

\bigskip

\appendix\setcounter{section}{0}

\section*{appendix}

\section{notation and definitions}

\label{Appendix notation}

Throughout this paper, the notation $\sim$ indicates equality in distribution.
We write `$x:=y$' and `$y=:x$' to mean that $x$ is defined by $y$. The
standard Gaussian cdf is denoted by $\Phi$; $%
%TCIMACRO{\TeXButton{U}{\mathscr{U}\!}}%
%BeginExpansion
\mathscr{U}\!%
%EndExpansion
_{[0,1]}$ and $%
%TCIMACRO{\TeXButton{P}{\mathscr{P}\!}}%
%BeginExpansion
\mathscr{P}\!%
%EndExpansion
(\lambda)$ are the uniform distribution on $[0,1]$ and the Poisson
distribution with mean $\lambda$, respectively. $\mathbb{I}_{\{\cdot\}}$ is
the indicator function. The space of c\`{a}dl\`{a}g functions $\mathbb{R}%
\rightarrow\mathbb{R}$ (equipped with its Skorokhod $J_{1}$-topology, unless
otherwise stated; see Kallenberg, 1997, Appendix A2) is denoted by $%
%TCIMACRO{\TeXButton{curly D}{\mathscr{D}}}%
%BeginExpansion
\mathscr{D}%
%EndExpansion
_{\mathbb{R}}$; similarly, $%
%TCIMACRO{\TeXButton{curly D}{\mathscr{D}}}%
%BeginExpansion
\mathscr{D}%
%EndExpansion
_{[0,1]}$ denotes the space of c\`{a}dl\`{a}g functions $[0,1]\rightarrow
\mathbb{R}$. For matrices $a,b,c$ with $n$ rows, $S_{ab}:=a^{\prime}b/n$ and
$S_{ab.c}:=S_{ab}-S_{ac}S_{cc}^{-1}S_{cb}$, assuming that $S_{cc}$ has full rank.

For a bootstrap sequence, say $Y_{n}^{\ast}$, we use $Y_{n}^{\ast}%
\overset{p^{\ast}}{\rightarrow}_{p}0$ or $Y_{n}^{\ast}=o_{p^{\ast}}(1)$ to
mean that, for any $\epsilon>0$, $\mathbb{P}^{\ast}(|Y_{n}^{\ast}%
|>\epsilon)\rightarrow_{p}0$, where $\mathbb{P}^{\ast}$ denotes the
probability measure conditionally on the original data $D_{n}$. We also write
$Y_{n}^{\ast}=O_{p^{\ast}}(1)$ to mean that $\mathbb{P}^{\ast}(|Y_{n}^{\ast
}|>M)\rightarrow_{p}0$ for some large enough $M$. Expectations and variance
under $\mathbb{P}^{\ast}$ are denoted with $\mathbb{E}^{\ast}$ and
$\mathbb{V}^{\ast}$, respectively. We use $Y_{n}^{\ast}\overset{d^{\ast}%
}{\rightarrow}_{p}\xi$ to mean that $\mathbb{E}^{\ast}[g(Y_{n}^{\ast}%
)|D_{n}]\overset{p}{\rightarrow}\mathbb{E[}g(\xi)]$ for all bounded continuous
functions $g:\mathbb{R}\rightarrow\mathbb{R}$. In terms of cdfs, $\hat{G}%
_{n}(u):=\mathbb{P}^{\ast}(Y_{n}^{\ast}\leq u)\rightarrow_{p}G(u):=\mathbb{P}%
(\xi\leq u)$ at all continuity points $u\in\mathbb{R}$ of $G$.

We refer to the fact that $\mathbb{E}[g(Y_{n}^{\ast})|D_{n}]\overset
{w}{\rightarrow}\mathbb{E}[g(\xi)|\mathcal{D}]$ for all bounded continuous
functions $g:\mathbb{R}\rightarrow\mathbb{R}$ (with $(Y_{n}^{\ast},D_{n})$ and
$(\xi,\mathcal{D})$ random elements of metric spaces $\mathbb{R}%
\times\mathcal{S}_{D_{n}}$ and $\mathbb{R}\times\mathcal{S}_{\mathcal{D}}$,
respectively) as `weak convergence in distribution', denoted as `$\overset
{d^{\ast}}{\rightarrow}_{w}$'. For the special case of scalar random variables
$Y_{n}^{\ast}$ and $\xi$, if the conditional distribution $\xi|\mathcal{D}$ is
diffuse (non-atomic), weak convergence in distribution is equivalent to the
weak convergence $\hat{G}_{n}(\cdot):=\mathbb{P}^{\ast}(Y_{n}^{\ast}\leq
\cdot)\overset{w}{\rightarrow}%
%TCIMACRO{\TeXButton{G}{\mathscr{G}\!}}%
%BeginExpansion
\mathscr{G}\!%
%EndExpansion
\mathcal{(\cdot}):=\mathbb{P}(\xi\leq\cdot|\mathcal{D})\ $in
$\mathscr{D}_{\mathbb{R}}$, where $%
%TCIMACRO{\TeXButton{G}{\mathscr{G}\!}}%
%BeginExpansion
\mathscr{G}\!%
%EndExpansion
$ is a random distribution function. For multivariate generalizations, see
Cavaliere and Georgiev (2020, Appendix A). Finally, if $G$ is a (random) cdf,
$G^{-1}$ denotes its right-continuous generalized inverse.

\section{proofs of the main results}

\label{Appendix main proofs}

In the proof of Theorem~\ref{Thm 1}, we make use of the following lemma, which
provides a bound on $a_{n,m}^{\ast}:=%
%TCIMACRO{\TeXButton{T}{\mathscr{T}}}%
%BeginExpansion
\mathscr{T}%
%EndExpansion
_{n,m}^{\ast}-\mathcal{Z}_{n,m}^{\ast}=m^{1/2}\Vert\hat{G}_{n,m}^{\ast}%
-\Phi\Vert-m^{1/2}\Vert\hat{G}_{n,m}^{\ast}-\hat{G}_{n}\Vert$ of
(\ref{eq Z* and a*}). Its proof follows directly from the triangle inequality
for (semi)norms.

\begin{lemma}
\label{Lemma an,b*}With $a_{n,m}^{\ast}$ defined in (\ref{eq Z* and a*}), it
holds that $|a_{n,m}^{\ast}|\leq\sqrt{m}\Vert\hat{G}_{n}-\Phi\Vert$.
\end{lemma}

\medskip

\noindent\textsc{Proof of Theorem~\ref{Thm 1}}. By Lemma~\ref{Lemma an,b*} we
have that $|a_{n,m}^{\ast}|\leq\sqrt{m}\Vert\hat{G}_{n}-\Phi\Vert
=O_{p}(m^{1/2}n^{-\alpha})$ under Assumption~\ref{Assn 1}. The limit of
$\mathcal{Z}_{n,m}^{\ast}$ can be found, e.g., as in Bickel and Freedman
(1981), Theorem 4.1. Specifically, let $\psi_{m}(F)$ denote the law of
$\sqrt{m}(\hat{F}_{m}(\cdot)-F(\cdot))$, where $\hat{F}_{m}$ is the edf of $m$
independent r.v.s with common law $F$; then, in the space of probability
measures on $%
%TCIMACRO{\TeXButton{curly D}{\mathscr{D}}}%
%BeginExpansion
\mathscr{D}%
%EndExpansion
_{\mathbb{R}}$ equipped with the weak topology, $\psi_{m}(F)$ tends to the law
of $W(F)$ as $m\rightarrow\infty$ by a standard invariance principle, where
$W$ is a standard Brownian bridge. Conditionally on $D_{n}$, $\sqrt{m}(\hat
{G}_{n,m}^{\ast}(\cdot)-\hat{G}_{n}(\cdot))$ has law $\psi_{m}(\hat{G}_{n})$;
since $\Vert\hat{G}_{n}-\Phi\Vert_{\infty}\rightarrow_{p}0$, by Proposition
4.1 of Bickel and Freedman (1981) it also holds that $\psi_{m}(\hat{G}_{n})$
becomes arbitrarily close (in probability) to $\psi_{m}(\Phi)$ as
$(n,m\rightarrow\infty)$, where $\psi_{m}(\Phi)$ converges (weakly)\ to the
law of $W(\Phi)$ as $m\rightarrow\infty$. Thus, the law of $\sqrt{m}(\hat
{G}_{n,m}^{\ast}(\cdot)-\hat{G}_{n}(\cdot))$ conditionally on $D_{n}$
converges weakly in probability to the law of $W(\Phi)$ as $(n,m\rightarrow
\infty)$. By the CMT it follows that, conditionally on the data, $\Vert
\sqrt{m}(\hat{G}_{n,m}^{\ast}(\cdot)-\hat{G}_{n}(\cdot))\Vert\overset
{w}{\rightarrow}_{p}\Vert W(\Phi)\Vert$ as $(n,m\rightarrow\infty)$. The
requirement that $m^{1/2}n^{-\alpha}\rightarrow0$ completes the proof.\hfill
$\square$

\medskip

\noindent\textsc{Proof of Theorem~\ref{Thm 2}.} By Skorokhod coupling,
consider a probability space where distributional copies of $\hat{G}_{n}$ and
$%
%TCIMACRO{\TeXButton{G}{\mathscr{G}\!}}%
%BeginExpansion
\mathscr{G}\!%
%EndExpansion
$ are defined (for simplicity, we denote also these copies by $\hat{G}_{n}$
and $%
%TCIMACRO{\TeXButton{G}{\mathscr{G}\!}}%
%BeginExpansion
\mathscr{G}\!%
%EndExpansion
$), such that $\hat{G}_{n}\overset{a.s.}{\rightarrow}%
%TCIMACRO{\TeXButton{G}{\mathscr{G}\!}}%
%BeginExpansion
\mathscr{G}\!%
%EndExpansion
$ in $%
%TCIMACRO{\TeXButton{curly D}{\mathscr{D}}}%
%BeginExpansion
\mathscr{D}%
%EndExpansion
_{\mathbb{R}}$ as $n\rightarrow\infty$. Fix an outcome $\omega$ in the
probability-one event $\{\hat{G}_{n}\rightarrow%
%TCIMACRO{\TeXButton{G}{\mathscr{G}\!}}%
%BeginExpansion
\mathscr{G}\!%
%EndExpansion
\}\cap\{\Vert%
%TCIMACRO{\TeXButton{G}{\mathscr{G}\!}}%
%BeginExpansion
\mathscr{G}\!%
%EndExpansion
-\Phi\Vert>0\}$. For this outcome there exists a sequence of increasing
continuous bijections $\lambda_{n}:[0,1]\rightarrow\lbrack0,1]$ such that
$\Vert\hat{G}_{n}(\omega)\circ\lambda_{n}-%
%TCIMACRO{\TeXButton{G}{\mathscr{G}\!}}%
%BeginExpansion
\mathscr{G}\!%
%EndExpansion
\Vert_{\infty}\rightarrow0$ and $\Vert\lambda_{n}-\mathrm{id}_{[0,1]}%
\Vert_{\infty}\rightarrow0$ as $n\rightarrow\infty$.

Consider further a product extension of the coupling probability space such
that on the added factor space $m$ i.i.d. draws from $\hat{G}_{n}(\omega)$ are
defined, with associated empirical process $\hat{G}_{n,m}^{\ast}(\omega)$, and
also $m$ i.i.d. draws of $%
%TCIMACRO{\TeXButton{G}{\mathscr{G}\!}}%
%BeginExpansion
\mathscr{G}\!%
%EndExpansion
(\omega)$ are defined, with associated empirical process $%
%TCIMACRO{\TeXButton{G}{\mathscr{G}\!}}%
%BeginExpansion
\mathscr{G}\!%
%EndExpansion
_{m}(\omega)$ ($m=1,2,\ldots$). Finally, let $\mathbb{P}^{\ast}$ denote the
probability measure on the added factor space.

As the triangle inequality yields $\Vert\hat{G}_{n,m}^{\ast}(\omega)-\Phi
\Vert\geq\Vert\hat{G}_{n}(\omega)-\Phi\Vert-\Vert\hat{G}_{n,m}^{\ast}%
(\omega)-\hat{G}_{n}(\omega)\Vert$, it holds that
\begin{align*}
\mathbb{P}^{\ast}(%
%TCIMACRO{\TeXButton{T}{\mathscr{T}}}%
%BeginExpansion
\mathscr{T}%
%EndExpansion
_{n,m}^{\ast}(\omega)  &  \geq c):=\mathbb{P}^{\ast}(\Vert\hat{G}_{n,m}^{\ast
}(\omega)-\Phi\Vert\geq cm^{-1/2})\\
&  \geq\mathbb{P}^{\ast}(\Vert\hat{G}_{n}(\omega)-\Phi\Vert-\Vert\hat{G}%
_{n,m}^{\ast}(\omega)-\hat{G}_{n}(\omega)\Vert\geq cm^{-1/2})\\
&  =\mathbb{P}^{\ast}(\Vert\hat{G}_{n,m}^{\ast}(\omega)-\hat{G}_{n}%
(\omega)\Vert\leq\Vert\hat{G}_{n}(\omega)-\Phi\Vert-cm^{-1/2}).
\end{align*}
Here, first, the rhs of the inequality satisfies $\Vert\hat{G}_{n}%
(\omega)-\Phi\Vert-cm^{-1/2}\rightarrow\Vert%
%TCIMACRO{\TeXButton{G}{\mathscr{G}\!}}%
%BeginExpansion
\mathscr{G}\!%
%EndExpansion
(\omega)-\Phi\Vert=:%
%TCIMACRO{\TeXButton{Y}{\mathscr{Y}}}%
%BeginExpansion
\mathscr{Y}%
%EndExpansion
(\omega)>0$ as $(n,m\rightarrow\infty)$, by the continuity of $\Vert\cdot
\Vert$. Second, the lhs satisfies $\Vert\hat{G}_{n,m}^{\ast}(\omega)-\hat
{G}_{n}(\omega)\Vert\overset{\mathbb{P}^{\ast}}{\rightarrow}0$ as
$(n,m\rightarrow\infty)$ by the following argument. On the one hand, as
$(n,m\rightarrow\infty)$, $\hat{G}_{n,m}^{\ast}(\omega)-\hat{G}_{n}%
(\omega)\overset{\mathbb{P}^{\ast}}{\rightarrow}0$ in $%
%TCIMACRO{\TeXButton{curly D}{\mathscr{D}}}%
%BeginExpansion
\mathscr{D}%
%EndExpansion
_{\mathbb{R}}$ if and only if $\hat{G}_{n,m}^{\ast}(\omega)\circ\lambda
_{n}-\hat{G}_{n}(\omega)\circ\lambda_{n}\overset{\mathbb{P}^{\ast}%
}{\rightarrow}0$ in $%
%TCIMACRO{\TeXButton{curly D}{\mathscr{D}}}%
%BeginExpansion
\mathscr{D}%
%EndExpansion
_{\mathbb{R}}$. On the other hand, regarding the latter, it is checked
directly that $\hat{G}_{n,m}^{\ast}(\omega)\circ\lambda_{n}$ is an empirical
process for the cdf $\hat{G}_{n}(\omega)\circ\lambda_{n}$. By Proposition 4.1
of Bickel and Freedman (1981), the convergence $\Vert\hat{G}_{n}(\omega
)\circ\lambda_{n}-%
%TCIMACRO{\TeXButton{G}{\mathscr{G}\!}}%
%BeginExpansion
\mathscr{G}\!%
%EndExpansion
(\omega)\Vert_{\infty}\rightarrow0$ as $n\rightarrow\infty$ implies that the
law of $\sqrt{m}(\hat{G}_{n,m}^{\ast}(\omega)\circ\lambda_{n}-\hat{G}%
_{n}(\omega)\circ\lambda_{n})$, as a probability measure on $%
%TCIMACRO{\TeXButton{curly D}{\mathscr{D}}}%
%BeginExpansion
\mathscr{D}%
%EndExpansion
_{\mathbb{R}}$, becomes arbitrarily close to the law of $\sqrt{m}(%
%TCIMACRO{\TeXButton{G}{\mathscr{G}\!}}%
%BeginExpansion
\mathscr{G}\!%
%EndExpansion
_{m}(\omega)-%
%TCIMACRO{\TeXButton{G}{\mathscr{G}\!}}%
%BeginExpansion
\mathscr{G}\!%
%EndExpansion
(\omega))$ as $(n,m\rightarrow\infty)$. As the latter law converges to the law
of $W(%
%TCIMACRO{\TeXButton{G}{\mathscr{G}\!}}%
%BeginExpansion
\mathscr{G}\!%
%EndExpansion
(\omega))$ as $m\rightarrow\infty$ by a standard invariance principle, so does
the law of $\sqrt{m}(\hat{G}_{n,m}^{\ast}(\omega)\circ\lambda_{n}-\hat{G}%
_{n}(\omega)\circ\lambda_{n})$ as $(n,m\rightarrow\infty)$. By the CMT, this
implies that $\Vert\hat{G}_{n,m}^{\ast}(\omega)\circ\lambda_{n}-\hat{G}%
_{n}(\omega)\circ\lambda_{n}\Vert_{\infty}=O_{\mathbb{P}^{\ast}}(m^{-1/2})$ as
$(n,m\rightarrow\infty)$. Hence, $\hat{G}_{n,m}^{\ast}(\omega)-\hat{G}%
_{n}(\omega)\overset{\mathbb{P}^{\ast}}{\rightarrow}0$ in $%
%TCIMACRO{\TeXButton{curly D}{\mathscr{D}}}%
%BeginExpansion
\mathscr{D}%
%EndExpansion
_{\mathbb{R}}$ as $(n,m\rightarrow\infty)$. By the continuity of $\Vert
\cdot\Vert$, it follows that $\Vert\hat{G}_{n,m}^{\ast}(\omega)-\hat{G}%
_{n}(\omega)\Vert\overset{\mathbb{P}^{\ast}}{\rightarrow}0$ as
$(n,m\rightarrow\infty)$, as asserted previously. Therefore,%
\[
\underset{(n,m\rightarrow\infty)}{\lim\inf}\mathbb{P}^{\ast}(%
%TCIMACRO{\TeXButton{T}{\mathscr{T}}}%
%BeginExpansion
\mathscr{T}%
%EndExpansion
_{n,m}^{\ast}(\omega)\geq c)\geq1.
\]
Since $\omega$ was chosen in an almost certain event, it can be concluded that
$\mathbb{P}^{\ast}(%
%TCIMACRO{\TeXButton{T}{\mathscr{T}}}%
%BeginExpansion
\mathscr{T}%
%EndExpansion
_{n,m}^{\ast}\geq c)\rightarrow1$ a.s. on the coupling probability space.
Therefore, on the original probability space, $\mathbb{P}^{\ast}(%
%TCIMACRO{\TeXButton{T}{\mathscr{T}}}%
%BeginExpansion
\mathscr{T}%
%EndExpansion
_{n,m}^{\ast}\geq c)\overset{p}{\rightarrow}1$.\hfill$\square$

\medskip

\noindent\textsc{Proof of Theorem~\ref{Thm pretest}.} (a) Say $%
%TCIMACRO{\TeXButton{T}{\mathscr{T}}}%
%BeginExpansion
\mathscr{T}%
%EndExpansion
_{n,m}^{\ast}\overset{w}{\rightarrow}_{p}%
%TCIMACRO{\TeXButton{T}{\mathscr{T}}}%
%BeginExpansion
\mathscr{T}%
%EndExpansion
_{\infty}^{\ast}$ as $(n,m\rightarrow\infty)$. By the law of iterated
expectations, it holds that%
\begin{align*}
\left\vert \mathbb{E[}f(D_{n})g(%
%TCIMACRO{\TeXButton{T}{\mathscr{T}}}%
%BeginExpansion
\mathscr{T}%
%EndExpansion
_{n,m}^{\ast})]-\mathbb{E[}f(D_{n})]\mathbb{E[}g(%
%TCIMACRO{\TeXButton{T}{\mathscr{T}}}%
%BeginExpansion
\mathscr{T}%
%EndExpansion
_{n,m}^{\ast})]\right\vert  &  =\left\vert \mathbb{E}\{f(D_{n})[\mathbb{E}%
^{\ast}(g(%
%TCIMACRO{\TeXButton{T}{\mathscr{T}}}%
%BeginExpansion
\mathscr{T}%
%EndExpansion
_{n,m}^{\ast}))-\mathbb{E}(g(%
%TCIMACRO{\TeXButton{T}{\mathscr{T}}}%
%BeginExpansion
\mathscr{T}%
%EndExpansion
_{n,m}^{\ast}))]\}\right\vert \\
&  \leq\Vert f\Vert_{\infty}\mathbb{E}\left\vert \mathbb{E}^{\ast}(g(%
%TCIMACRO{\TeXButton{T}{\mathscr{T}}}%
%BeginExpansion
\mathscr{T}%
%EndExpansion
_{n,m}^{\ast}))-\mathbb{E}(g(%
%TCIMACRO{\TeXButton{T}{\mathscr{T}}}%
%BeginExpansion
\mathscr{T}%
%EndExpansion
_{n,m}^{\ast}))\right\vert .
\end{align*}
It follows that as $(n,m\rightarrow\infty)$%
\begin{align*}
&  \sup\nolimits_{\Vert f\Vert_{\infty}\leq1}\left\vert \mathbb{E[}f(D_{n})g(%
%TCIMACRO{\TeXButton{T}{\mathscr{T}}}%
%BeginExpansion
\mathscr{T}%
%EndExpansion
_{n,m}^{\ast})]-\mathbb{E[}f(D_{n})]\mathbb{E[}g(%
%TCIMACRO{\TeXButton{T}{\mathscr{T}}}%
%BeginExpansion
\mathscr{T}%
%EndExpansion
_{n,m}^{\ast})]\right\vert \\
&  \hspace{2.54cm}\overset{}{\leq}\mathbb{E}\left\vert \mathbb{E}^{\ast}[g(%
%TCIMACRO{\TeXButton{T}{\mathscr{T}}}%
%BeginExpansion
\mathscr{T}%
%EndExpansion
_{n,m}^{\ast})]-\mathbb{E}[g(%
%TCIMACRO{\TeXButton{T}{\mathscr{T}}}%
%BeginExpansion
\mathscr{T}%
%EndExpansion
_{\infty}^{\ast})]\right\vert +|\mathbb{E}[g(%
%TCIMACRO{\TeXButton{T}{\mathscr{T}}}%
%BeginExpansion
\mathscr{T}%
%EndExpansion
_{n,m}^{\ast})]-\mathbb{E}[g(%
%TCIMACRO{\TeXButton{T}{\mathscr{T}}}%
%BeginExpansion
\mathscr{T}%
%EndExpansion
_{\infty}^{\ast})]|\rightarrow0
\end{align*}
because $\mathbb{E}^{\ast}[g(%
%TCIMACRO{\TeXButton{T}{\mathscr{T}}}%
%BeginExpansion
\mathscr{T}%
%EndExpansion
_{n,m}^{\ast})]\overset{p}{\rightarrow}\mathbb{E[}g(%
%TCIMACRO{\TeXButton{T}{\mathscr{T}}}%
%BeginExpansion
\mathscr{T}%
%EndExpansion
_{\infty}^{\ast})]$ and the dominated convergence theorem applies.

\noindent(b)\ It is sufficient to establish that%
\begin{equation}
\sup_{A_{n}\in\sigma(D_{n})}|\mathbb{P}(A_{n}\cap\{%
%TCIMACRO{\TeXButton{T}{\mathscr{T}}}%
%BeginExpansion
\mathscr{T}%
%EndExpansion
_{n,m}^{\ast}\in B\})-\mathbb{P}(A_{n})\mathbb{P}(%
%TCIMACRO{\TeXButton{T}{\mathscr{T}}}%
%BeginExpansion
\mathscr{T}%
%EndExpansion
_{n,m}^{\ast}\in B)|\rightarrow0 \label{eq independence}%
\end{equation}
as $(n,m\rightarrow\infty)$. This can be done using part (a) and
approximations of indicator functions with continuous functions, or directly
as done below. Then, as $\mathbb{P}^{\ast}(%
%TCIMACRO{\TeXButton{T}{\mathscr{T}}}%
%BeginExpansion
\mathscr{T}%
%EndExpansion
_{n,m}^{\ast}\in B)\overset{p}{\rightarrow}\mathbb{P}(%
%TCIMACRO{\TeXButton{T}{\mathscr{T}}}%
%BeginExpansion
\mathscr{T}%
%EndExpansion
_{\infty}^{\ast}\in B)=:\mathbb{P}_{\infty}(B)$ for the $\mathbb{P}_{\infty}%
$-continuity set $B$, also $\mathbb{P}(%
%TCIMACRO{\TeXButton{T}{\mathscr{T}}}%
%BeginExpansion
\mathscr{T}%
%EndExpansion
_{n,m}^{\ast}\in B)\rightarrow\mathbb{P}_{\infty}(B)>0$ by dominated
convergence. Thus, division by $\mathbb{P}(%
%TCIMACRO{\TeXButton{T}{\mathscr{T}}}%
%BeginExpansion
\mathscr{T}%
%EndExpansion
_{n,m}^{\ast}\in B)$, which is bounded away from zero under the assumption
that $\mathbb{P}_{\infty}(B)>0$, completes the proof.

We turn to (\ref{eq independence}). By the law of iterated expectations (and
the fact that $\mathbb{P}\left(  X\in\mathcal{E}\right)  =\mathbb{E}%
(\mathbb{I}_{\{X\in\mathcal{E}\}})$), it holds that%
\begin{align*}
&  \left\vert \mathbb{P}(A_{n}\cap\{%
%TCIMACRO{\TeXButton{T}{\mathscr{T}}}%
%BeginExpansion
\mathscr{T}%
%EndExpansion
_{n,m}^{\ast}\overset{}{\in}B\})-\mathbb{P}(A_{n})\mathbb{P}(%
%TCIMACRO{\TeXButton{T}{\mathscr{T}}}%
%BeginExpansion
\mathscr{T}%
%EndExpansion
_{n,m}^{\ast}\in B)\right\vert \\
&  \hspace{0.5in}\overset{}{=}\left\vert \mathbb{E}[\mathbb{E}^{\ast
}(\mathbb{I}_{A_{n}}\mathbb{I}_{\{%
%TCIMACRO{\TeXButton{T}{\mathscr{T}}}%
%BeginExpansion
\mathscr{T}%
%EndExpansion
_{n,m}^{\ast}\in B\}})]-\mathbb{E}[\mathbb{I}_{A_{n}}\mathbb{E}(\mathbb{I}_{\{%
%TCIMACRO{\TeXButton{T}{\mathscr{T}}}%
%BeginExpansion
\mathscr{T}%
%EndExpansion
_{n,m}^{\ast}\in B\}})]\right\vert \\
&  \hspace{0.5in}\overset{}{=}\left\vert \mathbb{E\{I}_{A_{n}}[\mathbb{E}%
^{\ast}(\mathbb{I}_{\{%
%TCIMACRO{\TeXButton{T}{\mathscr{T}}}%
%BeginExpansion
\mathscr{T}%
%EndExpansion
_{n,m}^{\ast}\in B\}})-\mathbb{E}(\mathbb{I}_{\{%
%TCIMACRO{\TeXButton{T}{\mathscr{T}}}%
%BeginExpansion
\mathscr{T}%
%EndExpansion
_{n,m}^{\ast}\in B\}})]\}\right\vert \\
&  \hspace{0.5in}\overset{}{\leq}\mathbb{E}\left\vert \mathbb{P}^{\ast}(%
%TCIMACRO{\TeXButton{T}{\mathscr{T}}}%
%BeginExpansion
\mathscr{T}%
%EndExpansion
_{n,m}^{\ast}\in B)-\mathbb{P}(%
%TCIMACRO{\TeXButton{T}{\mathscr{T}}}%
%BeginExpansion
\mathscr{T}%
%EndExpansion
_{n,m}^{\ast}\in B)\right\vert .
\end{align*}
It follows that%
\begin{align*}
&  \sup_{A_{n}\in\sigma(D_{n})}\left\vert \mathbb{P}(A_{n}\cap\{%
%TCIMACRO{\TeXButton{T}{\mathscr{T}}}%
%BeginExpansion
\mathscr{T}%
%EndExpansion
_{n,m}^{\ast}\in B\})-\mathbb{P}(A_{n})\mathbb{P}(%
%TCIMACRO{\TeXButton{T}{\mathscr{T}}}%
%BeginExpansion
\mathscr{T}%
%EndExpansion
_{n,m}^{\ast}\in B)\right\vert \\
&  \hspace{0.5in}\overset{}{\leq}\mathbb{E}\left\vert \mathbb{P}^{\ast}(%
%TCIMACRO{\TeXButton{T}{\mathscr{T}}}%
%BeginExpansion
\mathscr{T}%
%EndExpansion
_{n,m}^{\ast}\in B)-\mathbb{P}_{\infty}(B)\right\vert +|\mathbb{P}(%
%TCIMACRO{\TeXButton{T}{\mathscr{T}}}%
%BeginExpansion
\mathscr{T}%
%EndExpansion
_{n,m}^{\ast}\in B)-\mathbb{P}_{\infty}(B)|\rightarrow0
\end{align*}
as $(n,m\rightarrow\infty)$, because $\mathbb{P}^{\ast}(%
%TCIMACRO{\TeXButton{T}{\mathscr{T}}}%
%BeginExpansion
\mathscr{T}%
%EndExpansion
_{n,m}^{\ast}\in B)\overset{p}{\rightarrow}\mathbb{P}_{\infty}(B)$ and
dominated convergence applies.\hfill$\square$

\medskip

\noindent\textsc{Proof of Proposition~\ref{prop diagnostics}}. First note that
$\hat{\pi}_{n,m,K}^{\ast}(\cdot)$ with $\cdot\in\mathbb{R}$ is an empirical
process. Then, the proof is analogous to that of Theorem 14.3 in Billingsley
(1999). Specifically, first note that, if $\xi_{n,m:k}^{\ast}$, $k=1,\ldots
K$, are i.i.d. uniform conditionally on the data, and $\hat{\xi}_{n,m}^{\ast}$
is their empirical process, then the result $\sqrt{K}(\hat{\xi}_{n,m,K}^{\ast
}(\cdot)-(\cdot))\overset{w^{\ast}}{\rightarrow}_{p}W(\cdot)$ on $%
%TCIMACRO{\TeXButton{curly D}{\mathscr{D}}}%
%BeginExpansion
\mathscr{D}%
%EndExpansion
[0,1]$ as $(n,m,K\rightarrow\infty)$ is standard. Second, the representation
$\sqrt{K}(\hat{\pi}_{n,m,K}^{\ast}(\cdot)-(\cdot))=\sqrt{K}(\hat{\xi}%
_{n,m,K}^{\ast}(\Pi_{n,m}^{\ast}(\cdot))-(\cdot))$ can be used, where
$\Pi_{n,m}^{\ast}$ is the cdf of $p_{n,m}^{\ast}$ conditionally on the data.
That is, $\sqrt{K}(\hat{\pi}_{n,m,K}^{\ast}(\cdot)-(\cdot))$ can be
represented as the composition $\sqrt{K}(\hat{\xi}_{n,m,K}^{\ast}%
(\cdot)-(\cdot))$ after $\Pi_{n,m}^{\ast}(\cdot)$. Since the limiting law of
$\sqrt{K}(\hat{\xi}_{n,m,K}^{\ast}(\cdot)-(\cdot))$ is that of the continuous
process $W$ and since $\Pi_{n,m}^{\ast}(\cdot)$ converges to the identity
function in probability, by the CMT $\sqrt{K}(\hat{\pi}_{n,m,K}^{\ast}%
(\cdot)-(\cdot))$ converges to the composition $W(\cdot)$ after $id(\cdot)$,
which is $W(\cdot)$.\hfill$\square$

\section{additional proofs and results}

\label{Appendix additional proofs}

\noindent\textsc{Results in Section~\ref{Sec Example IV reprise}}. Write%
\[
T_{n}^{\ast}=\frac{\hat{\omega}_{n}^{-1}}{\left(  \hat{\pi}_{n}+S_{zv^{\ast}%
}\right)  ^{\prime}\left(  \hat{\pi}_{n}+S_{zv^{\ast}}\right)  }(\sqrt
{n\hat{\omega}_{n}}\hat{\pi}_{n}^{\prime}S_{zu^{\ast}}+\sqrt{n\hat{\omega}%
_{n}}S_{zv^{\ast}}^{\prime}S_{zu^{\ast}}).
\]
Here the term $\sqrt{n\hat{\omega}_{n}}\hat{\pi}_{n}^{\prime}S_{zu^{\ast}}$ is
$%
%TCIMACRO{\TeXButton{N}{\mathscr{N}\!}}%
%BeginExpansion
\mathscr{N}\!%
%EndExpansion
(0,1)$ conditionally on $D_{n}$. This fact and the inequality%
\begin{align*}
\left\vert \mathbb{P}\left(  \zeta(\xi+\eta)\leq x\right)  -\mathbb{P}(\xi\leq
x)\right\vert  &  \leq\mathbb{P}(\tfrac{x}{1-a\text{ }\mathrm{sgn}%
(x)}+b)-\mathbb{P}(\tfrac{x}{1+a\text{ }\mathrm{sgn}(x)}-b)\\
&  +\mathbb{P}\left(  |\zeta-1|\geq a\right)  +\mathbb{P}\left(  |\eta|\geq
b\right)  ,
\end{align*}
which holds for any r.v.'s $\xi,\eta,\zeta$ and any $a\in(0,1)$ and $b>0$,
yield%
\begin{align}
|\hat{G}_{n}(x)-\Phi(x)|  &  \leq\Phi(\tfrac{x}{1-a\text{ }\mathrm{sgn}%
(x)}+b)-\Phi(\tfrac{x}{1+a\text{ }\mathrm{sgn}(x)}-b)\label{tail}\\
&  +\mathbb{P}^{\ast}\left(  \left\vert \frac{\hat{\omega}_{n}^{-1}}{\left(
\hat{\pi}_{n}+S_{zv^{\ast}}\right)  ^{\prime}\left(  \hat{\pi}_{n}%
+S_{zv^{\ast}}\right)  }-1\right\vert \geq a\right)  +\mathbb{P}^{\ast}%
(\sqrt{n\hat{\omega}_{n}}|S_{zv^{\ast}}^{\prime}S_{zu^{\ast}}|\geq
b).\nonumber
\end{align}
Let $a\in(0,\frac{1}{2})$ in the following. Then, by the mean-value theorem
and the boundedness of $\left\vert x\right\vert \Phi^{\prime}(x)$ on
$\mathbb{R}$, it holds that%
\[
\Phi\left(  \tfrac{x}{1-a\mathrm{sgn}(x)}+b\right)  -\Phi\left(  \tfrac
{x}{1+a\mathrm{sgn}(x)}-b\right)  \leq C\left(  a+b\right)
\]
for some universal constant $C>0$. As $\left(  \hat{\pi}_{n}+S_{zv^{\ast}%
}\right)  ^{\prime}\left(  \hat{\pi}_{n}+S_{zv^{\ast}}\right)  \overset
{p}{\rightarrow}\pi^{\prime}\pi>0$, the event in the first tail probability in
(\ref{tail}) equals, with probability approaching one, the complement of the
event $\left\{  -a(1+a)^{-1}<\hat{\omega}_{n}(2\hat{\pi}_{n}^{\prime
}S_{zv^{\ast}}+S_{zv^{\ast}}^{\prime}S_{zv^{\ast}})<a(1-a)^{-1}\right\}  $,
such that the tail probability of interest does not exceed%
\begin{align*}
\mathbb{P}^{\ast}\left(  \hat{\omega}_{n}|2\hat{\pi}_{n}^{\prime}S_{zv^{\ast}%
}+S_{zv^{\ast}}^{\prime}S_{zv^{\ast}}|\geq2a\right)   &  \leq\mathbb{P}^{\ast
}(|2\hat{\pi}_{n}^{\prime}S_{zv^{\ast}}|\geq a\hat{\omega}_{n}^{-1}%
)+\mathbb{P}^{\ast}\left(  S_{zv^{\ast}}^{\prime}S_{zv^{\ast}}\geq
a\hat{\omega}_{n}^{-1}\right) \\
&  =\bar{\Phi}(\sqrt{n\hat{\omega}_{n}^{-1}}\tfrac{a}{2})+\bar{\Psi}_{k}%
(n\hat{\omega}_{n}^{-1}a),
\end{align*}
the last line because $\hat{\pi}_{n}^{\prime}S_{zv^{\ast}}\sim N(0,n^{-1}%
\hat{\omega}_{n}^{-1})$ and $nS_{zv^{\ast}}^{\prime}S_{zv^{\ast}}\sim\chi
^{2}(k)$; here $\bar{\Phi}=1-\Phi$ and $\bar{\Psi}_{k}$ denote resp. the
complementary cdfs of the $N(0,1)$ and the $\chi^{2}(k)$ distributions. To
discuss the second tail probability in (\ref{tail}), by using the
decomposition $v_{i}^{\ast}=\rho_{uv}u_{i}^{\ast}+\sqrt{1-\rho_{uv}^{2}%
}\varepsilon_{i}^{\ast}$ with $(u_{i}^{\ast},\varepsilon_{i}^{\ast})\sim
$i.i.d.$N(0,I_{2})$, it is found that $S_{zv^{\ast}}^{\prime}S_{zu^{\ast}%
}=\rho_{uv}S_{zu^{\ast}}^{\prime}S_{zu^{\ast}}+\sqrt{1-\rho_{uv}^{2}%
}S_{z\varepsilon^{\ast}}^{\prime}S_{zu^{\ast}}$ and%
\begin{align*}
\mathbb{P}^{\ast}\left(  \sqrt{n\hat{\omega}_{n}}|S_{zv^{\ast}}^{\prime
}S_{zu^{\ast}}|\geq b\right)   &  \leq\mathbb{P}^{\ast}\left(  \sqrt
{n\hat{\omega}_{n}}|\rho_{uv}|S_{zu^{\ast}}^{\prime}S_{zu^{\ast}}\geq\tfrac
{b}{2}\right) \\
&  +\mathbb{P}^{\ast}\left(  \sqrt{n\hat{\omega}_{n}(1-\rho_{uv}^{2}%
)}|S_{z\varepsilon^{\ast}}^{\prime}S_{zu^{\ast}}|\geq\tfrac{b}{2}\right) \\
&  \leq\bar{\Psi}_{k}\left(  \sqrt{n\hat{\omega}_{n}^{-1}}\tfrac{b}%
{2|\rho_{uv}|}\right)  +k\bar{\Pi}\left(  \sqrt{n\hat{\omega}_{n}^{-1}%
(1-\rho_{uv}^{2})^{-1}}\tfrac{b}{2k}\right)  ,
\end{align*}
the last line because $nS_{zu^{\ast}}^{\prime}S_{zu^{\ast}}\sim\chi^{2}(k)$
and $nS_{z\varepsilon^{\ast}}^{\prime}S_{zu^{\ast}}\sim\xi^{\prime}\eta$ with
$(\xi^{\prime},\eta^{\prime})^{\prime}\sim N(0,I_{2k})$;\ here $\bar{\Pi}$
stands for the complementary cdf of $\xi_{1}\eta_{1}$. Recapitulating, with
probability approaching one, it holds that%
\begin{align*}
\Vert\hat{G}_{n}-\Phi\Vert_{\infty}  &  \leq C(a+b)+\bar{\Phi}\left(
\sqrt{n\hat{\omega}_{n}^{-1}}\tfrac{a}{2}\right)  +\bar{\Psi}_{k}(n\hat
{\omega}_{n}^{-1}a)+\\
&  +\bar{\Psi}_{k}\left(  \sqrt{n\hat{\omega}_{n}^{-1}}\tfrac{b}{2|\rho_{uv}%
|}\right)  -k\bar{\Pi}\left(  \sqrt{n\hat{\omega}_{n}^{-1}(1-\rho_{uv}%
^{2})^{-1}}\tfrac{b}{2k}\right)
\end{align*}
for any $a\in(0,\frac{1}{2})$ and $b>0$. The rate of decay of $\bar{\Phi}$,
$\bar{\Psi}_{k}$ and $\bar{\Pi}$ is well known:%
\[
\bar{\Phi}(x)\sim Cx^{-1}e^{-x^{2}/2},\bar{\Psi}_{k}(x)\sim Cx^{k/2}%
e^{-x/2},\bar{\Pi}(x)\sim x^{-1/2}e^{-x}%
\]
as $x\rightarrow\infty$, each one with its own $C$; see eq. (2) in Leipus,
Siaulys, Dirma and Zove (2023) for the case of $\bar{\Pi}$. By choosing
$a=b=\epsilon n^{-1/2}\ln n$ with $\epsilon>0$ as small as convenient, it is
seen that $\Vert\hat{G}_{n}-\Phi\Vert_{\infty}=O_{p}(n^{-\alpha})$ for any
$\alpha\in(0,\frac{1}{2})$.\hfill$\square$\medskip

\noindent\textsc{Results in Section~\ref{Sec Example Jaco}}. Let first
$\dot{g}\neq0$ be fixed. Say that a sequence of possibly conditional cdf's
$F_{n}$ is square-root consistent if $\Vert F_{n}-\Phi\Vert_{\infty}%
=O_{p}(n^{-1/2})$. Interest is in establishing the fact that $\hat{G}%
_{n}(g^{\prime}(\hat{\theta}_{n})\hat{\sigma}_{n}\cdot)$ is square-root
consistent under the assumption that the conditional cdf's $\hat{F}_{n}$ of
$Z_{n}^{\ast}:=\sqrt{n}(\hat{\theta}_{n}^{\ast}-\hat{\theta}_{n})/\hat{\sigma
}_{n}$ do so.

Recall that $T_{n}^{\ast}=g^{\prime}(\hat{\theta}_{n})\hat{\sigma}_{n}%
Z_{n}^{\ast}+\frac{1}{2}g^{\prime\prime}(\hat{\theta}_{n})n^{-1/2}\hat{\sigma
}_{n}^{2}Z_{n}^{\ast2}$. Notice that the conditional cdf's of $Z_{n}^{\ast}$
and $-Z_{n}^{\ast}$ either both are, or both are not square-root consistent.
As square-root consistency of for the conditional cdf's of $Z_{n}^{\ast}$ is
the only property of $Z_{n}^{\ast}$ used in this proof, it follows that
$Z_{n}^{\ast}$ and $-Z_{n}^{\ast}$ are exchangeable for the purposes of the
proof. Hence, possibly upon replacing $Z_{n}^{\ast}$ and $g^{\prime}%
(\hat{\theta}_{n})$ by respectively $Z_{n}^{\ast}(1-2\mathbb{I}_{\{g^{\prime
}(\hat{\theta}_{n})<0\}})$ and $g^{\prime}(\hat{\theta}_{n})(1-2\mathbb{I}%
_{\{g^{\prime}(\hat{\theta}_{n})<0\}})$ in the expression for $T_{n}^{\ast}$,
it is legitimate to discuss $T_{n}^{\ast}$ under the assumption $g^{\prime
}(\hat{\theta}_{n})\geq0$. Under this assumption, and given that
$\mathbb{P}(g^{\prime}(\theta_{n})=0)\rightarrow0$, it holds that%
\[
\hat{G}_{n}(g^{\prime}(\hat{\theta}_{n})\hat{\sigma}_{n}x)=\mathbb{P}^{\ast
}(Z_{n}^{\ast}+\tfrac{1}{2}n^{-1/2}\gamma_{n}Z_{n}^{\ast2}\leq x)
\]
with probability approaching one, where $\gamma_{n}:=g^{\prime\prime}%
(\hat{\theta}_{n})\hat{\sigma}_{n}/g^{\prime}(\hat{\theta}_{n})$. Next, the
conditional cdf's of $\pm(Z_{n}^{\ast}+\frac{1}{2}n^{-1/2}\gamma_{n}%
Z_{n}^{\ast2})$ either both are, or both are not square-root consistent, and
since $Z_{n}^{\ast}$ and $-Z_{n}^{\ast}$ were said to be exchangeable for the
purposes of the proof, there is no loss of generality in imposing also
$\gamma_{n}\geq0$. With the signs imposed, it holds that%
\[
\hat{G}_{n}(g^{\prime}(\hat{\theta}_{n})\hat{\sigma}_{n}x)=\hat{F}%
_{n}(x)\mathbb{I}_{\{\gamma_{n}=0\}}+\mathbb{P}^{\ast}(\hat{q}_{-}(x)\leq
Z_{n}^{\ast}\leq\hat{q}_{+}(x))\mathbb{I}_{A_{n}}%
\]
with probability approaching one, where $A_{n}:=\{x\geq-\sqrt{n}\gamma
_{n}^{-1}/2,\gamma_{n}\neq0\}$ and%
\[
\hat{q}_{\pm}(x):=n^{1/2}\gamma_{n}^{-1}(\pm\sqrt{1+2n^{-1/2}x\gamma_{n}}-1).
\]
From the assumption that the conditional cdf's $\hat{F}_{n}$ of $Z_{n}^{\ast}$
are square-root consistent, it is further found that%
\begin{align*}
\hat{G}_{n}(g^{\prime}(\hat{\theta}_{n})\hat{\sigma}_{n}x)  &  =\hat{F}%
_{n}(x)\mathbb{I}_{\{\gamma_{n}=0\}}+(\hat{F}_{n}(\hat{q}_{+}(x))-\hat{F}%
_{n}(\hat{q}_{-}(x)))\mathbb{I}_{A_{n}}\\
&  =\Phi(x)\mathbb{I}_{\{\gamma_{n}=0\}}+\left(  \Phi(\hat{q}_{+}%
(x))-\Phi(\hat{q}_{-}(x))\right)  \mathbb{I}_{A_{n}}+O_{p}(n^{-1/2})
\end{align*}
uniformly in $x$. Turn now to%
\[
\Vert\hat{G}_{n}(g^{\prime}(\hat{\theta}_{n})\hat{\sigma}_{n}\cdot)-\Phi
\Vert_{\infty}\leq\sup_{x\in\mathbb{R}}|[\hat{G}(g^{\prime}(\hat{\theta}%
_{n})\hat{\sigma}_{n}x)-\Phi(x)]\mathbb{I}_{A_{n}}|+\sup_{x\in\mathbb{R}}%
|\Phi(x)\mathbb{I}_{A_{n}^{c}}\mathbb{I}_{\{\gamma_{n}\neq0\}}|.
\]
As $\hat{\theta}_{n}\overset{p}{\rightarrow}\theta_{0}$ and $g^{\prime\prime
}(\theta_{n})\ $(thus, $\gamma_{n}$) is bounded, it follows that
$\mathbb{P}(A_{n}^{c}\cap\{\gamma_{n}\neq0\})\rightarrow0$ and the previous
upper bound equals, with probability approaching one,%
\begin{align*}
&  \sup_{x\in\mathbb{R}}|[\hat{G}(g^{\prime}(\hat{\theta}_{n})\hat{\sigma}%
_{n}x)-\Phi(x)]\mathbb{I}_{A_{n}}|\overset{}{=}\sup_{x\in\mathbb{R}}%
|[\Phi(\hat{q}_{+}(x))-\Phi(\hat{q}_{-}(x))-\Phi(x)]\mathbb{I}_{A_{n}}|\\
&  \hspace{2cm}\overset{}{\leq}\sup_{x\in\mathbb{R}}|\Phi(\hat{q}%
_{-}(x))\mathbb{I}_{A_{n}}|+\sup_{4\gamma_{n}|x|\leq n^{1/2}}|[\Phi(\hat
{q}_{+}(x))-\Phi(x)]\mathbb{I}_{A_{n}}|\\
&  \hspace{2cm}+\sup_{4\gamma_{n}|x|>n^{1/2}}|[\Phi(\hat{q}_{+}(x))-\Phi
(x)]\mathbb{I}_{A_{n}}|.
\end{align*}
Here, by the monotonicity of $\Phi$,
\[
\sup_{x\in\mathbb{R}}|\Phi(\hat{q}_{-}(x))\mathbb{I}_{A_{n}}|\leq
\mathbb{I}_{A_{n}}\Phi(-n^{1/2}\gamma_{n}^{-1})|=O_{p}(n^{-1/2})
\]
because $\gamma_{n}=O_{p}(1)$. Further, for outcomes in $A_{n}$, $\left\vert
\hat{q}_{+}(x)-x\right\vert \leq2n^{-1/2}\gamma_{n}x^{2}$ by simple algebra.
Hence, by the mean-value theorem,%
\[
\left\vert \Phi(\hat{q}_{+}(x))-\Phi(x)\right\vert \leq2n^{-1/2}\gamma
_{n}x^{2}\Phi^{\prime}\left(  x+\zeta_{n}[\hat{q}_{+}(x)-x]\right)
\]
for some $\zeta_{n}\in\lbrack0,1]$, such that%
\begin{align*}
&  \sup_{4\gamma_{n}|x|\leq n^{1/2}}|[\Phi(\hat{q}_{+}(x))-\Phi(x)]\mathbb{I}%
_{A_{n}}|\overset{}{\leq}2n^{-1/2}\gamma_{n}\sup_{x\in\mathbb{R}}\left[
x^{2}\max_{2|y|\leq|x|}\Phi^{\prime}\left(  x+y\right)  \right] \\
&  \hspace{2cm}\overset{}{=}2n^{-1/2}\gamma_{n}\sup_{x\geq0}\left[  x^{2}%
\Phi^{\prime}\left(  \tfrac{x}{2}\right)  \right]  \overset{}{=}O_{p}%
(n^{-1/2})
\end{align*}
by using the exponential functional form of $\Phi^{\prime}$. Finally, by the
monotonicity of $\Phi$,%
\begin{align*}
&  \sup_{4\gamma_{n}|x|>n^{1/2}}|[\Phi(\hat{q}_{+}(x))-\Phi(x)]\mathbb{I}%
_{A_{n}}|\overset{}{\leq}\mathbb{I}_{\{\gamma_{n}\neq0\}}[\Phi(-\tfrac{1}%
{4}n^{1/2}\gamma_{n}^{-1})+\Phi(-\hat{q}_{+}(-\tfrac{1}{4}n^{1/2}\gamma
_{n}^{-1}))]\\
&  \hspace{2cm}\overset{}{\leq}\mathbb{I}_{\{\gamma_{n}\neq0\}}[\Phi
(-\tfrac{1}{4}n^{1/2}\gamma_{n}^{-1})+\Phi(-n^{1/2}\gamma_{n}^{-1})]\overset
{}{=}O_{p}(n^{-1/2}).
\end{align*}
By combining the previous results, $\Vert\hat{G}_{n}(g^{\prime}(\hat{\theta
}_{n})\hat{\sigma}_{n}\cdot)-\Phi\Vert_{\infty}=O_{p}(n^{-1/2})$.

Next, let instead $g^{\prime}(\theta_{0})=\lambda n^{1/2}$. Then $(\sqrt
{n}g^{\prime}(\hat{\theta}_{n}),\hat{G}_{n})\overset{w}{\rightarrow}(\ell,%
%TCIMACRO{\TeXButton{G}{\mathscr{G}\!}}%
%BeginExpansion
\mathscr{G}\!%
%EndExpansion
)$ on $\mathbb{R}\times%
%TCIMACRO{\TeXButton{curly D}{\mathscr{D}}}%
%BeginExpansion
\mathscr{D}%
%EndExpansion
_{\mathbb{R}}$ as shown in Section \textsc{\ref{Sec Example Jaco}}. Further,
introducing $\hat{\varsigma}_{n}:=g^{\prime}(\hat{\theta}_{n})\hat{\sigma}%
_{n}$, it holds that%

\[
L_{n}\leq\Vert\hat{G}_{n}(\hat{\varsigma}_{n}\cdot)-\Phi\Vert_{\infty}\leq
U_{n}%
\]
with%
\begin{align*}
L_{n}\overset{}{:=}  &  |\hat{G}_{n}(n\hat{\varsigma}_{n})-\Phi(n)|\mathbb{I}%
_{\{\hat{\varsigma}_{n}<0\}}\\
&  +\max\{|\hat{G}_{n}(\sqrt[3]{n}\hat{\varsigma}_{n})-\Phi(\sqrt[3]%
{n})|,|\hat{G}_{n}(-\sqrt[3]{n}\hat{\varsigma}_{n})-\Phi(-\sqrt[3]%
{n})\}\mathbb{I}_{\{\hat{\varsigma}_{n}\geq0\}}%
\end{align*}
obtained by evaluating $\hat{G}_{n}(\hat{\varsigma}_{n}\cdot)-\Phi$ at $n$ and
$\pm\sqrt[3]{n}$, and%
\[
U_{n}:=\mathbb{I}_{\{\hat{\varsigma}_{n}<0\}}+\max\{\hat{G}_{n}(0),1-\hat
{G}_{n}(0)\}\mathbb{I}_{\{\hat{\varsigma}_{n}\geq0\}}%
\]
obtained by considerations of monotonicity. As the sample paths of $%
%TCIMACRO{\TeXButton{G}{\mathscr{G}\!}}%
%BeginExpansion
\mathscr{G}\!%
%EndExpansion
$ are a.s. continuous and $\mathbb{P}(\ell=0)=0$, it follows from the CMT that%
\[
(U_{n},U_{n}-L_{n})\overset{d}{\rightarrow}(\mathbb{I}_{\{\ell<0\}}+\max\{%
%TCIMACRO{\TeXButton{G}{\mathscr{G}\!}}%
%BeginExpansion
\mathscr{G}\!%
%EndExpansion
(0),1-%
%TCIMACRO{\TeXButton{G}{\mathscr{G}\!}}%
%BeginExpansion
\mathscr{G}\!%
%EndExpansion
(0)\}\mathbb{I}_{\{\ell\geq0\}},0)=(1-%
%TCIMACRO{\TeXButton{G}{\mathscr{G}\!}}%
%BeginExpansion
\mathscr{G}\!%
%EndExpansion
(0)\mathbb{I}_{\{\ell\geq0\}},0)
\]
because $%
%TCIMACRO{\TeXButton{G}{\mathscr{G}\!}}%
%BeginExpansion
\mathscr{G}\!%
%EndExpansion
(0)<\frac{1}{2}$ a.s. From here the limit of $\Vert\hat{G}_{n}(\hat{\varsigma
}_{n}\cdot)-\Phi\Vert_{\infty}$ is obtained.

To see that the consistency conclusion of Theorem~\ref{Thm 2} remains valid,
some modifications are due in its proof. First, $\omega$ is taken as an
outcome in the almost certain event $\{\hat{G}_{n}\rightarrow%
%TCIMACRO{\TeXButton{G}{\mathscr{G}\!}}%
%BeginExpansion
\mathscr{G}\!%
%EndExpansion
\}\cap\{%
%TCIMACRO{\TeXButton{G}{\mathscr{G}\!}}%
%BeginExpansion
\mathscr{G}\!%
%EndExpansion
(0)\mathbb{I}_{\{\ell\geq0\}}\neq1\}$. Second,
\begin{align*}
\mathbb{P}^{\ast}  &  (%
%TCIMACRO{\TeXButton{T}{\mathscr{T}}}%
%BeginExpansion
\mathscr{T}%
%EndExpansion
_{n,m}^{\ast}(\omega)\geq c):=\mathbb{P}^{\ast}(\Vert\hat{G}_{n,m}^{\ast
}(\omega,\hat{\varsigma}_{n}(\omega)\cdot)-\Phi\Vert_{\infty}\geq cm^{-1/2})\\
&  \geq\mathbb{P}^{\ast}(\Vert\hat{G}_{n,m}^{\ast}(\omega,\hat{\varsigma}%
_{n}(\omega)\cdot)-\hat{G}_{n}(\omega,\hat{\varsigma}_{n}(\omega)\cdot
)\Vert_{\infty}\leq\Vert\hat{G}_{n}(\omega,\hat{\varsigma}_{n}(\omega
)\cdot)-\Phi\Vert_{\infty}-cm^{-1/2}).
\end{align*}
Here%
\begin{align*}
\Vert\hat{G}_{n,m}^{\ast}(\omega,\hat{\varsigma}_{n}(\omega)\cdot)-\hat{G}%
_{n}(\omega,\hat{\varsigma}_{n}(\omega)\cdot)\Vert_{\infty}  &  =|\hat
{G}_{n,m}^{\ast}(\omega,0)-\hat{G}_{n}(\omega,0)|\mathbb{I}_{\{\hat{\varsigma
}_{n}(\omega)=0\}}\\
&  +\Vert\hat{G}_{n,m}^{\ast}(\omega,\cdot)-\hat{G}_{n}(\omega,\cdot
)\Vert_{\infty}\mathbb{I}_{\{\hat{\varsigma}_{n}(\omega)\neq0\}}\\
&  \leq\Vert\hat{G}_{n,m}^{\ast}(\omega,\cdot)-\hat{G}_{n}(\omega,\cdot
)\Vert_{\infty}\overset{\mathbb{P}^{\ast}}{\rightarrow}0
\end{align*}
as $(n,m\rightarrow\infty)$, by the proof of Theorem~\ref{Thm 2}, whereas
$\Vert\hat{G}_{n}(\omega,\hat{\varsigma}_{n}(\omega)\cdot)-\Phi\Vert_{\infty
}-cm^{-1/2}\rightarrow1-%
%TCIMACRO{\TeXButton{G}{\mathscr{G}\!}}%
%BeginExpansion
\mathscr{G}\!%
%EndExpansion
(\omega,0)\mathbb{I}_{\{\ell(\omega)\geq0\}}>0$ as $(n,m\rightarrow\infty)$.
Hence, the conclusion. \hfill$\square$

\section*{references}

\begin{description}
\item \textsc{Andrews, D.W.K. }(2000). Inconsistency of the bootstrap when a
parameter is on the boundary of the parameter space. \emph{Econometrica}, 68, 399--405.

\item \textsc{Andrews, D.W.K. and M. Buchinsky }(2000). A three-step method
for choosing the number of bootstrap repetitions. \emph{Econometrica }67, 23--51.

\item \textsc{Andrews, D.W.K. and P. Guggenberger} (2009). Hybrid and
size-corrected subsampling methods. \emph{Econometrica }77, 721--762.

\item \textsc{Andrews, I. and J. Shapiro }(2025). Communicating scientific
uncertainty via approximate posteriors. NBER\ Working Paper 32038.

\item \textsc{Andrews, I., J.H. Stock and L. Sun }(2019). Weak instruments in
instrumental variable regression: theory and practice. \emph{Annual Review of
Economics }11, 727--753.

\item \textsc{Angelini, G., G. Cavaliere and L. Fanelli} (2022). Bootstrap
inference and diagnostics in state space models: with applications to dynamic
macro models. \emph{Journal of Applied Econometrics} 37, 3--22.

\item \textsc{Angelini, G., G. Cavaliere and L. Fanelli} (2024). An
identification and testing strategy for proxy-SVARs with weak proxies.
\emph{Journal of Econometrics} 238, 105604.

\item \textsc{Athreya, K.B. }(1987). Bootstrap of the mean in the infinite
variance case, \emph{Annals of Statistics }15, 724--731.

\item \textsc{B\aa rdsen, G. and L. Fanelli} (2015). Frequentist evaluation of
small DSGE models. \emph{Journal of Business \& Economic Statistics} 33, 307--322.

\item \textsc{Basawa, I.V., A.K. Mallik, W.P. McCormick, J.H. Reeves, and R.L.
Taylor }(1991). Bootstrapping unstable first-order autoregressive processes,
\emph{Annals of Statistics} 19, 1098--1101.

\item \textsc{Beran, R. }(1997). Diagnosing bootstrap success, \emph{Annals of
the Institute of Statistical Mathematics }49, 1--24.

\item \textsc{Bickel, P.J. and D.A. Freedman }(1981). Some asymptotic theory
for the bootstrap. \emph{Annals of Statistics }9, 1196--1217.

\item \textsc{Billingsley, P}. (1999). \emph{Convergence of probability
measures}. 2nd edition, Wiley: NY.

\item \textsc{Bound, J., A. Jaeger and R.M. Baker }(1995). Problems with
instrumental variables estimation when the correlation between the instruments
and the endogenous explanatory variable is weak. \emph{Journal of the American
Statistical Association }90, 443--450.

\item \textsc{Bose, A. (1988).} Edgeworth correction by bootstrap in
autoregressions. \emph{Annals of Statistics} 16, 1709--1722.

\item \textsc{Cavaliere, G., and I.\ Georgiev }(2020). Inference under random
limit bootstrap measures. \emph{Econometrica} 88, 2547--2574.

\item \textsc{Cavaliere, G., I.\ Georgiev and A.M.R. Taylor }(2013). Wild
bootstrap of the mean in the infinite variance case. \emph{Econometric
Reviews} 32, 204--219.

\item \textsc{Cavaliere, G., I.\ Georgiev and A.M.R. Taylor }(2016).
Sieve-based inference for infinite-variance linear processes. \emph{Annals of
Statistics }44, 1467--1494.

\item \textsc{Cavaliere, G., H.B. Nielsen and A.\ Rahbek }(2015). Bootstrap
testing of hypotheses on co-integration relations in vector autoregressive
models. \emph{Econometrica},\emph{\ }83, 813--831.

\item \textsc{Datta, S. }(1995). On a modified bootstrap for certain
asymptotically non-normal statistics. \emph{Statistics and Probability Letters
}24, 91--98.

\item \textsc{Davidson, R. }(2017). Diagnostics for the bootstrap and fast
double bootstrap. \emph{Econometric Reviews} 36, 1021--1038.

\item \textsc{Davidson, R. and J.G. Mackinnon }(2010). Wild bootstrap tests
for IV regression. \emph{Journal of Business \& Economic Statistics} 28, 128--144.

\item \textsc{de Chaisemartin, C. and X. D'Haultfoeuille }(2024). Under the
null of valid specification, pre-tests cannot make post-test inference
liberal. \textsc{arXiv}:2407.03725; R\&R \emph{American Economic Review.}

\item \textsc{Feller, W.} (1971). \emph{An introduction to probability theory
and its applications}, Vol. 2, 2nd edition, Wiley: NY.

\item \textsc{Gertler, M. and P. Karadi} (2015). Monetary policy surprices,
credit costs, and economic activity. \emph{American Economic Journal:
Macroeconomics} 7, 44-76.

\item \textsc{Hall, P. }(1992). \emph{The Bootstrap and Edgeworth Expansion},
Springer-Verlag: Berlin.

\item \textsc{Hahn, J. and Z. Liao }(2021). Bootstrap standard error estimates
and inference. \emph{Econometrica} 89, 1963--1977.

\item \textsc{Han, S. and A. McCloskey }(2019). Estimation and inference with
a (nearly) singular Jacobian. \emph{Quantitative Economics }10, 1019--1068.

\item \textsc{Hansen, B.E. }(2025). Jackknife Standard Errors for Clustered
Regression. Under revision at the \emph{Review of Economic Studies}.

\item \textsc{Jentsch, C. and K.C. Lunsford} (2019). The dynamic effects of
personal and corporate income tax changes in the United States: Comment.
\emph{American Economic Review }109, 2655--2678.

\item \textsc{Jentsch, C. and K.C. Lunsford} (2022). Asymptotic valid
bootstrap inference for Proxy SVARs. \emph{Journal of Business and Economic
Statistics }40, 1876--1891.

\item \textsc{K\"{a}nzig, D.R. }(2021). The macroeconomic effects of oil
supply news: evidence from OPEC announcements. \emph{American Economic Review
}111, 1092-1125.

\item \textsc{Kallenberg, O. }(1997). \emph{Foundations of modern
probability}. Springer:\ NY.

\item \textsc{Knight, K. }(1989). On the bootstrap of the sample mean in the
infinite variance case. \emph{Annals of Statistics} 17, 1168--1175.

\item \textsc{Leipus R., J. Siaulys, M. Dirma and R. Zove} (2023). On the
distribution-tail behaviour of the product of normal random variables.
\emph{Journal of Inequalities and Applications}, 32

\item \textsc{Lepage, R., M. Woodroofe and J. Zinn }(1981). Convergence to a
stable distribution via order statistics. \emph{Annals of Probability }9, 624--632.

\item \textsc{Montiel Olea, J.L., J.H. Stock, and M.W. Watson} (2021).
Inference in SVARs identified with an external instrument. \emph{Journal of
Econometrics} 225, 74-87.

\item \textsc{Moreira, M.J., J.R. Porter and G.A. Suarez }(2009). Bootstrap
validity for the score test when instruments may be weak. \emph{Journal of
Econometrics} 149, 52--64.

\item \textsc{Newey, W.K. and D. McFadden} (1994). Large sample estimation and
hypothesis testing. In R.F. Engle and D.L. McFadden (eds.), \emph{Handbook of
Econometrics }IV, Chap. 36.

\item \textsc{Phillips, P.C.B.} (1987). Towards a unified asymptotic theory
for autoregression. \emph{Biometrika }74, 535--547\textsc{.}

\item \textsc{Phillips, P.C.B. and H.R. Moon} (1999). Linear regression limit
theory for nonstationary panel data. \emph{Econometrica }67, 1057--1111.

\item \textsc{Politis, D.N., J.P. Romano, and M. Wolf} (1999).
\emph{Subsampling}. Springer: NY.

\item \textsc{Singh, K. }(1981). On the asymptotic accuracy of Efron's
bootstrap. \emph{Annals of Statistics }9, 1187--1195.

\item \textsc{Staiger, D. and J.H. Stock} (1997). Instrumental variables
regression with weak instruments. \emph{Econometrica} 65, 557--586.

\item \textsc{Stock, J.H., and M.W. Watson }(2018). Identification and
estimation of dynamic causal effects in macroeconomics using external
instruments. \emph{Economic Journal} 128, 917--948.

\item \textsc{Stock, J.H. and M. Yogo} (2005). Testing for weak instruments in
linear IV regression. In \emph{Identification and Inference for Econometric
Models: Essays in Honor of Thomas Rothenberg}, eds. D.W.K. Andrews, J.H.
Stock, pp. 80--108. Cambridge, UK: Cambridge University Press.

\item \textsc{Wang, A.} (2025). A general diagnostic of the normal
approximation in GMM models. \emph{Econometrics Journal} 28, 261--275.

\item \textsc{Young, A. }(2019). Channeling Fisher: randomization tests and
the statistical insignificance of seemingly significant experimental results.
\emph{Quarterly Journal of Economics} 134, 557--598.

\item \textsc{Zhan, Z.} (2018). Detecting weak identification by bootstrap.
Accessed on July 1, 2025. \textsf{https://sites.google.com/site/zhaoguozhan/}
\end{description}

\end{document}